%% file: new_theories.tex
\documentclass[12pt]{article}

\pdfoutput=1
\usepackage{ifpdf}
\usepackage[a4paper,top=3.2cm,width=17cm, bottom=3.2cm, left=2cm]{geometry}
\usepackage{graphicx}
\usepackage{microtype}
\usepackage[utf8x]{inputenc}
\usepackage[T1]{fontenc}
\usepackage{ae}
\usepackage{aecompl}
\usepackage{amsmath}
\usepackage{amssymb}
\usepackage{mathdots}
\usepackage{setspace}
\def\bfseries{\fontseries \bfdefault \selectfont \boldmath}

\usepackage[english]{babel}
\usepackage{datetime}
\shortdate 
\usepackage{bm}
\usepackage{subfigure}
\usepackage[colorlinks=true,linktocpage=true,linkcolor=blue,citecolor=blue]{hyperref}

\usepackage[normalem]{ulem} 
\usepackage{soul} 
\usepackage{nicefrac}
\usepackage{float}
\usepackage[titletoc,title]{appendix}
\usepackage{authblk}
\usepackage{cite}

\numberwithin{equation}{section}

\usepackage{titlesec}

\titleformat*{\section}{\large\bfseries}
\titleformat*{\subsection}{\bfseries}
\titleformat*{\subsubsection}{\bfseries}

\newcommand{\be}{\begin{eqnarray*}}
\newcommand{\ee}{\end{eqnarray*}}

\newcommand{\beq}{\begin{eqnarray}}
\newcommand{\eeq}{\end{eqnarray}}

\newcommand{\bel}[1]{\begin{eqnarray}\label{#1}}
\newcommand{\eel}{\end{eqnarray}}

\newcommand{\beal}[1]{\bel{#1}}
\newcommand{\eea}{\eel}

\newcommand{\rf}[1]{Eq.~(\ref{#1})}
\newcommand{\rfn}[1]{~(\ref{#1})}
\newcommand{\rfs}[1]{Sec.~\ref{#1}}
\newcommand{\rff}[1]{Fig.~\ref{#1}}
\newcommand{\rfc}[1]{Ref.~\cite{#1}}

\newcommand{\nn}{\nonumber}

\newcommand{\p}{\partial}

\newcommand{\D}{{\mathcal D}}

\newcommand{\f}[2]{\frac{#1}{#2}}

\newcommand{\half}{\f{1}{2}}

\def\vx{{\bf x}}
\def\vy{{\bf y}}
\def\vp{{\bf p}}
\def\vk{{\bf k}}
\def\kdotx{{{\bf k} \cdot {\bf x}}}
\def\pdotU{{p \cdot U}}
\def\pdotZ{{p \cdot Z}}

\def\AH{\small \hbox{AHYDRO }}

\newcommand{\sym}{${\mathcal N}=4$}
\newcommand{\symm}{${\mathcal N}=4$ SYM}

\newcommand{\ed}{{\cal E}}       
\newcommand{\sd}{{\cal S}}       
\newcommand{\peq}{{\cal P}}     
\newcommand{\pT}{{\cal P}_T}   
\newcommand{\pL}{{\cal P}_L}   
\newcommand{\cR}{{\cal R}}       
\newcommand{\pa}{{\cal A}}
\newcommand{\pac}{{a}}  

\def\umu{U^\mu}                          
\def\umul{U_\mu}
\def\unu{U^\nu}
\def\unul{U_\nu}
\def\zmu{Z^\mu}                           
\def\znu{Z^\nu}

\newcommand{\WA}{\mathcal{A}}

\def\pt{p_T}                                  
\def\pl{p_L}                                  

\def\gmunu{g^{\mu\nu}}

\def\Dmunu{\Delta^{\mu\nu}}

\def\Tmunu{T^{\mu \nu}}

\def\TmunuEQ{T^{\mu\nu}_{\rm EQ}}
\def\Tmunueq{T^{\mu\nu}_{\rm eq}}

\def\Tmunua{T_{\rm a}^{\mu \nu}}

\def\Pimunu{{\Pi^{\mu \nu}}}
\def\pimunu{{\pi^{\mu \nu}}}
\def\pimunul{{\pi_{\mu \nu}}}
\def\pinumu{{\pi^{\nu \mu}}}
\def\pimumu{\pi^\mu_{\,\,\,\mu}}
\def\pimunut{{\tilde \pi}^{\mu\nu}}

\def\Pit{{\tilde \Pi}}

\def\sigmamunu{{\sigma^{\mu \nu}}}

\def\ximunu{{\xi^{\mu \nu}}}

\def\ximumu{\xi^\mu_{\,\,\,\mu}}
\def\Ximunu{\Xi^{\mu\nu}}
\def\Ximunul{\Xi_{\mu\nu}}

\def\fa{f_{\rm a}}
\def\fRS{f_{\rm RS}}

\def\trelm{\tau_{\mathrm rel}}
\def\trel{\tau_{\mathrm rel}}
\newcommand{\tpi}{\tau_{\pi}}

\newcommand{\bort}{\mathcal{B}}
\newcommand{\borta}{\tilde{\mathcal{B}}}

\begin{document}

\title{{\bf New theories of relativistic hydrodynamics\\ in the LHC era}}

\renewcommand\Authfont{\scshape\small}
\renewcommand\Affilfont{\itshape\small}

\author[1,2,3]{Wojciech Florkowski\thanks{wojciech.florkowski@ifj.edu.pl}}

\author[4,5]{Michal P. Heller\thanks{michal.p.heller@aei.mpg.de}}

\author[5,6]{Micha\l{} Spali\'nski\thanks{m.spalinski@uwb.edu.pl}}

\affil[1]{Institute of Nuclear Physics, Polish Academy of Sciences, PL-31-342 Krakow, Poland}
\affil[2]{Jan Kochanowski University, PL-25-406 Kielce, Poland}
\affil[3]{ExtreMe Matter Institute EMMI, GSI, D-64291 Darmstadt, Germany}
\affil[4]{Max Planck Institute for Gravitational Physics
D-14476 Potsdam-Golm, Germany}
\affil[5]{National Center for Nuclear Research, PL-00-681 Warsaw, Poland}
\affil[6]{Physics Department, University of Bia\l{}ystok, PL-15-245 Bia\l{}ystok, Poland}

\date{}
\maketitle
\thispagestyle{empty}

\begin{abstract}
The success of relativistic hydrodynamics as an essential part of the phenomenological description of
heavy-ion  collisions at RHIC and  the LHC has motivated a significant body of theoretical work concerning its
fundamental aspects. Our review presents these developments from the perspective of the underlying microscopic
physics, using the language of quantum field theory, relativistic kinetic theory, and holography. We discuss
the gradient expansion, the phenomenon of hydrodynamization, as well as several models of hydrodynamic
evolution equations, highlighting the interplay between collective long-lived and transient modes in
relativistic matter. Our aim to provide a unified presentation of this vast subject -- which is naturally
expressed in diverse mathematical languages -- has also led us to include several new results on the
large-order behaviour of the hydrodynamic gradient~expansion.
\end{abstract}

\vspace{5pc}

\newpage

\tableofcontents

\newpage

\input sect-intro.tex

\input sect-urhic.tex

\input sect-micro.tex

\input sect-adscft.tex

\input sect-hydrod.tex

\input sect-fund.tex

\input sect-eff.tex

\input sect-kint.tex

\input sect-resu.tex

\input sect-out.tex

\bigskip

\section*{Acknowledgements}
We would like to thank our collaborators and peers for numerous discussions on
the topics covered in this review. We are especially grateful to Saso
Grozdanov and Nikolaos Kaplis for providing us with plots for
Fig.~\ref{fig:qnmskGB} and bringing several useful insights. We would also
like to thank Ines Aniceto for pointing out that in Fig.~\ref{fig.borelNeq4}
one can see traces of higher singularities. We would like to thank Viktor
Svensson for proof-reading the manuscript and valuable comments.  We are also
very grateful to Bill Zajc for many insightful remarks and suggestions which
have helped us to improve the clarity of the original manuscript.
M.P.H. would like to thank participants of the \emph{Canterbury Tales of Hot
  QFTs in the LHC Era} workshop held at University of Oxford in July 2017 for
stimulating comments that helped to improve the first version of this
manuscript. W.F. was supported by in part by the ExtreMe Matter Institute EMMI
at the GSI Helmholtzzentrum f\"ur Schwerionenforschung, Darmstadt,
Germany. M.P.H. acknowledges support from the Alexander von Humboldt
Foundation and the Federal Ministry for Education and Research through the
Sofja Kovalevskaja Award. M.S. was supported by the National Science Centre
Grant No. 2015/19/B/ST2/02824.

\newpage

\input sect-app.tex

\newpage

\bibliography{hydro_review}{}
\bibliographystyle{utphys}

\end{document}

%% file: sect-intro.tex
\section{Introduction and Outline}
\label{sect:intro}

The heavy-ion collision program pursued in recent years at the Relativistic Heavy Ion Collider
(RHIC)\footnote{For the list of acronyms, symbols, and notation conventions used in this work see
Secs.~\ref{app:acro}, \ref{app:nota}, and \ref{app:conv}.}  and the Large Hadron Collider (LHC) explores
properties of nuclear matter under extreme conditions, close to those existing shortly after the Big Bang. The
theoretical interpretation of its results requires a wide variety of methods which are needed to bridge the
gap between the fundamental theory of the strong interactions in the form of Quantum Chromodynamics (QCD) and
the experimentally accessible observables.

It was established at RHIC and confirmed at the LHC that the nuclear matter produced in heavy-ion collisions at
ultrarelativistic energies exhibits clear signatures of collective behaviour. They are interpreted as experimental
evidence for the creation of strongly-coupled Quark-Gluon Plasma (QGP), an equilibrium phase of QCD formed of deconfined
quarks and gluons. The successful phenomenological description of collective behaviour in the soft observables sector is
based on relativistic hydrodynamics~\cite{LLfluid} with a small viscosity to entropy density ratio, with initial
conditions set very early, perhaps as soon as a fraction of fermi/c after the collision.

The unfolding of this story throughout the last 15 years or so has led to a great deal of progress in the theoretical aspects of
relativistic hydrodynamics. This period constitutes a veritable golden age for this discipline. Despite many
excellent review articles \cite{CasalderreySolana:2011us,DeWolfe:2013cua,Romatschke:2009im,Jeon:2015dfa,Jaiswal:2016hex}
and books~\cite{Yagi:2005yb,Vogt:2007zz,Florkowski:2010zz,rezzolla2013relativistic} devoted to these developments, we believe that
there is a need for a systematic presentation of new ideas in the approach to relativistic hydrodynamics.

The key novelty of our present review is to recognize at the outset that hydrodynamic behaviour is a property
of the underlying, microscopic descriptions of physical systems evolving toward equilibrium. This behaviour is
captured by the truncated gradient expansion of the expectation value of the energy-momentum tensor, and
possibly other conserved currents. The role of hydrodynamics is to mimic this behaviour at the level of a
classical effective theory. The goal in this respect is to seek formulations of hydrodynamics which
incorporate such degrees of freedom and such dynamics that they capture in the best way critical aspects of
the evolution toward equilibrium. An important point is that, in general, such an effective description is not
{\em derived} from a microscopic theory. Rather, it is posited in accordance with very general principles such as
symmetry and conservation laws and is then reconciled with the underlying microscopic model by matching the
gradient expansion at the hydrodynamic level with the gradient expansion at the microscopic level.

The inspiration for this attitude is the effective field theory paradigm which dominates our thinking in
high-energy physics. An important milestone was the formulation of the Baier, Romatschke, Son, Starinets,
Stephanov (BRSSS) theory in 2007~\cite{Baier:2007ix}, which guarantees that the hydrodynamic description can
be matched with any microscopic dynamics up to second order in the gradient expansion. Since then, the
importance of effects  which govern the applicability of hydrodynamics has come to be appreciated. This
development has led to the realization that the hydrodynamic description can be accurate under much more
extreme conditions than earlier seemed reasonable~\cite{Chesler:2009cy,Chesler:2010bi,Heller:2011ju}. In particular, from this perspective it is not outrageous to apply models of relativistic hydrodynamics even for
very anisotropic, inhomogeneous or small systems naturally created in collisions. It has in fact puzzled
practitioners for a while that hydrodynamics can be used successfully in conditions which cannot plausibly be viewed as close even to local equilibrium. This has led to usage of the term ``hydrodynamization'' to mean the onset of the regime where a hydrodynamic description is useful. An extensive recent discussion of this and its phenomenological consequences can be found in \rfc{Romatschke:2016hle}.

Our goal is to review attempts to formulate new hydrodynamic theories which try to capture effects at the edge
of what would traditionally be considered the domain of applicability of hydrodynamics. Important questions
here concern the role of higher order terms in the gradient expansion~\cite{Hiscock:1985zz,Lublinsky:2007mm,
Denicol:2011fa,Heller:2013fn,Heller:2015dha}, as well as the role of collective degrees of freedom not
explicitly included in hydrodynamics and their relation to causality~\cite{Baier:2007ix,Heller:2014wfa}. We
discuss approaches to formulating effective hydrodynamic descriptions taking inspiration from kinetic theory,
quantum field theory and string theory.

Historically, methods based on the AdS/CFT correspondence, or gauge-gravity duality, more generally referred
to as holography, have played an important role in understanding many of the points discussed in this article.
On the other hand, most of these ideas can also be understood without reference to string theory. In
particular, many essential aspects of the story can also be seen from the perspective of kinetic theory, even
though some are more complex in that setting. We aim at presenting a unified picture which makes it easier to see similarities and differences between
different microscopic frameworks in the context of applicability of hydrodynamics and hydrodynamic theories.
Because of this, and also because there are many excellent specialized reviews, we have refrained from a
comprehensive presentation of each individual framework or aspect. This has led to necessary omissions.
Certainly, among the most important ones are:  the anomalous transport phenomena reviewed in
Ref.~\cite{Landsteiner:2016led}, progress on understanding the entropy production constraint in the
hydrodynamic gradient expansion~\cite{Haehl:2015pja}, the question of transport in the vicinity of a critical
point (see, e.g., Ref.~\cite{Stephanov:2017wlw}), detailed analysis of the effects of conformal symmetry
breaking in hydrodynamics and beyond (see, e.g.,
Refs.~\cite{Buchel:2008uu,Buchel:2015saa,Janik:2015waa,Ishii:2015gia,Janik:2016btb,Attems:2016tby,Attems:2017zam}),
the issue of thermal fluctuations in hydrodynamics (see, e.g.,
Ref.~\cite{Kovtun:2011np,Kapusta:2011gt,Akamatsu:2016llw}), entropy production by horizons in holography (see,
e.g., Refs.~\cite{Bhattacharyya:2008xc,Figueras:2009iu,Booth:2010kr,Booth:2011qy}) and holographic collisions
(see Ref.~\cite{Chesler:2015fpa} for a comprehensive picture of early developments and
Ref.~\cite{Chesler:2016ceu} for a state-of-the-art presentation).

This review is structured as follows. We begin with an overview of the theoretical challenges raised by the
heavy-ion collisions programme in \rfs{sect:urhic}. In \rfs{sect:micro} we use linear response theory to
introduce the notion of a mode of equilibrium plasma and describe how modes manifest themselves in quantum
field theory and kinetic theory. We introduce basic kinetic theory notions and signal the
importance of string theory methods for the development of the field. In \rfs{sect:adscft} we
describe the relevant details on how holography works and demonstrate the connection between quasinormal modes
of black branes and modes of strongly-coupled plasmas. Certainly, these developments  significantly influenced
the way we shaped our presentation in the preceding section. The main point made in Secs.~\ref{sect:micro}
and~\ref{sect:adscft} is the idea of a separation between imaginary parts of frequencies of long-lived
(hydrodynamic) modes and transient (non-hydrodynamic) modes, which makes the former dominate the late time
dynamics. This leads up to \rfs{sect:hydrod}, which is devoted to a detailed presentation of the application
of microscopic frameworks to the case of Bjorken flow, which simplifies the problem immensely while keeping
many essential features. We review the results of holographic calculations which have led to the idea of
hydrodynamization, that is, the emergence of quasi-universal features in the late-time behaviour at times
when, superficially, the system is still far from local equilibrium. We also review subsequent results leading
to similar conclusions obtained in the framework of kinetic theory. In \rfs{sect:fund} we turn to the issue of
finding an effective description of late time behaviour in the language of relativistic hydrodynamics. We
emphasise the need to maintain causality, which leads to the appearance of  non-hydrodynamic modes also at the
level of the effective theory. \rfs{sect:effective} reviews the notion of the gradient expansion in
hydrodynamics and the idea that it may be used to match microscopic calculations. This is the central point
which we emphasise in this review: hydrodynamic theories are engineered to match aspects of calculations in
microscopic theories. We discuss this notion in detail for the case of M\"uller-Israel-Stewart (MIS) theory and its generalizations which attempt to capture some features of early
time behaviour. This line of reasoning is continued in \rfs{sect:kint}, which discusses how models of kinetic
theory can be used as a guide in constructing hydrodynamic theories. An important example is the case of
anisotropic hydrodynamics which is presented in detail. The question of large order behaviour of gradient
expansions is reviewed in \rfs{sect:resu}, where we explain how the gradient series encodes information about
both the hydrodynamic and non-hydrodynamic sectors at the level of fundamental theories as well as for their
hydrodynamic descriptions. We close with an outlook in \rfs{sect:outlook}.

This review contains also some new results which have not been published earlier, but significantly complement or
improve earlier  presentations of the reviewed material. Among the ones that we would like to highlight are a unified
presentation of holographic and RTA kinetic theory calculations of the gradient expansion at large orders in
\rfs{sect:resu:micro} and the demonstration of the presence of a subleading transient mode in the Borel transform of the
gradient expansion in holography in Fig.~\ref{fig.borelNeq4}.

%% file: sect-urhic.tex
\section{Motivation: ultrarelativistic heavy-ion collisions}
\label{sect:urhic}

\subsection{Overview}

The physical system which is the subject of this review is a lump of hot, dense, strongly interacting matter consisting of quarks and gluons.
This type of matter existed in the Early Universe but at about 10 microseconds after the Big Bang, due to the expansion of the Universe and cooling, it transformed itself into hadrons. Similar physical  conditions are now realized in Earth laboratories by colliding heavy atomic nuclei at the highest available energies.

The first experiments with ultra-relativistic heavy ions (i.e., with energies exceeding 10 GeV per nucleon in the projectile beam) took place at BNL and at CERN in 1986. In 2000 the first data from RHIC at BNL was  analysed. The LHC at CERN  completed its first heavy-ion running period in the years 2010--2013. At the moment, the second run is taking place (2015--2018), while the third one is planned for the years 2021--2023.

Of particular importance in the heavy-ion program are experimental searches for theoretically predicted new phases of
matter, the description of transitions between such phases (deconfinement, chiral symmetry restoration) and, ultimately,
a possible reconstruction of the entire phase diagram of strongly interacting matter in a wide range of thermodynamic
parameters such as temperature and baryon chemical potential. In this context, new experiments done at lower energies
and with different colliding systems are also very  important, as this allows us to study the beam energy and baryon
number density dependence of many aspects of particle production.

\subsection{Phenomenology: the standard model of heavy ion collisions}
\label{sect:standardmod}

The current understanding of heavy-ion collisions typically separates their evolution into three stages: i) the initial or pre-equilibrium stage, presumably dominated by gluons;  ii) the hydrodynamic stage, in which the dynamics can be successfully described by relativistic viscous hydrodynamics and where the phase transition back to hadronic matter takes place; iii) the freeze-out stage where hadrons form a gas, first dense and then dilute, and in the end, the final state particles are created.  Different physical processes play a role as the system evolves and it is one of the challenges in the field to identify the dominant effects at each stage. From the theoretical point of view, the fact that the collision process can be modeled as a sequence of distinct stages is attractive, since it allows for independent modifications and/or improvements in the theoretical description of each stage. For example, different versions of hydrodynamics can be used for the second stage of evolution, such as switching from perfect-fluid hydrodynamics to viscous hydrodynamics.

\subsubsection{The initial stage}

The very early stages of heavy-ion collisions are most often described with the help of microscopic models which refer
to the presence of coherent colour fields at the moment when the two nuclei pass through each other. Such models refer
directly to QCD and to the phenomenon of gluon saturation, which allows for an effective treatment of gluons in terms of
classical fields obeying Yang-Mills (YM) equations~\cite{McLerran:1993ni,McLerran:1993ka}. When quantum effects are
incorporated, this framework is generally known as the Color-Glass-Condensate (CGC) model. An alternative to QCD-based
models, which are now being intensively studied (see for example~\cite{Gelis:2014qga}),are approaches based on the
AdS/CFT
correspondence, which will be widely discussed later in this review.

Any microscopic model of the early stages requires assuming certain initial conditions which usually refer to the initial distribution of matter in the colliding nuclei. Such geometric concepts are very often introduced in the framework of the Glauber model where the nucleon distributions in nuclei are random and given by the nuclear density profiles, whereas the elementary nucleon-nucleon collision is characterized by the total inelastic cross section $\sigma_{\rm in}$. The Glauber model allows for introducing the concepts of participants or wounded nucleons (nucleons that at least once interacted inelastically) and of binary nucleon-nucleon collisions~\cite{Bialas:1976ed}. Densities of the numbers of wounded nucleons and binary collisions (in the transverse plane with respect to the beam axis) serve to make the estimates of the initial energy-density profiles of the colliding system.

Early applications of relativistic hydrodynamics to model the RHIC data very often used the Glauber model estimates
 as a direct input for the subsequent hydrodynamic stage. With the use of perfect-fluid codes, this approach means
that one assumes (implicitly) local thermalisation of matter at the moment of initialisation of
hydrodynamic evolution. Since a successful description of the data was achieved with initialisation times on
the order of a fraction of fermi/c, the conclusion drawn from these calculations was that matter
produced in heavy-ion collisions undergoes a process of very fast thermalisation~\cite{Heinz:2004pj}.
Note that, as we discuss in \rfs{sect:hydrod}, at the moment of writing our review this
conclusion is being questioned~\cite{Romatschke:2016hle}.

\subsubsection{The hydrodynamic stage}

The presence of a hydrodynamic stage with a low shear viscosity to entropy density ratio in the space-time
evolution of matter produced in heavy-ion collisions is crucial for the explanation of several physical effects,
including the elliptic-flow phenomenon~\cite{Romatschke:2007mq}. In this case, the hydrodynamic expansion
explains the momentum anisotropy of the final-state hadrons, which turns out to be the hydrodynamic response to an
initial spatial anisotropy of matter.

An attractive feature of using the hydrodynamic approach is that it easily and consistently incorporates the phase transition into a global picture of the collisions. The phase transition is included directly by the use of a specific equation of state (EOS). Different forms of EOS can be used in model calculations and one can check which one leads to the best description of the data. Interestingly, such studies support the lattice QCD EOS indicating the presence of a crossover phase transition at finite temperature and zero baryon chemical potential~\cite{Borsanyi:2010cj}.

\subsubsection{The freeze-out of hadrons}

In the late stages of evolution, the system density decreases and the mean free path of hadrons increases. This process eventually leads to the decoupling of particles which become non-interacting objects moving freely toward the detectors. Since after this stage the momenta of particles do not change anymore, this process is referred to as the thermal freeze-out  (the momenta of particles become frozen). Apart from the decrease in density, also the growing rate of the collective expansion favours the process of decoupling. If the expansion rate is much larger than the scattering rate, the freeze-out may occur even at relatively large densities. Generally speaking, the process of decoupling is a complicated non-equilibrium process which should be studied with the help of the kinetic equations. In particular, different processes and/or different types of particles may decouple at different times, so one often introduces a hierarchy of different freeze-outs. In particular, one freqently distinguishes between the chemical and thermal freeze-outs. The former is the stage where the hadronic abundances are established. The chemical freeze-out precedes the thermal freeze-out.

\subsubsection{RHIC vs. LHC}
\label{sect:RHICvsLHC}

The first hydrodynamic models used to interpret the heavy-ion data from RHIC were based on 2+1~\footnote{The 2+1 denotes
in this case two space and one time dimensions in (3+1)-dimensional spacetime, since the 2+1 codes assume that the
longitudinal dynamics is determined completely by the boost symmetry.} perfect-fluid hydrodynamics (Huovinen, Kolb,
Heinz, Ruuskanen, and Voloshin \cite{Huovinen:2001cy}, Teaney and Shuryak \cite{Teaney:2001av}, Kolb and Rapp
\cite{Kolb:2002ve}). These works assumed the bag equation of state for QGP and the resonance gas model for the hadronic
phase. The two phases were connected by a first-order phase transition with  the latent heat varying from 0.8 GeV/fm$^3$
in \rfc{Teaney:2001av} to 1.15 GeV/fm$^3$ in \rfc{Huovinen:2001cy,Kolb:2002ve}. The initialization time for
hydrodynamics was 1 fm/c in \rfc{Teaney:2001av} and 0.6 fm/c  in \cite{Huovinen:2001cy,Kolb:2002ve}. The models differed
also in their treatment of the hadronic phase. In \rfc{Teaney:2001av} the hydrodynamic evolution was coupled to the
hadronic rescattering model RQMD, while in \rfc{Kolb:2002ve} partial chemical equilibrium was incorporated into the
hydrodynamic framework. On the other hand, in \rfc{Huovinen:2001cy} full chemical equilibrium was assumed. The use of a
very short initialization time for perfect-fluid hydrodynamics triggered ideas about early equilibration time of matter
produced in heavy-ion collisions and shaped our way of thinking about these processes in the following years.

As the experimental program was continued at RHIC, theoretical models based on the use of viscous hydrodynamics with a
realistic QCD equation of state at zero baryon density have been developed. This allowed for the first quantitative
estimates of the shear viscosity to entropy density ratio, $\eta/\sd$, which turned out to be very close to the value
$\hbar/(4\pi k_B)$ obtained from the AdS/CFT correspondence~\cite{Kovtun:2004de}. Recent comparisons between
hydrodynamical calculations and the data lead to the range  $1 \leq \eta/\sd  \leq 2.5$ in units of $\hbar/(4\pi
k_B)$~\cite{Song:2010mg}. This value is smaller than that of any other known substance, including superfluid liquid
helium.

Although the $\eta/\sd$ ratio is small in viscous codes describing experimental data, dissipative corrections
to equilibrium values of thermodynamic variables turn out to be quite large in such models at early stages of
the collisions. This is so, because initially there exist large gradients of flow in the produced systems and
non-equilibrium corrections are proportional to the products of transport coefficients and such gradients. Large
values of the latter compensate smallness of the former. This results in substantial values of the shear
stress tensor and modification of the pressure components — with the transverse component of the pressure much
larger than the longitudinal one. Such a pressure anisotropy has motivated investigations aiming at
generalising the standard viscous hydrodynamic framework on one hand and at extending the validity of
viscous hydrodynamics on the other. All these issues are discussed in our review.

The initial energy density in central PbPb collisions at the LHC, inferred from the number of produced particles via Bjorken's formula~\cite{Bjorken:1982qr} at the beam energy $\sqrt{s_{\rm NN}}$~=~2.76~TeV, is more than an order of magnitude larger than that of the deconfinement transition predicted by lattice QCD. The viscous hydrodynamic codes including fluctuating initial conditions showed very remarkable agreement with the measured flow harmonic coefficients $v_n$. The odd flow harmonics~\cite{Alver:2010gr} were found to have a weak centrality dependence, which is typical  for initial state geometric fluctuations. In the coming years, the bulk viscosity to entropy ratio, $\zeta/\sd$, may be estimated from experimental data, so that the two main viscosity coefficients of QGP can be determined~\cite{Noronha-Hostler:2013gga}. In the future, the shear and bulk viscosities might be also found directly from QCD and one will use these values in the hydrodynamic calculations in order to check the overall consistency of the theoretical frameworks.

%% file: sect-micro.tex
\section{Microscopic approaches}
\label{sect:micro}

The point of departure for the theoretical progress reviewed in this article is the fact that the basic
condition for a description in terms of hydrodynamic variables to be useful is that the underlying
microscopic theory should display quasi-universal behaviour at late stages of its dynamical
evolution. Such behaviour is a sign of a significant reduction of the number of degrees of freedom and
constitutes the final stretch on the way to local thermodynamic equilibrium. In this section we review, in general terms, various approaches to late-time behaviour relevant for the dynamics of QGP. We start
with perhaps the most general approach to this problem in the context of quantum field theory, which is the theory of linear response.

\subsection{Linear response and degrees of freedom of relativistic collective states}
\label{sect:linresp}

From the point of view of phenomenological applications to hydrodynamic
evolution in heavy-ion collisions, the most important quantity to consider is
the one-point function of the energy-momentum tensor, $\langle {\hat T}^{\mu
  \nu} \rangle$, of a microscopic model\footnote{Although our presentation
  here aims at QFT applications (in particular, we treat the energy-momentum
  tensor as an operator), the methods presented in this section also apply
  directly in the context of relativistic kinetic theory, see
  \rfs{sect:RTAmodes}, or hydrodynamic theories, see \rfs{sect:mis}.} in a
non-equilibrium state. The simplest such states can be described within linear
response theory, i.e., starting with an equilibrium state and subjecting it to
a weak perturbation.
\begin{figure}
\begin{center}
\includegraphics[height = .3\textheight]{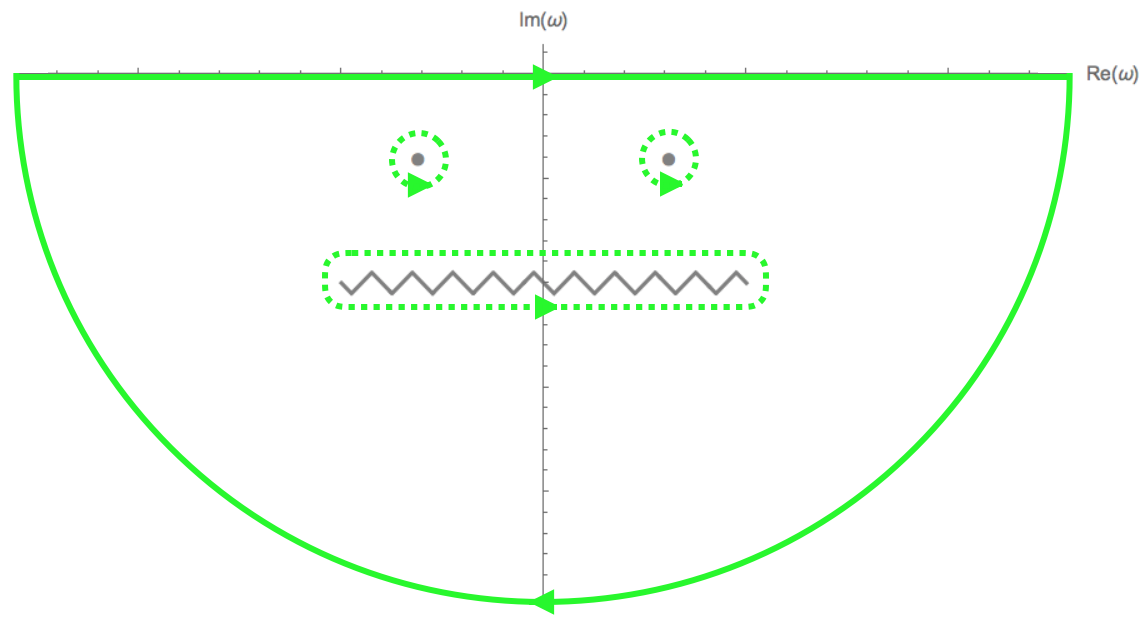}
\caption{The contour integral computing the integral over frequencies
  in~\rf{eq.LRTf}, for a generic value of momentum $\vk$, is a sum of
  contributions from singularities, single poles and branch points, of
  $G_{R}^{\mu \nu, \, \alpha \beta} (\omega, \vk)$ in the lower half of
  frequency plane. Each such contribution we call a mode. Depending on a
  microscopic model there can be a finite or an infinite number of modes and
  at late times the one which corresponds to the singularity closest to the
  real axis dominates the response, since it is the least-damped.}
\label{fig:contour}
\end{center}
\end{figure}

The source that directly couples to the energy-momentum tensor is the
background metric~$g_{\mu \nu}$. Within linear response theory one has, e.g.,
see \cite{Kovtun:2005ev},
\bel{eq.LRT}
\delta \langle {\hat T}^{\mu \nu} \rangle (x) = - \frac{1}{2} \int d^{4} y \, G_{R}^{\mu \nu, \, \alpha \beta} (x^0 - y^0, \vx - \vy) \, \delta g_{\alpha \beta} (y),
\eel
where $\delta \langle {\hat T}^{\mu \nu} \rangle (x)$ is the change in the
expectation value of the energy-momentum tensor after perturbing the flat
background metric $\eta_{\alpha \beta}$ with $\delta g_{\alpha \beta} (y)$ and
$G_{R}^{\mu \nu, \, \alpha \beta} (x^0 - y^0, \vx - \vy)$ is the retarded
two-point correlator of the energy-momentum tensor evaluated in global thermal
equilibrium with temperature $T$,
\be
G_{R}^{\mu \nu, \, \alpha \beta} (x^0 - y^0, \vx - \vy) = - i \, \theta (x^{0}
- y^{0}) \langle [ {\hat T}^{\mu \nu} (x),  {\hat T}^{\alpha \beta}(y) ]
\rangle_{T}. 
\ee
Let us now rewrite the right-hand side of \rf{eq.LRT} in Fourier space
\bel{eq.LRTf}
\delta \langle  {\hat T}^{\mu \nu} \rangle (x)
= -  \frac{1}{2 (2\pi)^{4}} \int d^{3} k  \int d \omega \, e^{- i \, \omega \, x^{0} + i \, \kdotx} \, G_{R}^{\mu \nu, \, \alpha \beta} (\omega, \vk)
\, \delta g_{\alpha \beta} (\omega, \vk), \quad \quad \quad
\eel
where the Fourier-transformed quantities can be recognized by their arguments. In \rf{eq.LRTf}, the integral over
momenta $\vk$ is taken over~$\mathbb{R}^3$ and the integral over frequencies $\omega$ is taken over~$\mathbb{R}$.
Furthermore, due to the rotational symmetry of the thermal state, the retarded two-point function of the energy-momentum
tensor decomposes into a sum of three independent terms, see e.g. Ref.~\cite{Kovtun:2005ev}. This results in three
decoupled sets of components of the energy-momentum tensor evolving independently. Assuming that the momentum $\vk$ is
aligned along the $x^{3}$ direction, these are referred to as the:
\begin{itemize}
\item scalar channel: $\delta \langle  {\hat T}^{12} \rangle$);
\item shear channel: $\delta \langle  {\hat T}^{0a} \rangle$ and $\delta \langle  {\hat T}^{3a} \rangle$ with $a = 1, \, 2$;
\item sound channel: $\delta \langle  {\hat T}^{00} \rangle$, $\delta \langle  {\hat T}^{03}\rangle$ and $\delta \langle  {\hat T}^{33} \rangle$.
\end{itemize}
For vanishing spatial momentum, $\vk = 0$, or at zero temperature all three channels are equivalent to each other. In the following, to keep the presentation as compact as possible, we review the physics of the sound channel in detail and we refer the reader to the relevant literature for
results pertaining to the other channels.

The basic idea is to use the technique of contour integration to express the frequency integral for each value of $\vk$ in terms of the singularities of $G_{R}^{\mu \nu, \, \alpha \beta} (\omega, \vk)$ on the corresponding lower half complex-$\omega$ plane, see Fig.~\ref{fig:contour}. Based on case studies within holography~\cite{Kovtun:2005ev,Grozdanov:2016vgg,Buchel:2015saa,Janik:2015waa,Janik:2016btb}, free QFT~\cite{Hartnoll:2005ju} and kinetic theory~\cite{Romatschke:2015gic}, in general we expect singularities to come in two varieties: single poles or branch-points, and we allow for an infinite number of either type. On the same grounds, apart from the trivial free theory case when there is no equilibration (see Ref.~\cite{Hartnoll:2005ju}), we also expect them to lie away from the origin for $\vk \neq 0$. Note that the singularities are located in a way symmetric with respect to the imaginary axis. At late times either type of singularity at a given value of $\omega = \omega_{\rm sing}(k)$ will give rise to a contribution of the form
\bel{eq.singcontrib}
\delta \langle  {\hat T}^{\mu \nu} \rangle (x) \sim e^{- i \, \omega_{\rm sing}(k) \, x^{0} + i \, \kdotx },
\eel
where the imaginary part of $\omega_{\rm sing}(k)$ is responsible for
dissipation (here $k=|\vk|$) and we suppressed possible subleading power-like
fall-off with time occurring for branch-points singularities. The contribution
given by Eq.~(\ref{eq.singcontrib}) generically, see also below, gives rise to
exponential decay with time. A possible real part of $\omega_{\rm sing}(k)$ is
then responsible for oscillations in time during the approach of $\langle
{\hat T}^{\mu \nu} \rangle$ to equilibrium. \emph{Each such contribution we
  call a mode -- an~excitation of equilibrium plasma}. Clearly, the same type
of analysis applies to operators other than $\hat{T}^{\mu \nu}$ in an
underlying microscopic model, but whenever possible\footnote{Note that the
  analysis of modes in free SU($N_{c}$) at finite temperature in
  Ref.~\cite{Hartnoll:2005ju}, reviewed in the next section, has been
  performed only for the simplest scalar operator.} we will specialize to the
$\hat{T}^{\mu \nu}$ case due to its direct importance in the context of
heavy-ion collisions.

The equilibration time for each mode is set by the inverse of
$\Im{[\omega_{\rm sing}(k)]}$ and, clearly, the long-lived modes are those for
which $\Im{[\omega_{\rm sing}(k)]} \rightarrow 0$. In the absence of second
order phase transition or spontaneous symmetry breaking, which is the
situation we specialize to in this review, the only long-lived modes appear
for operators being conserved currents,
see~e.g.~Ref.~\cite{Arnold:1997gh}. These modes are called \emph{hydrodynamic}
since, as we shall see in~\rfs{sect:fund}, they can be modelled by solutions
of linearized hydrodynamic equations. For these special modes also the real
part of the frequency approaches zero as momentum vanishes,
see~\rfs{sect:mis}. In the absence of conserved charges other than
$\hat{T}^{\mu \nu}$, which is the situation considered in this review, there
are only two kinds of hydrodynamic modes: one kind in the shear channel and
one kind in the sound channel. All remaining modes are transient (short-lived)
modes which become negligible after the longest timescale among $1 /
\Im{[\omega_{\rm sing}(k)]}$ for the values of $\vk$ giving nontrivial
contribution to~\rf{eq.LRTf}. Such modes are referred to as
\emph{nonhydrodynamic} or \emph{transient modes}.

Although heuristic in flavour, the discussion above is actually quite
general. The differences between microscopic models seem to manifest
themselves in different singularity structures for the transient modes, as
well as in the detailed way hydrodynamic modes approach the origin as momentum
vanishes.

Excitations for which the imaginary part of the frequency is small relative to the real part are often referred to as
quasiparticles. It is important to stress at this point that unless we are explicitly talking about kinetic theory we
will not assume that we are dealing with systems for which excitations of the equilibrium state are quasiparticles. The
latter case is, however, quite important and we shall discuss some of its general aspects in the following section.

Finally, one practical remark is the following. Consider the
relation~(\ref{eq.LRTf}) in Fourier space. Let us assume that in the absence
of any metric perturbation (source) for a given value of $\vec{k}$ we have
obtained a solution to the relevant microscopic equations of motion, say a
variant of relativistic kinetic theory or holography, of the form of
\rf{eq.singcontrib}. In such problems, solutions in Fourier space turn out to
exist only for specific values of frequencies $\omega(k)$. According to
Eq.~(\ref{eq.LRTf}), such a non-zero result is possible only if the retarded
two-point function of the energy-momentum tensor in Fourier space has a
singularity there. This justifies identifying such solutions with modes, which
we formally defined as singularities of retarded two-point functions.

The most important notion introduced in this section is the idea of a mode
defined as a contribution to $\langle T^{\mu \nu} \rangle$ from a particular
singularity in $\omega$ of $G_{R}^{\mu \nu, \, \alpha \beta} (\omega, \vk)$ at
a given value of $\omega = \omega_{\rm sing}(k)$. It is further of fundamental
importance to distinguish between two types of modes: the universal long-lived
modes and all the rest, which are transient. This distinction is the reason
why an effective hydrodynamic description of late time non-equilibrium
behaviour is possible. In the following, we proceed with an overview of modes
of an equilibrium relativistic matter as described by free SU($N_{c}$) gauge
theory (\rfs{sect:retardedfQFT}), RTA kinetic theory (\rfs{sect:ktrta}) and
strongly coupled QFTs captured by holography (\rfs{sect:qnm}). To the best of
our knowledge, these are the only examples of systems in which such analysis
has been performed at the moment of writing this review.


\subsection{Free SU($N_{c}$) Yang-Mills theory}
\label{sect:retardedfQFT}

Motivated by developments in holography which we describe in~\rfs{sect:qnm}, Ref.~\cite{Hartnoll:2005ju}
considered free SU($N_{c}$) Yang-Mills theory and calculated the retarded two-point function at finite
temperature of the scalar glueball operator $\mathrm{tr}\, F_{\mu \nu} F^{\mu \nu}$, where $F_{\mu \nu}$ is
the field strength of the SU($N_{c}$) gauge field\footnote{In fact, this two-point function is the same as of
the $\mathrm{tr} \, F_{\mu \nu} \tilde{F}^{\mu \nu}$ operator.}. The Fourier-transformed result takes the form
\begin{eqnarray}
\label{eq.GRfree}
G_{R}^{\mathrm{tr}\, F_{\mu \nu} F^{\mu \nu}} (\omega, \, \vk)= &-& \frac{N_{c}^{2}}{\pi^{2}} \left( k^2 - \omega^{2} \right)^{2}
\left\{
\frac{1}{2} + \left(i \, \frac{\pi \, T}{2 \, k} - \frac{\omega}{4 \, k} \right) \log\frac{\omega + k}{\omega - k} + i \, \frac{\pi T}{k} \log{\frac{\Gamma\left(\frac{- i (\omega + k)}{4 \pi T}\right)}{\Gamma\left(\frac{- i (\omega - k)}{4 \pi T}\right)}}
\right\} \nonumber\\
&+&
\frac{N_{c}^{2}}{\pi^{2}}
\left\{
\frac{2 \pi^{2} T^{2}}{3} \left(\omega^2 - k^2\right) + \frac{16 \pi^4 \, T^{4}}{15} + \frac{k^{2}}{6} \left( \frac{7 k^{2}}{5} - \omega^{2} \right)
\right\}.
\end{eqnarray}
Its singularity structure, originating from the term $\log{\frac{\Gamma\left(\frac{- i (\omega + k)}{4 \pi
T}\right)}{\Gamma\left(\frac{- i (\omega - k)}{4 \pi T}\right)}}$, is given by an infinite series of branch-cuts
extending between $\omega = - 4 \pi \, i \, T \, n - k$ and $\omega = - 4 \pi \, i \, T \, n + k$, where $n =
1, 2, \ldots$. As explained in~\rfs{sect:linresp}, these branch cuts are responsible for the exponential
fall-off of $\delta \langle \mathrm{tr}\, F_{\mu \nu} F^{\mu \nu} \rangle $. Furthermore, at vanishing
momentum, $\vk = 0$, branch cuts transform into single poles located at imaginary axis at $\omega = - 4 \pi \,
i \, T \, n$. On top of that there is another branch-cut extending between $\omega = - k$ and $\omega = k$
that gives rise to oscillatory power law behaviour. All these features can be explicitly seen upon Fourier
transforming Eq.~(\ref{eq.GRfree}) to real time $t = x^{0} - y^{0}$:
\bel{eq.GRfreet}
G_{R}^{\mathrm{tr}\, F_{\mu \nu} F^{\mu \nu}} (t, \, \vk) = \theta(t) \, \frac{N_{c}^{2}}{\pi} \,
\frac{1}{k} \left( k^{2} + \partial_{t}^{2}\right)^{2} \left\{
T\, \frac{\sin{k \, t}}{t} \coth{\left( 2 \pi T \, t \right)}
-
\frac{\sin{k\, t} - k \, t \, \cos{k \, t}}{2 \pi \,   t^{2}}
\right\}.
\eel
In the above expression, branch-points positions $\omega = - 4 \pi \, i \, T \, n \pm k$ give rise to the
behaviour $\sin{k t} \coth{2 \pi \, T \, t}$ whereas the branch-cut itself leads to the subleading power-law
fall-off $1/t$. The other term, $\frac{\sin{k\, t} - k \, t \, \cos{k \, t}}{2 \pi\, T\, t^{2}}$,
originates from the branch-cut between $\omega = \pm k$. It is apparent in \rf{eq.GRfreet} that the
latter effect is present also in the vacuum (i.e. for $T = 0$), whereas the former comes from the presence of
a medium.

This analysis applies in the absence of any interactions and, as noted in the original~\rfc{Hartnoll:2005ju}, the
exponential decay seen in correlators must come from interference between different partial contributions to
\rf{eq.GRfreet}. At the moment of writing this review and to the best of authors' knowledge there are no quantitative
weak-coupling results on the general structure of singularities of correlators upon inclusion of interactions. A
toy-model of this situation is the RTA kinetic theory for which the correlators of the energy-momentum tensor and
conserved current were recently computed in \rfc{Romatschke:2015gic} and we discuss this important recent development
in~\rfs{sect:RTAmodes}.

\subsection{Kinetic theory}
\label{sect:kt}

In this section we introduce elements of relativistic kinetic theory, as a weak-coupling language appropriate for QCD at
asymptotically high temperatures in which some of the problems of interest for this review, such as values of transport
coefficients~\cite{Jeon:1995zm,Arnold:2000dr,Arnold:2002zm,Arnold:2003zc,Czajka:2017bod} or emergence of hydrodynamic
behaviour~\cite{Kurkela:2014tea,Kurkela:2015qoa,Keegan:2015avk}, can be phrased and investigated. Touching upon these
important developments in the context of the so-called effective kinetic theory~\cite{Jeon:1995zm}, we will devote most
of our attention to a model much simpler, yet rich in physics:  relativistic kinetic theory in the relaxation time
approximation~\cite{Bhatnagar:1954zz,Anderson:1974a}.

\subsubsection{Boltzmann kinetic equation}

The fundamental object used in kinetic theory is the one-particle distribution function $f(x,p)=f(t,{\bf x},{\bf p})$, giving the number of particles $\Delta N$ in the phase-space volume element $\Delta^{3}x\Delta^{3}p$ placed at the phase-space point $({\bf x},{\bf p})$ and the time $t$~\cite{deGroot:1980},
\bel{fdef}
\Delta N= f(x,p)\,\Delta ^{3}x\Delta ^{3}p,
\eel
where the four-momentum argument of the distribution function is taken to be on-shell. The main task of kinetic theory is to formulate the time evolution equation for $f(x,p)$. In the non-relativistic case it satisfies the famous Boltzmann equation derived in 1872.

Knowing the distribution function allows us to calculate several important macroscopic quantities, in particular the particle number four-current
\bel{covcur}
n^\mu (x)=\int dP\,p^\mu f(x,p)
\eel
and the energy-momentum tensor
\bel{enmomten}
\Tmunu(x) = \int dP\,p^\mu p^\nu \,f(x,p).
\eel
It is the time-dependence of the latter quantity that is of central importance in the context of hydrodynamics. In~Eqs.~(\ref{covcur}) and~(\ref{enmomten}) we have introduced the Lorentz invariant momentum measure
\bel{dP}
dP = {\frac{d^3p}{p^0}}.
\eel
One may check that the phase space distribution function $f(x,p)$ transforms like a scalar under Lorentz transformations, hence, (\ref{covcur}) and (\ref{enmomten}) transform indeed like a four-vector and a second rank tensor.

The energy and momentum conservation laws have the form
\bel{enmomcon}
\p_\mu \Tmunu(x)=0.
\eel
Expression (\ref{enmomten}) includes only the mass and the kinetic energy of particles. Note also that through the definition of the energy-momentum tensor given by \rf{enmomten}, in kinetic theory one never encounters negative pressures.

For systems where the effects of collisions are negligible, the relativistic Boltzmann equation is reduced to the continuity equation expressing the conservation of the number of particles
\begin{equation}
p^\mu \p_\mu f(x,p)=0.
\label{BE1a}
\end{equation}
To account for collisions, the kinetic equation is written in the form
\begin{equation}
p^\mu \p_{\mu }f(x,p) = C(x,p),
\label{BE}
\end{equation}
where the collision term (integral) $C(x,p)$ on the right-hand side of (\ref{BE}) is
\beq
C(x,p) &=& \frac{1}{2}\int dP_1 dP^\prime dP^\prime_1
\left[
f^{\prime }f_{1}^{\prime }W(p^{\prime },p_{1}^{\prime }|p,p_{1}^{\ })
 -f\,f_{1}W(p,p_{1}|p^{\prime },p_{1}^{\prime \ })\right].
\label{collterm}
\eeq
In \rf{collterm} we use the notation
\bel{efs}
f^{\prime }=f(x,p^{\prime }),\ f_{1}^{\prime }=f(x,p_{1}^{\prime }),\
f=f(x,p),\ f_1=f(x,p_1).
\eel
In the similar way we define the measures $dP_1$, $dP^\prime$, and $dP^\prime_1$. The transition rate $W$ is defined by the formula
\bel{trW0}
W(p,p_{1}|p^{\prime },p_{1}^{\prime})  \equiv  F_i \, p^{\prime \,0} p^{\prime \,0}_1
{\Delta \sigma(p,p_{1}|p^{\prime },p_{1}^{\prime \ })
\over \Delta^3p^\prime \Delta^3p^\prime_1},
\eel
where $F_i$ is the invariant flux
\bel{Fi}
F_i = \sqrt{(p \cdot p_1)^2 - m^4}
\eel
and $\Delta \sigma(p,p_{1}|p^{\prime },p_{1}^{\prime})$ is the differential cross section.

The form (\ref{BE}) is valid for identical particles obeying classical statistics. It can be easily generalised to the case where different types of particles scatter on each other. The generalisation to the quantum statistics case is obtained by the introduction of the so called Uehling-Uhlenbeck corrections to the collision integral. They have the form of the factors $1 - \epsilon \, f$, where $\epsilon = +1 $ for bosons, $\epsilon = -1$ for fermions.



\subsubsection{Effective kinetic theory}
\label{sect:ekt}

In a weakly-interacting QFT at high temperatures and densities, particles acquire effective masses and widths. In this case it is possible to construct the effective kinetic theory which uses the quasiparticle distribution function $f(x,p)$ ~\cite{Jeon:1995zm,Arnold:2002zm,Baier:2000sb}. The latter depends on an on-shell four-momentum $p$, but now the quasiparticle energy $E$ is a function of the spatial momentum and an effective, thermal mass, $E=(\vp^2 + m_{\rm th}(q(x))^2)^{1/2}$. The effective mass depends on a space-time dependent auxiliary field $q(x)$ which, in turn, depends self-consistently on the distribution function
\bel{qofx}
q(x) = \int dP\,  f(x,p).
\eel
The modified quasiparticle Boltzmann equation can be written in the form
\begin{equation}
p^\mu \p_{\mu }f(x,p)  + m_{\rm th} \, \partial^\nu m_{\rm th} \, \partial^{(p)}_\nu f(x,p) = C(x,p).
\label{BERTA}
\end{equation}
Here $\partial^{(p)}_\nu = \partial/\partial_{p^\nu}$ denotes the derivative with respect to momentum and $C(x,p)$ describes now scattering of quasiparticles. Since quasiparticles are on the mass shell, the term $\partial^{(p)}_0 f(x,p) $ vanishes and we see that the spatial gradient of the effective mass acts like an external force, changing the momentum of propagating quasiparticles.

The formalism of effective kinetic theory allows for the calculations of transport coefficients, although the results depend on the approximations made and the processes included in the collision integral. In an SU(3) gauge theory, the shear viscosity at high temperature has the leading-log form
\bel{etaQCD}
\eta = \chi \f{T^3}{g^4 \ln g^{-1}},
\eel
where $g$ is the temperature dependent coupling constant and  $\chi$  is a factor depending on the number of fermion species ($\chi = 106.664$ for $N_f=3$ \cite{Arnold:2000dr}).

Note that the presence of a small dimensionless coupling constant introduces a characteristic hierarchy of timescales for equilibration in weakly-coupled models. Using the EKT framework outside its regime of validity, i.e. for intermediate values of the coupling constant such as those expected in heavy-ion collisions at RHIC and LHC energy, destroys this hierarchy~\cite{Keegan:2015avk}. Furthermore, as we discuss in the following, it leads to results very similar to those of kinetic theory with a simple collisional kernel in the relaxation time approximation or the results of strong-coupling calculations using holography.

\subsubsection{Relaxation time approximation}
\label{sect:ktrta}

Due to the complicated form of the collision integral, in practical applications one very often uses a simplified version of the kinetic equation~\cite{Bhatnagar:1954zz,Anderson:1974a}
\bel{rta}
p^\mu \p_\mu f(x,p) = \pdotU (x) \, \frac{f(x,p)-f_{\rm eq}(x,p)}{\trel},
\eel
where $\umu(x)$ is the flow vector of matter (defined as the unique, normalized timelike eigenvector of the energy-momentum tensor $T^\mu_{\,\,\,\nu}(x)$, see Eq.~(\ref{udef})); $\trel$ is referred to as the relaxation time, and $f_{\rm eq}(x,p)$ is the equilibrium distribution function. Equation (\ref{rta}) has a simple physical  interpretation -- the non-equlibrium distribution function $f(x,p)$ approaches the equilibrium form $f_{\rm eq}(x,p)$ at a rate set by the relaxation time $\trel$. This is the reason why the theory defined by Eq.~(\ref{rta}) is often referred to as to the relaxation time approximation (RTA).


The equilibrium distribution  $f_{\rm eq}(x,p)$ has the standard (Bose-Einstein or Fermi-Dirac) form
\begin{equation}
f_{\rm eq}(x,p) = \frac{1}{(2\pi)^3} \left\{
\exp\left[ -\frac{\pdotU}{T(x)} \right] - \epsilon \right\}^{-1}.
\label{feq}
\end{equation}
where, again, $\epsilon = +1 $ for bosons, $\epsilon = -1 $ for fermions and the classical Boltzmann definition corresponds to the limit $\epsilon \rightarrow 0$


The function $T(x)$ defines the local (effective) temperature of the system. The value of $T(x)$ is determined at each
space-time point $x$ from the condition that  $f_{\rm eq}(x,p)$ yields the same local energy density as $f(x,p)$. Note
that this condition, together with the definition of $U$, make Eq.~(\ref{rta}) highly nonlinear. We shall refer to
these conditions as the Landau matching conditions. Only if the system is very close to local equilibrium can $T$ be
treated as a thermodynamic quantity and used in thermodynamic identities.


The relaxation time $\trel$ which appears in \rf{rta} is a priori a scalar function of the phase space variables $x$ and $p$ and one may formulate different models where this function may take different forms. For conformal systems without any conserved currents, $\trel$ should scale inversely with the effective temperature of the system, namely
\begin{equation}
\trel = \frac{\gamma}{T},
\label{conf-teq}
\end{equation}
where $\gamma$ is a dimensionless constant. As discussed below in \rfs{sect:hydrod} this coefficient can be related to the shear viscosity in an effective hydrodynamic description of the system. Also, see Ref.~\cite{Romatschke:2015gic}, one can use this relation together with Eq.~(\ref{etaQCD}) to view the RTA kinetic theory as a simple model of the EKT kinetic theory of QCD.

\subsubsection{Modes of the conformal RTA kinetic theory}
\label{sect:RTAmodes}

In order to determine structure of singularities of the retarded correlator of
the energy-momentum tensor in kinetic theory, it is convenient to write down the
Boltzmann equation in an arbitrary curved background and use
Eq.~(\ref{eq.LRT}). As already anticipated in~\rfs{sect:linresp}, we will
adopt a slightly different strategy in order to reproduce the results of
Ref.~\cite{Romatschke:2015gic}. We work directly in Fourier space and focus
entirely on the sound channel, i.e. perturbations of $T$, $U^{3}$ and $f$
exhibiting harmonic dependence on $x^{0}$ and $x^{3}$: $e^{- i \, \omega \,
  x^{0} + i k x^{3}}$. Looking for solutions of the flat space RTA Boltzmann
equation
one obtains the following condition for $\omega$ as a function of $k$
\small
\bel{eq.condsingRTA}
2 \, k \, \trelm \, \left\{ (k \, )^{2} + 3 \, i \, \omega \, \trelm  \right\} + i \left\{ (k \, \trelm)^{2} + 3 \, i \, \omega \, \trelm + 3 \,  (\omega \, \trelm)^2 \right\}\, \log{\frac{\omega \, \trelm - k \, \trelm + i}{\omega  \, \trelm + k \, \trelm + i}} = 0. \quad
\eel
\normalsize
Indeed, as one can check this is equivalent to the expression for singularities of the retarded two-point function of
the energy-momentum tensor obtained in Ref.~\cite{Romatschke:2015gic}. One immediately notices a branch cut-singularity
given by $\log{\frac{\omega \, \trelm - k \, \trelm + i}{\omega  \, \trelm + k \, \trelm + i}}$ with branch points at
$\omega = - i \, \frac{1}{\trelm} \pm k$. Note that in the absence of interactions in this model, i.e. in the limit of
$\trelm \rightarrow \infty$, this branch-cut coincides with the branch-cut surviving the $T \rightarrow 0$ limit of the
free SU($N_{c}$) gauge theory calculation of Ref.~\cite{Hartnoll:2005ju} described in~\rfs{sect:retardedfQFT}. One
should then think of it as a modification of the free propagation of particles
by the effects of interactions captured by the
RTA collisional kernel. For massive particles, we expect a branch-cut due to a factor of $\log{\frac{\omega \, \trelm -
\sqrt{k^{2} + m^{2}} \, \trelm + i}{\omega  \, \trelm +  \sqrt{k^{2} + m^{2}} \, \trelm + i}}$ and when $m \, \trelm\gg
1$ we obtain quasiparticle excitations. Note also that the singularity of  $\log{\frac{\omega \, \trelm - k \, \trelm +
i}{\omega  \, \trelm + k \, \trelm + i}}$ becomes a pole in the limit $\vk \rightarrow 0$.


%
\begin{figure}[!t]%
\begin{center}
 \includegraphics[height = .25\textheight]{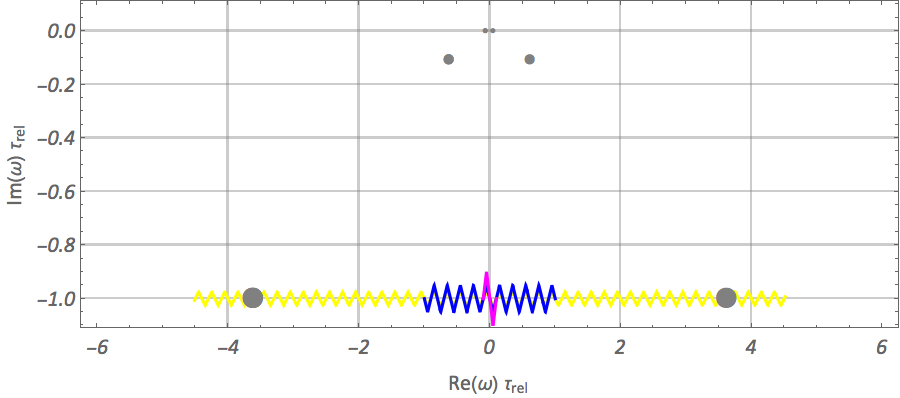}
\caption{The structure of singularities in the complex frequency plane at fixed spatial momentum $k$ of the retarded two-point function of $\hat{T}^{\mu\nu}$ in the sound channel in the RTA kinetic theory. Gray dots represent single poles corresponding to the sound wave mode at $k \, \trelm = 0.1$ (the smallest dots), $k \, \trelm = 1$ (medium dots) and at the verge of its existence at $k \, \trelm = 4.531$ (the largest dots). Spiked lines represent branch cut singularities which connect branch points at corresponding values of momenta: $k \, \trelm = 0.1$ (magenta with the largest amplitude), $k \, \trelm = 1$ (blue with medium amplitude) and $k \, \trelm = 4.531$ (yellow with the smallest amplitude). See the main text for a discussion of this.}
\label{fig:singRTA}
\end{center}
\end{figure}
Regarding other excitations, it turns out that in this model there is only one, and at low momenta it is a
long-lived mode corresponding to a hydrodynamic sound wave, see Fig.~\ref{fig:singRTA}. More precisely, for each
value of momentum between $k \, \trelm = 0$ and $k \, \trelm \approx 4.531$ there
are two poles in the complex frequency symmetrically located with respect to the imaginary axis whose
positions at low momenta approach the sound wave dispersion relation $\omega = \pm \frac{k}{\sqrt{3}} +
O(k^{2})$, see \rfs{sect:mis} for the hydrodynamic interpretation. This type of singularity, a single pole, is
also seen at strong coupling in holography, see \rfs{sect:qnm}. The key difference here is that for $k \,
\trelm >  4.531\ldots$ there is no solution to Eq.~(\ref{eq.condsingRTA}) and this mode ceases to
exist. This is the value of momentum for which $\Im(\omega)$ approaches $-\frac{1}{\trelm}$. For
higher momenta, momentum transfer in the sound channel occurs solely through the transient mode represented by the
branch-cut singularity in~the~function~$\log{\frac{\omega \, \trelm - k \, \trelm +
i}{\omega  \, \trelm + k \, \trelm + i}}$. This means that there are no hydrodynamic modes for such momenta and this fact can be interpreted as the breakdown of hydrodynamics in this system.

Note that there seems to be no physical principle why more singularities in the two-point function of the energy-momentum tensor are not present and one should expect that for more complicated collisional kernels there would be a more intricate mode structure. At the moment of writing this review, this issue remains an important open problem.

\subsection{String theory}
\label{sect:holography}

A very influential approach to the issues studied in this review grew out of
string theory and is often referred to as gauge-gravity duality or
holography. The fundamental insight behind this set of ideas is the AdS/CFT
correspondence~\cite{ Maldacena:1997re,Gubser:1998bc,Witten:1998qj}, which is
ultimately due to the fact that closed strings (which describe gravity) and
open strings (which describe Yang-Mills degrees of freedom) are made of the
same ``stuff''. This leads to a representation (in a sense a reformulation) of
string theory using supersymmetric Yang-Mills theory. In the 't Hooft
limit~\cite{tHooft:1973alw} (defined as $N_c\rightarrow\infty$ with the 't
Hooft coupling $g_{\rm YM}^{2} N_{c} \equiv \lambda$ fixed and large), the
observables of this Yang-Mills theory in flat four-dimensional Minkowski space
become expressible in terms of classical gravity in five dimensions.  In
consequence, this duality geometrizes states of certain QFTs in the form of
solutions of gravity (and more generally string theory) in a
higher-dimensional spacetime.  The fact that the additional non-compact
dimension plays a key role justifies the commonly used term holography. This
gravitational representation involves a negative cosmological constant, which
implies that the asymptotic behaviour of the geometry is not flat Minkowski
space, but rather anti-de Sitter space (at least locally).

The archetypical example of a QFT which possesses such a geometrical formulation is the maximally supersymmetric Yang-Mills theory in 3+1 dimensions, known as the \sym\ super Yang-Mills (SYM) theory. It arises by replacing the quark sector of QCD by a specially crafted matter sector\footnote{It consists of 6 scalar fields and 4 Weyl spinor fields in the adjoint representation of the gauge group, see~e.g.~Ref.~\cite{DHoker:2002nbb} for a review.} which also makes this theory conformal. The rank of the SU($N_{c}$) gauge group and the coupling constant are parameters specifying the theory. The nature of holography is such that when $N_{c}$ is large and interactions are strong then a large class of observables, in particular many observables of interest in the context of heavy-ion collisions, can be calculated by solving classical gravity equations coupled to appropriate matter fields in one dimension higher.

All the results used in this review were obtained in the context of \symm, which is the original setting of the AdS/CFT correspondence, intuited by Maldacena by considerations of the dynamics of coincident D3-branes in type IIB string theory. By now, however, it has been understood that there are many more examples of holography involving both conformal (as ${\cal N}$=\,4 SYM) and non-conformal QFTs. Such theories are called holographic and we will refer to them as hQFTs. When they are conformal, we called them holographic conformal field theories (hCFTs). In our review we will focus on the simplest case of hCFTs.

Due to its asymptotic freedom, it is clear that QCD does not fall into the
class of QFTs for which the holographic description is general relativity
(because -- at least from today's perspective -- classical gravity emerges
when the gauge coupling is strong) and it is not known how to extend AdS/CFT
to cover such cases. The promise of such a generalisation is so great however,
that despite the lack of a firm grounding in string theory, much effort has
gone into guessing what such a description might look like. In this spirit, in
\rfs{sect:adscft} we will try to give a picture of holography which focuses on
the aspects which one may expect to be relevant in such a wider context.

On the other hand, some of the properties of QGP above but not far above the crossover temperature are in qualitative agreement with the properties of deconfined phases of hQFTs. The prime example here is the value of the shear viscosity to entropy density ratio obtained from a holographic calculation, which qualitatively matches the value used in successful phenomenological descriptions of experimental data from RHIC and LHC (see in particular \rfc{CasalderreySolana:2011us} for an extensive overview of holography applied to QCD). It should  perhaps be noted that in the regime of weak coupling (and so outside the regime in which holography is realized through general relativity), effective similarities at long wavelengths were also observed between deconfined phases of QCD and ${\cal N}$=\,4 SYM, see Ref.~\cite{Czajka:2012gq,Czajka:2014gaa}. Finally, the usefulness of AdS/CFT calculations for real-life QCD may be due to the gluon sector being dominant -- the gluons are, after all, the same in QCD as in \symm.

Regardless of whether there are some holographic results which directly apply to QCD due to universality of some sort, there is another sense in which holographic calculations have proved tremendously useful: they have provided a reliable means of observing how hydrodynamic behaviour appears in a non-equilibrium system in a fully ab initio manner. This has prompted a number of crucial developments in the field of relativistic hydrodynamics which apply to any system, including QGP. These developments are the focal point of our review.

%% file: sect-adscft.tex
\section{Lessons from holography and the strong coupling picture}
\label{sect:adscft}

In this section we examine the lessons for relativistic hydrodynamics which follow from holographic
calculations at the linearized level.

\subsection{Gravitational description of strongly-coupled quantum field theories}
\label{sect.gravity}

In holography, the gravitational description of hQFTs of relevance for QCD is captured by solutions of equations of
motion originating from the following higher dimensional gravitational action
\bel{eq.SEH}
S = \frac{1}{2 l_{P}^3} \int d^{\,5} x \sqrt{-g} \left\{ R - 2 \times \left(-\frac{6}{L^{2}}\right) + \ldots \right\},
\eel
where $R$ is the Ricci scalar of a five-dimensional geometry, $-6/L^{2}$ stands for a negative cosmological constant,  and the ellipsis contain boundary terms as well as possible five-dimensional matter fields obeying two-derivative equations of motion. The most symmetric solution of these equations of motion is anti-de Sitter (AdS) space given by
\bel{eq.AdS}
ds^{2} = g_{a b} \, dx^{a} dx^{b} = \frac{L^{2}}{u^{2}} \left\{ du^{2} + h_{\mu \nu} (u, x) dx^{\mu} dx^{\nu} \right\},
\eel
with $h_{\mu \nu}(u,x)$ given by the Minkowski space metric $\eta_{\mu \nu}$ and all matter fields set equal to zero.
Other solutions involve non-trivial profiles of $h_{\mu \nu}(u,x)$ as well as nontrivial matter fields. Generically, one
most often neglects higher derivative terms that could potentially appear in~\rf{eq.SEH}, e.g. $R^{2}$, since when
treated exactly they give rise to unphysical effects and when treated as small perturbations they do not change the
qualitative features of results derived from~\rf{eq.SEH}. It is interesting to note that the issue of higher derivative
terms here is mathematically very similar~\cite{Cayuso:2017iqc} to the task  of making sense of the truncated gradient
expansion of hydrodynamics, as discussed in \rfs{sect:effective}. Furthermore, it could be the case that higher
derivative terms treated as leading order corrections may be trustworthy even when large. In the context of relativistic
hydrodynamics this has turned out to be the case, as discussed in \rfs{sect:hydrod}.

It is important to make a distinction between features of~\rf{eq.SEH} that
characterise a given hQFT and those that characterise a particular state
within this hQFT. When it comes to the former, the parameter that controls the
scaling of observables with the number of microscopic degrees of freedom is
the ratio of the curvature scale of AdS, $L$, and the five-dimensional Planck
scale, $l_{P}$, that is related to the so-called central
charge~$c$~\footnote{In CFTs in even spacetime dimensions central charges are
  coefficients appearing at independent contributions to the energy-momentum
  tensor two-point function in the vacuum or, equivalently, independent
  contributions to the conformal anomaly $\langle \hat{T}^{\mu}_{\mu} \rangle
  \neq 0$ when a CFT is placed on a generic curved background. For theories in
  four spacetime dimensions there are two such central charges, denoted by $a$
  and $c$. However, in the case when the action \rfn{eq.SEH} in the
  holographic description contains only two-derivative terms, the central
  charges are equal and it is then customary to use the letter $c$ to denote
  both of them.} through the formula
\bel{eq.LoverlP}
\frac{L^{3}}{l_{P}^{3}} = \frac{c}{\pi^{2}}.
\eel
For ${\cal N}=4$~SYM with $N_{c}$ colours the central charge is
\bel{eq.cNeq4}
c =  \frac{N_{c}^{2}}{4}
\eel
and in most of the results discussed in what follows we use this value of $c$. Of course, in order to trust the physical
description in terms of classical gravity, one must have $L \gg l_{P}$, which, as follows from
Eqs.~(\ref{eq.LoverlP})~and~(\ref{eq.cNeq4}), translates to $N_{c} \gg 1$, i.e. \mbox{the 't Hooft planar limit
\cite{tHooft:1973alw}}. The other parameter is the interaction strength between microscopic constituents and it turns
out that for infinite interaction strength, i.e. 't Hooft coupling constant $g_{\rm YM}^{2} N_{c} \equiv \lambda \gg 1$,
this parameter does not appear in the two-derivative part of~\rf{eq.SEH}. This observation has already an important
phenomenological bearing because it shows that none of the features of processes at infinite coupling captured
by~\rf{eq.SEH} will be coupling-enhanced.

Another set of parameters characterizing an underlying theory lies in the choice of infinitely many matter fields in~\rf{eq.SEH} and the associated choice of infinitely many dimensionless (when scaled by appropriate factors of~$L$) parameters in front of their kinetic and interaction terms. In the case of \symm\ these terms as well as their coefficients are determined by type IIB string theory, but in the spirit of this section they may be regarded in a phenomenological way as something to be chosen at will. Ideally one would like to have a ``top-down'' understanding of any such terms (i.e. one whereby they arise from a string theory calculation), but at this stage requiring such an understanding may be too restrictive.

Finally, since the AdS geometry acts effectively as a box\footnote{By
  analyzing propagation of light rays in the metric given by~\rf{eq.AdS} with
  $h_{\mu \nu}(u,x) = \eta_{\mu \nu}$ one concludes that geodesics can reach
  the plane of $u = 0$ in a finite time. As a result, their further
  propagation depends on a boundary condition imposed at $u = 0$. Similar
  results hold for fields propagating on the AdS geometry.} with a boundary at
$u = 0$, c.f.~\rf{eq.AdS}, for each field one needs to prescribe a boundary
condition at $u = 0$. These boundary conditions are interpreted as sources for
single trace gauge-invariant local operators in an underlying hQFT and
complete the specification of the latter.

As for the parameters which specify the states, these are all the other
parameters appearing in solutions of equations of motion derived
from~\rf{eq.SEH} that cannot be removed by diffeomorphisms involving the $u$
and $x$ coordinates. Such transformations need to vanish at $u = 0$, since
otherwise they would alter the given set of boundary conditions, and thus the
physical content, the definition of the underlying hQFT. Let us stress here
that the same geometry, i.e. the same state in an underlying QFT, can be
described using different coordinates, all of which are clearly on the same
footing. Note however that these coordinate systems might cover different
parts of the geometry and might break down (in the sense of components of the
metric or its inverse diverging) even when that geometry itself is perfectly
regular.

In the vast majority of cases studied in this review, we will be concerned with solutions of Einstein's equations with negative cosmological constant,
\bel{eq.Einstein}
R_{a b} - \frac{1}{2} R \, g_{a b} + \left(-\frac{6}{L^{2}}\right) \,  g_{a b} = 0,
\eel
viewed as a consistent truncation of the equations of motion coming from~\rf{eq.SEH} when the sources associated with all fields other than the five-dimensional metric are set to zero. These solutions have an interpretation as states in strongly-coupled CFTs for which the only local operator with a nontrivial expectation value is the energy-momentum tensor. These equations possess few analytic solutions and tackling the problem of time-dependent processes in strongly-coupled QFTs in general requires solving them using elaborate numerical methods.

Using the parametrization of the five-dimensional metric from~\rf{eq.AdS}, the expectation value of the energy-momentum tensor of an underlying strongly-coupled CFT can be obtained from the near-boundary ($u = 0$) behaviour of $h_{\mu \nu}(u,x)$ as
\bel{eq.nearbdry}
h_{\mu \nu}(u,x) = \eta_{\mu \nu} + \frac{\pi^{2}}{2\,c} \langle \hat{T}_{\mu \nu} \rangle \, u^{4} + \ldots\, ,
\eel
where the expression is provided for a CFT in a Minkowski space and all the suppressed contributions are
proportional to $\frac{\pi^{2}}{2\,c} \langle \hat{T}_{\mu \nu} \rangle$, as well as its powers and
derivatives. CFT states of interest will be then characterized by $\frac{\pi^{2}}{2\,c} \langle \hat{T}_{\mu
\nu} \rangle$ as a function of $x$ and one can obtain them by constructing appropriate solutions
of~\rf{eq.Einstein} using~\rf{eq.nearbdry}. It is also interesting to note that a given form of
$\frac{\pi^{2}}{2\,c} \langle \hat{T}_{\mu \nu} \rangle$ as a function of $x$ characterizes simultaneously
many holographic CFTs, out of which ${\cal N} = 4$~SYM corresponds to
setting~$c$ to the value 
given in~\rf{eq.cNeq4}. This analysis also makes it apparent that empty AdS space represents the vacuum state of a hCFT,
characterized by vanishing expectation values of all local operators including $\langle \hat{T}_{\mu \nu} \rangle$.
A generalization of~\rf{eq.nearbdry} to a hCFT in an arbitrarily curved but fixed
background $h_{\mu \nu}(x)$ is straightforward albeit tedious and starts by replacing the Minkowski metric~$\eta_{\mu \nu}$ in~\rf{eq.nearbdry} by~$h_{\mu \nu}(x)$. The details of this can be found in
Ref.~\cite{deHaro:2000vlm}.

The considerations presented here also make it apparent why $x$ act as coordinates in an underlying hQFT. For the
reasons outlined above, one often encounters in literature the statement that a hQFT ``lives on the boundary'' (at $u =
0$) which is simply means that boundary conditions at $u = 0$ define a given hQFT. Furthermore, the interior
of the higher dimensional geometry described by~\rf{eq.AdS} is often referred to as the bulk (of the 5-dimensional
spacetime). Note that the radial coordinate $u$ has an interpretation of the inverse of an energy scale in a hCFT, since
for the vacuum AdS solution the change of a physical distance in a dual CFT, $\Delta x \rightarrow \gamma \, \Delta x$
can be compensated by the change of~$u$, $u \rightarrow \gamma \, u$. Finally, it should be kept in mind that the bulk
geometry and other fields contain only information about one-point functions of local operators. Higher-point functions
are not specified by the geometry alone and provide a set of independent observables.

Solutions of~\rf{eq.Einstein} describe the simplest, yet very rich sector of dynamics of a class of the most symmetric
(because conformal) hQFTs. There are many possible and well-studied generalizations. Perhaps the most interesting
possibility from the phenomenological standpoint is to break the conformal symmetry, which can be achieved by
introducing a nonzero constant boundary condition for one of the scalar fields suppressed in~\rf{eq.SEH} which
represents a source for a relevant operator. We do not review such models in detail. Another interesting possibility
which we do not consider here is to allow for the U(1) conserved current to acquire a nonzero expectation value by
coupling gravity from~\rf{eq.Einstein} to a gauge field sector. These two generalizations bring up the following
important point. Whereas considering vacuum Einstein's equations with negative cosmological constant
alone,~\rf{eq.Einstein}, is a consistent truncation of equations of motion following from the actions~(\ref{eq.SEH})
containing a priori infinitely many bulk fields (as determined by string theory), it is not always clear if a reduction
to a small set of fields including gravity and other bulk matter makes sense. This problem has led to the so-called
``bottom-up'' perspective in which one postulates an action coupling gravity to a desired set of fields and deriving the
consequences of such a bulk matter sector for a putative hQFT. As mentioned earlier, such exploratory approaches have a
role to play given the difficulty in finding string theory constructions of holographic duals of QFTs which capture
physically important features of QCD.

Finally, as we anticipated earlier, adding a finite number of higher curvature terms to the Einstein-Hilbert action usually leads to an ill-defined initial value problem unless we treat them as small perturbations. Such higher curvature terms are present
(certainly from a string theory perspective) and certainly become important outside the strict $N_{c} = \infty$,
$\lambda = \infty$ limit, but we do not have a controllable way to account for them. This is why it is useful and
illuminating to consider as a toy-model adding the so-called Gauss-Bonnet term to the Einstein-Hilbert part of the
action~(\ref{eq.SEH}),
\bel{eq.SGB}
S_{GB} = \frac{1}{2 l_{P}^3} \int d^{5} x \sqrt{-g} \, \frac{\lambda_{GB}}{2} L^{2} \left\{  R^{2} - 4 R_{a b} R^{a b} + R_{a b c d} R^{a b c d} \right\},
\eel
for which the equations of motion for the bulk metric~$g_{a b}$ turn out to be of a two-derivative type and
avoid aforementioned problems~\cite{Myers:1987yn}. By changing the real dimensionless parameter $\lambda_{GB}$
in~\rf{eq.SGB} one can try to model the effects of relaxing the strict limit of $N_{c} = \infty$, $\lambda =
\infty$ on both equilibrium and non-equilibrium properties in hQFTs, see, e.g., Refs.~\cite{Kats:2007mq,Brigante:2007nu,Buchel:2008vz,Grozdanov:2016vgg,Grozdanov:2016fkt}. Naively, consistency of the gravitational picture requires $\lambda_{GB} \in (-\infty,\frac{1}{4}]$. An important caveat though is that other
considerations demonstrate that gravity with a nonvanishing cosmological constant and only the Gauss-Bonnet term cannot be consistent with the full self-consistency of a hQFT~\cite{Camanho:2014apa} (see, however, Ref.~\cite{Papallo:2015rna}), which means that we can fully
trust~\rf{eq.SGB} only for $|\lambda_{GB}| \ll 1$. Let us also stress that supplementing the Einstein-Hilbert
action with negative cosmological constant, Eq.~(\ref{eq.SEH}), with the Gauss-Bonnet term,
Eq.~(\ref{eq.SGB}), does not affect the expectation values of any operators~other~than~$\hat{T}^{\mu \nu}$.

In the rest of this section we will survey the simplest solutions of~\rf{eq.Einstein} with a view to make contact with the material from \rfs{sect:linresp}. Regarding other holographic results in this review, in \rfs{sect:sym} we will discuss $\langle \hat{T}^{\mu \nu} \rangle$ obtained from holography in a fully nonlinear manner for the simplest model of heavy-ion collisions -- one-dimensionally expanding plasma system of a Bjorken type. In \rfs{sect:matching} we discuss the so-called fluid-gravity duality which accounts in the bulk for the hydrodynamic gradient expansion in hQFTs. In \rfs{sect:resu} we overview holographic calculation of the gradient expansion at large orders for the Bjorken flow (late-time expansion) and make contact with the results of \rfs{sect:qnm}.

\subsection{Black branes and equilibrium states}
\label{sect:blackbranes}

Collective equilibrium states of hQFTs are represented on the gravity side by static black hole solutions~\cite{Witten:1998zw}. Because solutions corresponding to equilibrium states of CFTs in flat space have planar (rather than topologically spherical) horizons, they are referred to in the literature as black branes. Equilibrium states are characterized by conserved charges (or associated potentials, depending on the choice of ensemble) and in the case of interest -- strongly-coupled CFTs described by Eq.~(\ref{eq.Einstein}) -- the relevant quantity is the energy density or the temperature~$T$. Because of the underlying conformal symmetry there is no other dimensionful parameter associated with the thermal state.

The relevant solution, known as the AdS-Schwarzschild black brane,
takes the form
\bel{eq.AdS-Schw}
ds^{2} = \frac{L^{2}}{u^{2}} \left\{ du^{2} - \frac{\left(1-\frac{u^{4}}{u_{0}^{4}}\right)^{2}}{ 1 + \frac{u^{4}}{u_{0}^{4}} } \left(dx^{0}\right)^{2} +\left( 1 + \frac{u^{4}}{u_{0}^{4}} \right) d\vx^{2}\right\},
\eel
with a static horizon located at $u = u_{0}$. In the context of holography, horizon thermodynamics is a reflection of the thermodynamics of the dual strongly-coupled hCFT. The Hawking temperature~\cite{Hawking:1974sw} of the AdS-Schwarzschild black brane, equal to $\frac{\sqrt{2}}{\pi \, u_{0}}$, is equal to the temperature $T$ of the corresponding thermal state of the dual, strongly-coupled hCFT, whereas the Bekenstein-Hawking entropy~\cite{Bekenstein:1973ur,Hawking:1976de} (here: density) of the black hole (here: brane), equal to
\be
\sd = \frac{\mathrm{area \, density}}{4 G_{N}} = \frac{\pi^{2}}{2} N_{c}^{2} T^{3},
\ee
is simply the entropy density of the hCFT (here ${\cal N} = 4$ SYM in the
holographic regime). Comparing this result, valid for $\lambda \rightarrow
\infty$, with the thermodynamic entropy evaluated at vanishing coupling
$\lambda$ (i.e. for an ideal gas gluons and the remaining degrees of freedom
appearing in ${\cal N} = 4$ SYM), Ref.~\cite{Gubser:1998nz} pointed out that
they differ by only $25\%$. On the gravity side, this follows from the
aforementioned observation that the coupling constant $\lambda$ does not enter
the holographic calculations in two-derivative gravity at all.  Furthermore, a
small change in thermodynamic properties of quark-gluon plasma from
weak-coupling to the strong-coupling regime (albeit away from the crossover)
has also been observed in lattice studies of QCD, see e.g
Ref.~\cite{Petreczky:2012rq}.

For completeness, it will also be useful to describe the AdS-Schwarzschild geometry given by Eq.~(\ref{eq.AdS-Schw}) in coordinates that are regular at the horizon,
\bel{eq.AdS-Schw-EF}
ds^{2} = \frac{L^{2}}{u^{2}} \left\{ - 2 \, dx^{0} \, du - \left( 1 - \pi^{4} T^{4} u^{4} \right) \left(dx^{0}\right)^2 + d\vx^{2} \right\},
\eel
where one should note that despite using the same names, the $x^{0}$ and $u$ coordinates are now different from the ones in Eq.~(\ref{eq.AdS-Schw}) (in the sense that the same coordinate values correspond to different bulk points).

The key feature of the AdS-Schwarzschild black brane solution is the presence of the horizon, which acts as a surface of no return for all physical signals. To make this statement clear: in the approximation in which it does not backreact on the AdS-Schwarzschild geometry, once a wave-packet passes the hypersurface of $u = u_{0}$, it cannot influence whatever is happening between $u = 0$ and $u = u_{0}$. This notion extends also outside equilibrium with the horizon evolution obeying the second law of thermodynamics~\cite{Bekenstein:1974ax,Bardeen:1973gs}. One can thus say that it is the presence of the horizon that is the holographic manifestation of dissipation (entropy production) in QFTs.

\subsection{Excitations of strongly-coupled plasmas as black branes' quasinormal modes}
\label{sect:qnm}

The simplest non-equilibrium phenomenon to study holographically is the dynamics of linearized perturbations on top of the AdS-Schwarzschild black brane. Because the background solution~\rfn{eq.AdS-Schw-EF} is translationally invariant in both $x^{0}$ and~$\vx$, it is natural to seek for solutions in terms of Fourier modes,
\bel{eq.Zgrav}
Z(u,\,x^{0},\,\vx) = \int d\omega \, d^{3}k \, e^{- i \, \omega \, x^{0} + i \, \vk \cdot \vx}  \, Z(u,\omega,\vk),
\eel
where we make use of the fact that in the case of interest perturbations can be recast into a set of decoupled functions.

\begin{figure}[!t]%
\begin{center}
 \includegraphics[height = .30\textheight]{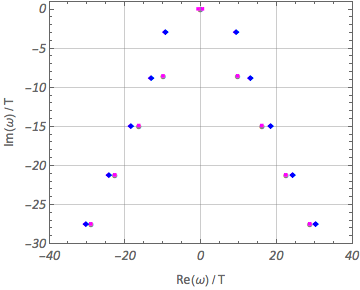}
\caption{The structure of the lowest lying singularities in the complex frequency plane at fixed spatial momentum $k$ of the retarded two-point function of $\hat{T}^{\mu\nu}$ in the sound channel in hCFTs for $k = 0.1 \, T$ (dots), $k = T$ (squares) and $k = 10 \, T$ (diamonds). One can clearly see that singularities are single poles (QNMs) and all of them apart from one retain fast decay time of the order of $1/T$. The remaining singularity can be made arbitrary long-lived by considering low-enough momentum, see also Fig.~\ref{fig:qnmsk}, and, as we discuss in \rfs{sect:mis}, is a manifestation of the hydrodynamic behaviour.}
\label{fig:xmastree}
\end{center}
\end{figure}

In order to make contact with the linear response theory introduced in
\rfs{sect:linresp}, see Eqs.~(\ref{eq.LRT})~and~(\ref{eq.LRTf}), we are
interested in the bulk metric perturbations, $\delta g_{a b}$ (or, after
fixing the bulk coordinates freedom as in \rf{eq.AdS}, $\delta h_{\mu \nu}$),
solving Eqs.~(\ref{eq.Einstein}) linearized around the AdS-Schwarzschild
metric. The reason for it is that via Eq.~(\ref{eq.nearbdry}) they correspond
to perturbations of the expectation value of the energy-momentum tensor around
its equilibrium value and can be generated by perturbing the background metric
of the dual hCFT.

The decomposition of the retarded two-point function of the energy-momentum tensor (and the resulting change in its
expectation value) into shear, sound and scalar channels has, as it must, a direct counterpart on the gravity side.
Focusing on the sound channel and assuming that the momentum $\vk$ is aligned along the $x^{3}$-direction, the relevant
bulk metric perturbations, see \rf{eq.AdS}, are $\delta h_{00}$, $\delta h_{03}$, $\delta h_{33}$ and the sum of $\delta
h_{11}$ and $\delta h_{22}$. They can be combined into a single bulk variable $Z(u, \omega, \vk)$ which obeys a second
order ordinary differential equation in $u$. Its behaviour at $u = 0$ is precisely such as dictated by \rf{eq.nearbdry}
and following the general discussion about the definition of modes within linear response theory from
\rfs{sect:linresp}, we will be interested in turning off the source, i.e. the situation in which $Z(u, \omega, \vk) =
{\cal O}(u^{4})$. The solution depends on two integration constants and this requirement sets one of them to zero. Given
that $Z$ can be freely rescaled, demanding that the horizon acts as a surface of no return imposes a condition on
frequencies $\omega$, which turn out for each value of $k$ to have infinitely many discrete complex solutions. In the
language of black hole physics these solutions are called quasinormal modes (QNMs) and are responsible for the approach
of the  horizon to its equilibrium form after nonlinear effects become negligible, see e.g. \cite{Berti:2009kk} for a
comprehensive review and Ref.~\cite{Horowitz:1999jd} for early discussion of QNMs in the context of holography.

As a result, we see that singularities of the retarded two-point function of the energy-momentum tensor of the strongly-coupled hCFT plasma at fixed $\vk$ are single poles located at the frequencies $\omega(k)$ of QNMs of the AdS-Schwarzschild black brane. The same holds in the scalar and shear channels and, more generally, occurs for any other operator in strongly-coupled plasma and extends to situations without conformal symmetry. Note that apart from the position of a pole, its residue is also very important. We refer the reader to Ref.~\cite{Amado:2007yr} for a discussion of residues of retarded two-point functions in holographic plasmas.

Several further comments are in order. First, the QNM frequencies are symmetric with respect to the imaginary axis and typically contain both real and imaginary parts. Second, the only scale in the problem is the temperature $T$ and QNM  frequencies of strongly-coupled hCFTs are necessarily linear in $T$. Hence, apart from the observation that  singularities of the retarded two-point function of the energy-momentum tensor are single poles, the only nontrivial feature is their distribution in the complex frequency plane as a function of $k$ measured in the units of $T$. Regarding the former, Fig.~\ref{fig:xmastree} displays the position of the lowest five QNM frequencies in the sound channel for three different values of $k/T$. From this plot one can infer that the frequencies are continuous functions of momentum $k$ and as a result, what one then calls a mode is the set of all excitations associated with a given $\omega(k)$.  Furthermore, one sees that for all QNMs apart from one their amplitude decreases by an order of magnitude over at most a time scale of $1/T$. The number of oscillations of each QNM over this decay time is of the order of unity. This feature of QNMs needs to be drastically contrasted with the quasiparticle behaviour occurring in weakly-coupled systems, in which there are many long-lived excitations. In the present case, in the sound channel there is only one long-lived excitation, which is the least-damped mode at low momenta. This special excitation is to be interpreted in \rfs{sect:mis} as the hydrodynamic sound wave -- one of the two independent solutions of the equations of hydrodynamics linearized on top of static plasma. The same situation occurs in the shear channel in which also one hydrodynamic mode appears, the shear wave, and all the remaining QNMs are exponentially damped over a time scale of order $1/T$. In the so-called scalar channel, all the modes are fast decaying. More generally, the presence of such long-lived, slowly evolving excitations can be inferred from the conservation of the energy-momentum tensor $\hat{T}^{\mu \nu}$, see Ref.~\cite{Nickel:2010pr}.
\begin{figure}[!t]%
\begin{center}
\includegraphics[height = .25\textheight]{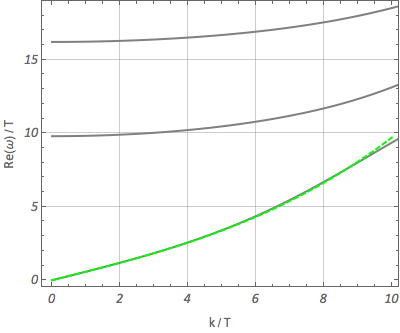} \quad \quad
\includegraphics[height = .25\textheight]{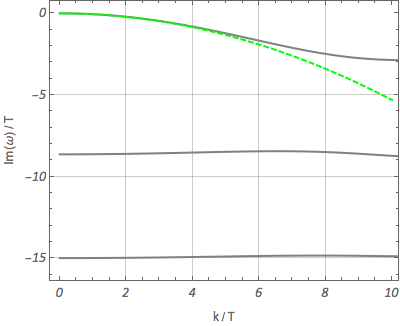}
\caption{Momentum-dependence of the lowest three QNMs in the sound channel. One clearly sees one long-lived excitation at low momentum whose frequency matches the dispersion relation (dashed green curve) for the hydrodynamic sound wave evaluated up to second order in gradient expansion. Furthermore, one can also notice a rather mild momentum-dependence of the transient modes at low enough momentum. This observation is one of the starting points for the HJSW theory discussed in \rfs{sect:beyond}. See Ref.~\cite{Kovtun:2005ev} for more information about QNMs in holography and recent Ref.~\cite{Fuini:2016qsc} for their behaviour at large momenta.}
\label{fig:qnmsk}
\end{center}
\end{figure}
\begin{figure}
\begin{center}
\includegraphics[height = .33\textheight]{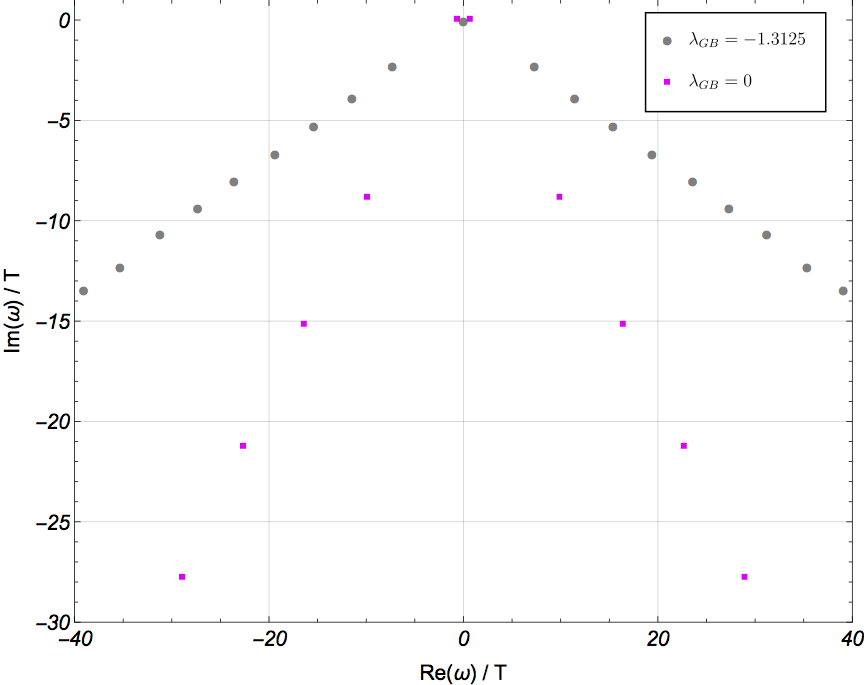} \\
\vspace{20 pt}
\includegraphics[height = .33\textheight]{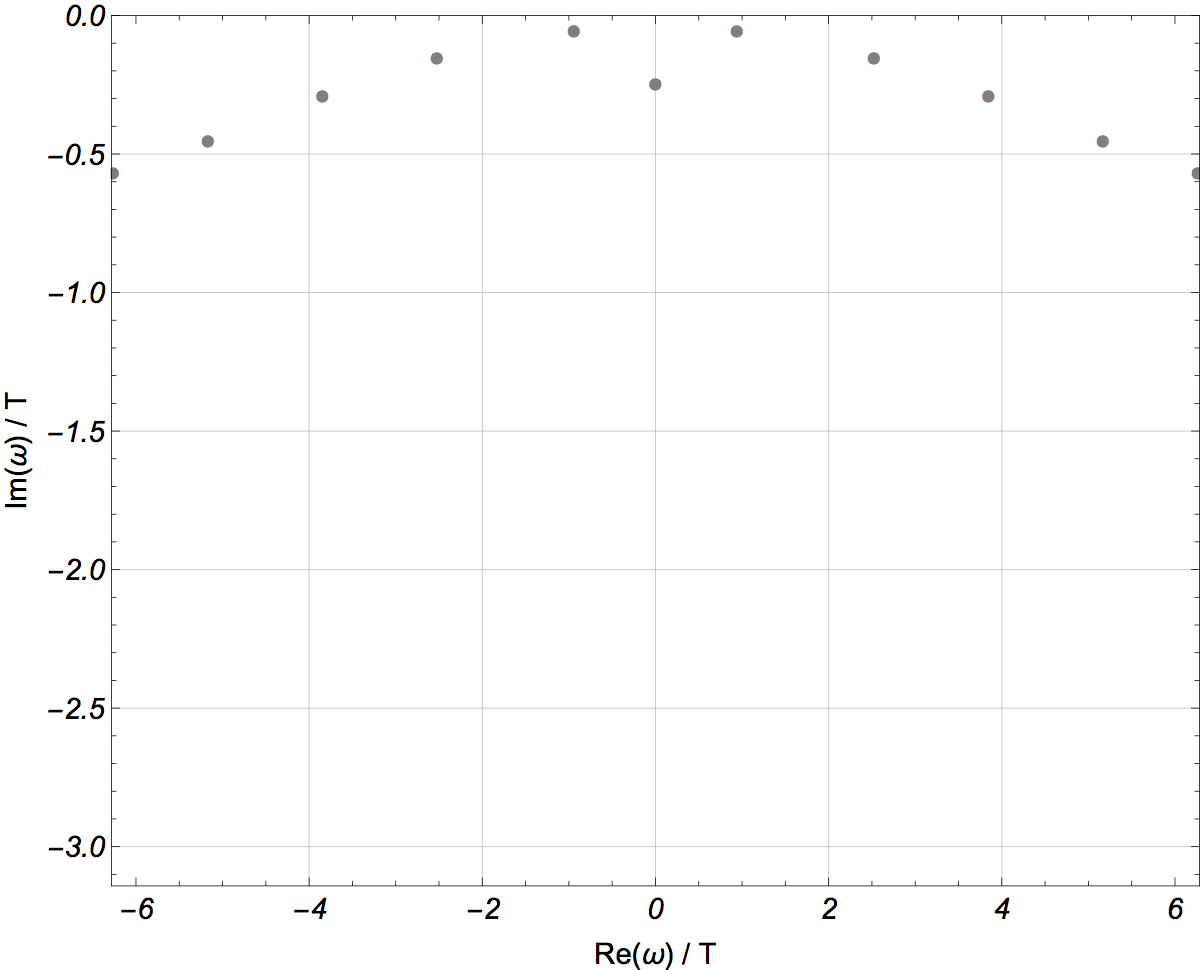}
\caption{Modes in sound channel in Gauss-Bonnet gravity at $k \approx 0.63 \, T$ for two different values of the Gauss-Bonnet coupling: $\lambda_{GB} = - 1.3125$ (top) and $\lambda_{GB}=-24.75$~(bottom). One sees a qualitative change in the structure of modes as compared to the Einstein gravity case ($\lambda_{GB} = 0$). In particular at non-zero values of $\lambda_{GB}$ new modes appear and at fixed momentum one can make them dominate late time dynamics instead of the long-lived $\lambda_{GB} = 0$ mode, as happens in both plots. Furthermore, in the bottom plot one sees new modes approaching the real axis (becoming long-lived), reminiscent of quasiparticles. There is a numerical evidence that poles condense and form a branch-cut as $\lambda_{GB} \rightarrow -\infty$, which would result in a picture somewhat similar to free SU($N_{c}$) gauge theory discussed in \rfs{sect:retardedfQFT}. The plot is based on data and methods from Refs.~\cite{Grozdanov:2016vgg,Grozdanov:2016fkt}.}
\label{fig:qnmskGB}
\end{center}
\end{figure}
Finally, let us also remark that qualitative changes in the spectrum of QNMs can occur by considering more complicated holographic equations of motion than Eqs.~(\ref{eq.Einstein}). A particularly interesting case to consider is supplementing Einstein-Hilbert action with negative cosmological constant with the Gauss-Bonnet term~(\ref{eq.SGB}), as recently considered in Refs.~\cite{Grozdanov:2016vgg,Grozdanov:2016fkt}. As can be seen in Fig.~\ref{fig:qnmskGB}, this theory is characterized by new excitations which in certain regions of parameter space dominate late-time response response and can even exhibit quasiparticle features.

%% file: sect-hydrod.tex
\section{Bjorken flow ab initio and hydrodynamization}
\label{sect:hydrod}

In this section we review first-principles-based studies of microscopic
models, revealing a transition to hydrodynamic behaviour starting from highly
non-equilibrium initial situations. We focus on the simplest available setup
by assuming that systems under consideration are conformally-invariant, as
well as transversely homogeneous and invariant under longitudinal Lorentz
boosts. In the context of hydrodynamics, one frequently speaks of
boost-invariant, or Bjorken flow~\cite{Bjorken:1982qr}.

\subsection{Boost invariance and the large proper-time expansion}
\label{sect:bif}

Boost invariance is a natural symmetry of systems produced at extremely high energies. Suppose that in the initial state
we deal with two colliding objects that approach each other with the rapidities\footnote{See Appendix~\ref{app:conv} for
the definition of rapidity.} $y$ and $-y$. In a Lorentz boosted reference frame, these two objects have rapidities
$y-y_f$ and $-y-y_f$, where $y_f$ is the rapidity of the new Lorentz reference frame with respect to the original one.
As long as $y_f \ll y$, the initial and final states look approximately the same in the two reference frames, which is
the origin of boost invariance. Strictly speaking, boost invariance is reached only in the limit $y \to \infty$. In
practice, one deals with a very large $y$ and a finite range of boosts $y_f$, for which the system may be regarded as
approximately boost invariant. For exactly boost-invariant systems, the rapidity distribution of produced particles,
$dN/dy$, is a (rapidity independent) constant. For systems which are approximately boost invariant, one expects that the
rapidity distribution exhibits a finite range plateau.

In practice it is often convenient to introduce the proper time $\tau$ and spacetime rapidity~$Y$ as new coordinates
replacing lab-frame time, $t$, and the spacial coordinate along the beam axis, $z$, see Eqs.~(\ref{pandx}), (\ref{rap})
and (\ref{srap}).\footnote{One should in general distinguish between the rapidity of a particle (measure of its
longitudinal velocity) and spacetime rapidity (spacetime coordinate).}  The statement of boost invariance then boils
down to saying that dependence of scalar quantities on the four Minkowski coordinates is reduced to dependence on the
proper time $\tau$ alone.

The expectation value of the boost-invariant energy-momentum tensor in a local rest frame takes the form
\bel{Tab}
T^{\mu}_{\,\,\, \nu} = \mathrm{diag}(\ed, \pL, \pT, \pT)^{\mu}_{\,\,\, \nu} \ ,
\eel
where the eigenvalues -- the energy density $\ed$ and the longitudinal and transverse pressures  $\pL$ and $\pT$
-- are functions of the proper time $\tau$ alone. Furthermore, in a conformal theory, the conditions of tracelessness
and conservation lead to the relations~\cite{Janik:2005zt}
\bel{PLT}
\pL = - \ed - \tau\, \dot{\ed}\ , \quad
\pT =  \ed + \f{1}{2} \tau\, \dot{\ed}.
\eel
The departure of these quantities from the equilibrium pressure at the same energy density
\bel{confEOS}
\peq \equiv \f{\ed}{3},
\eel
is a measure of how far a given state is from local equilibrium. It is therefore natural to study the approach to local equilibrium  by examining the behaviour of the normalized pressure anisotropy
\bel{rdef}
\pa \equiv \f{\pT-\pL}{\peq} .
\eel
This will be done in the following sections both for holography and RTA kinetic theory.

Conformal invariance also implies that energy density scales with the fourth power of temperature, that is,
\bel{efftemp}
\ed \sim T^4 .
\eel
For example, for  the equilibrium  \sym\ theory at strong coupling one finds~\cite{Witten:1998zw}
\bel{epsym}
\ed =\frac{3}{8} \pi^{2} N_c^2 \, T^{4},
\eel
see also \rfs{sect:blackbranes}. Similarly as in the case of the RTA kinetic theory discussed in Sec.~\ref{sect:ktrta}, the quantity $T$ in \rfn{efftemp} should be interpreted as an effective temperature, which is a measure of the local energy density~$\ed$. Note that this definition does not assume local equilibrium. In what follows we denote the effective temperature by $T$, since this introduces no ambiguity.~\footnote{Note that it may not be possible to do this in more general situations \cite{Casalderrey-Solana:2013aba,Arnold:2014jva}.}

In a boost invariant configuration all the physics is encoded in the dependence of the energy density (or, equivalently, the effective temperature) on $\tau$. This is a dramatic simplification compared to the general case, but it retains the key physical feature of power law relaxation to equilibrium, characteristic of hydrodynamics. This is  the prime reason why we decided to focus on this particular flow.

In conformal theories, such as those considered in this section, the late-time behaviour of the temperature is strongly
restricted by symmetry. In all microscopic models considered in the present section, the late-time behaviour is captured
by the following formal expression
\bel{thydro}
T(\tau) = \frac{\Lambda}{\left( \Lambda \tau \right)^{1/3}} \left(1 +
\sum_{k=1}^\infty \f{t_k}{\left( \Lambda \tau \right)^{2 k/3}} \right),
\eel
where $\Lambda$ is a dimensionful parameter which depends on the initial conditions. Indeed, it is the only trace of
initial conditions to be found in the late-time behaviour of the system.

The existence of such solutions in microscopic theories is of fundamental importance, since it describes the manner in
which the system approaches local equilibrium at late times.\footnote{Note this is a one-dimensionally expanding
systems, which in the limit of infinite proper time completely dilutes.} These late-time states will differ in
temperature at a given finite value of proper time; this dependence on the initial conditions is captured by the
dimensionful constant~$\Lambda$. When the system behaves to a good accuracy as described by a truncation of \rf{thydro}
to a finite number of terms (usually 2 or 3), we say that it has reached the hydrodynamic stage of its
evolution. The reason for this nomenclature is that the leading behaviour seen in \rf{thydro} appears in perfect fluid
hydrodynamics of Bjorken flow~\cite{Bjorken:1982qr} and subleading power-law corrections capture dissipative
hydrodynamic effects such as viscosity -- this will be discussed systematically in  \rfs{sect:brsss}.

It is interesting and useful to examine more closely how one can technically establish, given a numerically
calculated temperature history, that the hydrodynamic stage has been reached. To~this end it is very
convenient to formulate a criterion which is independent of the parameter~$\Lambda$,
whose value is different for each solution~\cite{Heller:2011ju} . First let us define the dimensionless variable
\bel{wdef}
w=\tau \, T(\tau),
\eel
which can be thought of as the proper time in units of inverse effective temperature; it has in fact a much
more profound significance, which is revealed in \rfs{sect:gradexp}. The key idea is to consider not the
temperature as a function of proper time, but rather a dimensionless observable such as the pressure
anisotropy \rfn{rdef} as a function of the dimensionless evolution parameter $w$. In a conformal theory, as the
hydrodynamic regime is reached, such a function is guaranteed to tend to a universal form, independent of the
initial, non-equilibrium state. This is because (as seen in \rf{thydro}) in the  hydrodynamic regime the only
trace of the initial state is the scale $\Lambda$, but this dimensionful quantity cannot appear in $\pa(w)$ at late times.

%
In most papers written about this subject the discussion was formulated in terms of the the dimensionless function (not
to be confused with the distribution function of kinetic theory!) defined as $f(w) \equiv \f{\tau}{w}
\f{dw}{d\tau}$. Using \rf{PLT} one can check that this object is trivially connected to the pressure anisotropy by the
relation $\pa(w)=18 (f(w)-2/3)$. Since in the context of Bjorken flow the pressure anisotropy $\pa(w)$ is a natural
observable with a clear physical significance, here we focus on this particular quantity rather than $f(w)$.
In practice, the relation 
\bel{fR}
\pa = 18 \left(\f{\tau}{w} \f{dw}{d\tau} -\frac{2}{3} \right)
\eel
can be used to calculate the pressure anisotropy given $T(\tau)$. 

Expressed in terms of the dimensionless variable $w$, the large proper-time expansion of $\pa$ takes the~form
\begin{equation}
\label{rgradex}
\pa(w) = \sum_{n=1}^{\infty} \pac_n \, w^{-n},
\end{equation}
i.e., it is a series in negative integer powers of $w$. This fact follows directly from \rf{fR} and
\rf{thydro} and, as we discuss in \rfs{sect:resu}, holds up to corrections that are exponentially suppressed
at large times (large $w$). The coefficients $\pac_n$ appearing in \rf{rgradex} are pure numbers, i.e., they are
independent of the parameter $\Lambda$ which distinguishes different histories of the system. The large-$w$
expansion\rfn{rgradex} is therefore a universal solution, determined only by the microscopic parameters of the
system under study.

\subsection{Strong coupling analysis using holography}
\label{sect:sym}

In the context of holography, the imposition of boost invariance reduces the
gravitational problem to one involving just two coordinates: the ``radial''
coordinate $u$ and a time coordinate identified with the proper-time $\tau$ on
the boundary. This has made it possible to carry out explicit calculations on
the gravity side of holography and translate the results into the language of
quantum field theory in four-dimensional Minkowski space.

Initially, analytic
calculations were performed in an expansion in inverse powers of the
proper-time. These calculations, reviewed below in Sections
\ref{sect:bifsym:grad} and \ref{sect:resu:holo}, reproduced first, as the
leading order late-time behaviour, the Bjorken description of boost-invariant
flow in the framework of perfect-fluid hydrodynamics (see
\rfs{app:bif}). Viscous as well as higher order corrections were subsequently
computed, placing the hydrodynamic description on the very firm ground of {\em
  ab initio} calculations in a strongly coupled Yang-Mills theory. This was
subsequently supplemented by numerical calculations of early time evolution
using methods of numerical general relativity -- these developments are
reviewed in \rfs{sect:bifsym:hdz} below.

\subsubsection{Large proper-time expansion}
\label{sect:bifsym:grad}

Einstein's equations~(\ref{eq.Einstein}) determining the bulk geometry possess an approximate analytic (semi-analytic at
high orders) solution in the form of a power series in inverse powers of $\tau$ first found in Ref.~\cite{Janik:2005zt}
and later developed in Refs.~\cite{Janik:2006ft,Nakamura:2006ih,Heller:2007qt,Heller:2008mb,Booth:2009ct,Heller:2013fn}.
We discuss what is perhaps a more streamlined version of this framework in \rfs{sect:resu:holo}.

Using the methodology of \rfs{sect:adscft}, see~\rf{eq.nearbdry}, one can read off the expectation value of hCFT's
energy-momentum tensor from the asymptotic behaviour of these solutions near the boundary. Proceeding in this manner
terms up to third order have been calculated analytically~\cite{Heller:2007qt,Janik:2006ft,Booth:2009ct}. These results
determine the energy density of hCFTs as a function of proper time. Expressed in terms of the effective temperature the
result reads
%
\small
\bel{tsym}
T(\tau) = \frac{\Lambda}{\left( \Lambda \, \tau \right)^{1/3}} \Big\{ 1 -
\frac{1}{6 \pi \left( \Lambda \, \tau \right)^{2/3}} +
\frac{-1 + \log{2}}{36 \pi^{2} \left( \Lambda \, \tau \right)^{4/3}} + \frac{-21 + 2\pi^{2} + 51 \log{2} - 24 (\log{2})^2}{1944 \pi^{3} \left(\Lambda \, \tau\right)^{2}}
+\dots\Big\}.
\eel
\normalsize
The form of this expansion matches \rf{thydro}. It can be translated into the large-$w$ expansion
of $\pa(w)$. At very late times the pressure anisotropy approaches zero, which
one interprets as reaching local equilibrium. The approach is governed by
\rf{rgradex}, with
%
\bel{rcoeffsholo}
\pac_1 = \f{2}{\pi}, \quad
\pac_2 = \f{2-2\log 2}{3\pi^2},\quad
\pac_3 = \f{15-2\pi^2-45\log 2+24 \log^2 2}{54 \pi^3}.
\eel
As discussed earlier in \rfs{sect:bif}, these coefficients are dimensionless and independent of the scale $\Lambda$ appearing in \rf{thydro}. The truncation of \rf{rgradex} keeping only the three leading terms with the above values of the expansion coefficients will be denoted by $\pa_H(w)$.

Up to this point, we have only discussed hydrodynamic behaviour at the microscopic level, but have not begun to develop the effective description of this regime. However, anticipating the developments discussed below in \rfs{sect:fund}, we wish to point out that the leading coefficient, $\pac_1$, is a multiple of the shear viscosity to entropy ratio,
\bel{rone}
\pac_1 =  8\,\f{\eta}{\sd},
\eel
so the value quoted in \rf{rcoeffsholo} reproduces the result $\eta/\sd=1/4\pi$ obtained in holographic models involving two-derivative gravity actions.

To summarise: the importance of \rf{tsym} lies in the fact that it represents
an {\em ab initio} calculation in a specific microscopic theory, which is
consistent with the form expected on the basis of hydrodynamics. This point
will be explored in greater detail in \rfs{sect:effective}, where we will also
discuss the significance of the coefficients appearing in the series
\rfn{tsym}. Note also that the large-time expansion in \rf{tsym}, or more
directly its counterpart in the language of $\pa(w)$, is a manifestation of
the fluid-gravity duality~\cite{Bhattacharyya:2008jc} which we discuss in
\rfs{sect:matching}.

\subsubsection{Hydrodynamization: emergence of hydrodynamic behaviour}
\label{sect:bifsym:hdz}

It was to be expected that starting from any initial state the system should evolve in such a way that at late
times it will be described by \rf{thydro} for some value of the constant $\Lambda$. The quantitative study of
this issue in a holographic setting translates into solving a set of nonlinear partial differential equations
with two independent variables. This task cannot be carried out analytically, but numerical calculations can be
set up in a relatively straightforward way. They involve three stages:
\begin{enumerate}
\item finding initial metrics which satisfy constraints implicit in
  the Einstein equations~(\ref{eq.Einstein});
\item evolving the geometry using a suitable numerical scheme, see e.g. Refs.~\cite{Heller:2012je,Chesler:2013lia,Cardoso:2014uka};
\item calculating the expectation value of the energy-momentum tensor in hCFT using \rf{eq.nearbdry}.
\end{enumerate}
The first numerical holographic calculations in this and similar contexts were
presented in Refs.~\cite{Chesler:2008hg,Chesler:2009cy,Chesler:2010bi}. Our
discussion here follows slightly later developments of
Refs.~\cite{Heller:2011ju,Heller:2012je,Jankowski:2014lna}.
The outcome of these efforts can be summarised by saying that it has been
verified directly in numerical simulations that the behaviour described by
\rf{tsym} emerges at late times for all initial states which were studied. The
cleanest way to see it will be through the use of the pressure anisotropy
$\pa$ as a function of the dimensionless clock variable $w$.

\begin{figure}[!t]%
\begin{center}
\includegraphics[height = .35\textheight]{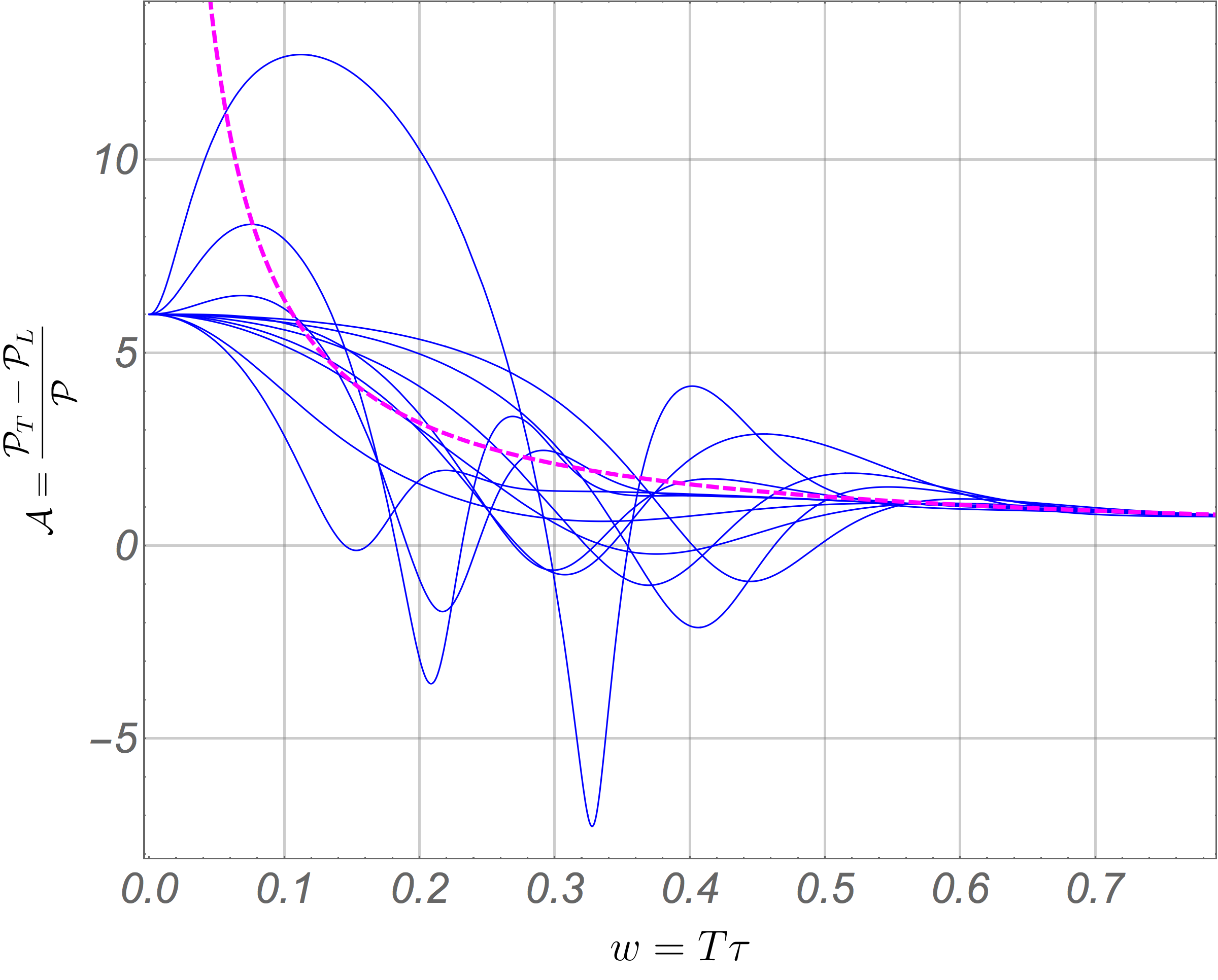} \caption{The pressure anisotropy for a small sample of solutions,
  represented by the blue curves. The dashed magenta curve is the first order truncation of $\pa_H(w)$. As will become
  more apparent in Fig.~\ref{fig:tilde}, the blue curves join the magenta curve when the anisotropy in the system is very
  sizable, much more than previously anticipated. The figure is based on numerical data taken from
  Ref.~\cite{Jankowski:2014lna} (where the pressure anisotropy was defined as $\pa/3$).}
\label{fig:asym}
\end{center}
\end{figure}

One should mention at this point, that it was shown in~\cite{Beuf:2009cx} that for small proper times only
even powers of $\tau$ can appear in a power series expansion of the energy density:
\bel{epstau}
\ed(\tau) = \sum_{n=0}^{\infty} \ed_n \tau^{2 n} ,
\eel
but there is no known constraint on what the leading power should be. Both groups
\cite{Heller:2011ju,Heller:2012je} and \cite{Jankowski:2014lna} assumed that the energy density approaches a nonzero
constant value $\ed_0\neq 0$ at $\tau=0$ where the initial conditions were set. Note that with this assumption it follows from \rf{rdef} that $R=6$ at $w=0$ and this value can be indeed seen in Fig.~\ref{fig:asym}. The same value of the initial pressure anisotropy is obtained within the CGC framework, see Ref.~\cite{Lappi:2006fp}, and necessarily involves negative longitudinal pressure ${\cal P}_{L} = - {\cal P}_{T}$. More generally for $\ed(\tau) \sim \tau^{2 n }$ at small $\tau$ one obtains $\pa(w = 0) = 6 + 9 \, n$, see Ref.~\cite{Wu:2011yd} for results motivated by Ref.~\cite{Grumiller:2008va} which assumes $n = 1$.

A given numerical solution for the energy density $\ed(\tau)$ starts out far from equilibrium, but the damping
of transient modes ensures, see \rfs{sect:qnm}, that after a sufficiently long time only hydrodynamic modes remain. Given the result of a numerical simulation for $\ed(\tau)$ one can check whether the hydrodynamic regime has been
reached by comparing  the pressure anisotropy $\pa(w)$ calculated from the simulation data with the hydrodynamic
form given by the truncated large-$w$ expansion $\pa_H(w)$.  Despite significant differences at early times, all
numerical solutions studied in \cite{Heller:2011ju,Heller:2012je,Jankowski:2014lna} exhibit this behaviour:
over 600 initial conditions have been evolved and in all cases it was found that hydrodynamics become a good
description at some $w<1$. This can be seen qualitatively in a plot of the pressure anisotropy in
\rff{fig:asym}, where a small set of histories is plotted.


In more quantitative terms, for a given solution $\ed(\tau)$ one can evaluate the resulting pressure
anisotropy $\pa(w)$, and then calculate the difference between this and the asymptotic form given by a
truncation of \rf{rgradex} with leading coefficients taken from \rf{rcoeffsholo}. This difference converges to
zero, and one can define the hydrodynamization ``time'' $w_{H}$ as the value beyond which the difference is
smaller than some threshold.
The distribution of hydrodynamization times $w_{H}$ obtained in
\cite{Jankowski:2014lna} on the basis of 600 histories is clustered around $w=0.6$, which is consistent with
the evidence found earlier in \rfc{Heller:2011ju}.

The essential physical message, seen already in \rff{fig:asym}, is that the truncated large-$w$ expansion
$\pa_H(w)$ typically becomes a good approximation at a time when the pressure anisotropy is still substantial, of the
order of 50\% of what would be the equilibrium pressure. For this reason it has been proposed that instead of speaking of thermalization which requires approximate local equilibrium, in particular approximate local isotropy, one should use a term such as hydrodynamization to describe this effect.

\subsection{Kinetic theory in the RTA}
\label{sect:bifrta}

The calculations reviewed above present a compelling picture of the emergence of universal behaviour at late
times when the coupling is strong. It is also possible to carry out analogous calculations in the weak coupling framework of kinetic theory.

As explained in \rfs{sect:ktrta}, one can simplify the potentially very complex collision kernel appearing in
the Boltzmann equation  by replacing it with a simplified expression, linearized  around the equilibrium
distribution which we take to be of the Boltzmann form (neglecting quantum statistics for simplicity).
Assuming the expansion along the $x^{3}$-axis and introducing
the boost-invariant variables~\cite{Bialas:1984wv,Bialas:1987en}
\bel{vardefs}
u = p_L \, x^{0} - E \, x^{3}, \quad v = E x^{0} - p_L \, x^{3},
\eel
the equilibrium distribution takes the form
\bel{boldist}
f_0(\tau,u,\pt) = \frac{1}{(2\pi)^3} \exp\left[ - \frac{\sqrt{u^2+\pt^2 \tau^2}}{\tau \, T(\tau)}  \right],
\eel
where the quantity $T(\tau)$ appearing here is identified with the effective
temperature and determined self-consistently by imposing the Landau-matching
condition discussed below.

The boost-invariant RTA Boltzmann equation reads~\cite{Baym:1984np,Florkowski:2013lza,Florkowski:2013lya}
\bel{bebif}
\trel \, \frac{\partial f(\tau, u, \pt)}{\partial \tau} = f_0(\tau, u, \pt)-f(\tau, u, \pt).
\eel
To ensure conformal symmetry we take the relaxation time to be of the form
\rfn{conf-teq}. The dependence of the temperature on the proper time is
determined dynamically by imposing the Landau matching condition
\bel{lmc}
\ed(\tau) = \f{6}{\pi^2} \,T^4(\tau),
\eel
where~\footnote{The
  value of the constant factor in \rf{lmc} is a consequence of adopting
  Boltzmann statistics.}
\begin{equation}
\label{eq.edkin}
\ed(\tau) = 4 \int d^4p \, \delta \left( p^2\right) \theta (p^0)
\, \frac{v^2}{\tau^2}\,  f(\tau,u,\pt) =
2 \int dP \, \frac{v^2}{\tau^2}\,  f(\tau,u,\pt)
\end{equation}
is the energy density (the factor 2 accounts for the spin degeneracy). Note again
that despite apparent linearity of the Boltzmann equation~\rfn{bebif}, the
Landau matching condition~\rfn{lmc} renders the problem of determining the
distribution function $f$ strongly nonlinear.

\subsubsection{Large proper-time expansion}
\label{sect:bifrta:grad}

To determine the coefficients $\pac_n$ in \rf{rgradex} we can proceed in a number of ways. What is perhaps the most streamlined formulation providing access to the large-$n$ behaviour of $\pac_{n}$ is presented in \rfs{sect:resu:rta}. The least efficient, but conceptually simplest way is to compute an iterative solution of the Boltzmann equation \rfn{bebif}, which can be done by setting up the iteration:
\bel{iter}
f_{n+1}(\tau, u, \pt) =  - \trel \, \frac{\partial f_{n}(\tau,
  u, \pt)}{\partial \tau}, \quad n \geq 0 ,
\eel
with $f_0$ given by \rf{boldist}. This is the Chapman-Enskog iteration discussed in the same context
in~\rfc{Jaiswal:2013npa}; the generated series of approximations is precisely the large proper-time
expansion. At each order one obtains an explicit expression for the distribution function at that order of
approximation, but one still needs to impose the Landau matching condition \rfn{lmc}, which determines the
proper-time dependence of the temperature $T(\tau)$ up to that order.

At order zero the Landau condition \rfn{lmc} identifies the function $T$ appearing in \rf{boldist} with the
effective temperature appearing in \rf{lmc}. Imposing \rf{lmc} at first subleading order leads to the equation
\bel{rta1}
3\, \tau  \, \dot{T} + T = 0,
\eel
whose solution is
\bel{rtatemp1}
T = \frac{\Lambda}{\left( \Lambda \tau \right)^{1/3}}
\eel
The integration constant appearing here was denoted by $\Lambda$
in order to allude to the solution\rfn{thydro}.

At order $K$ in place of \rf{rta1} one finds a differential equation of order
$K$. These equations  can be solved at large values of the proper time $\tau$
and posess solutions of the form given in \rf{thydro},
%
%
where the coefficients $t_k$ appearing there must be determined by imposing the
Landau matching condition\rfn{lmc}. It is important to note that matching at order
$K>1$ determines the coefficient $t_{K-1}$ of \rf{thydro}.
The coefficients $t_k$ for $k\geq K$ are not reliably determined by
matching at order $K$, since they will receive corrections from matching
at orders higher than $K$.

Once the late-time solution is determined to some order, one can calculate the expansion coefficients of the pressure anisotropy \rfn{rdef} in powers of the dimensionless variable $w$ (as described in \rfs{sect:bif}). The leading  coefficients read
\bel{rcoeffsrta}
\pac_1=8/5\ \gamma, \quad \pac_2=32/105\ \gamma^2, \quad \pac_3 = - 416/525 \, \gamma^3 .
\eel
As already noted in \rf{rone}, when matched to viscous hydrodynamics, the first coefficient is directly proportional to the shear viscosity to entropy ratio. The first relation in \rfn{rcoeffsrta} then translates to
\bel{etaos.rta}
\f{\eta}{\sd} = \f{\gamma}{5}.
\eel
This reveals the dual physical significance of the parameter~$\gamma$. On the one hand it determines the relaxation time, but on the other it reflects the viscosity of the fluid once hydrodynamic behaviour emerges. This is an important physical property of the RTA: it correlates the microscopic relaxation time and the viscosity at the level of the effective, hydrodynamic description.

\subsubsection{Emergence of hydrodynamic behaviour}
\label{sect:bifrta:hdz}

The numerical solution of the Boltzmann equation \rfn{bebif} has been extensively discussed in Refs.~\cite{Florkowski:2013lza,Florkowski:2013lya} following earlier work by Baym \cite{Baym:1984np}. This is done by reformulating the calculation of the energy density as the following integral equation for the function $T(\tau)$,
\begin{equation}
\label{eq.baym}
T^4(\tau) = D(\tau,\tau_0) \frac{\pi^2 {\cal E}^0(\tau)}{6}  +
\int_{\tau_0}^\tau d\tau^\prime \, \left( \frac{T(\tau')}{\gamma}
D(\tau,\tau^\prime) \right) \times \left(  T^4(\tau^\prime) \, H\left(\frac{\tau^\prime}{\tau}\right) \right),
\end{equation}
where we assumed the relaxation time $\tau_{\rm rel}$ of a conformal theory, see \rf{conf-teq},
%
\bel{ddef}
D(\tau_{2},\tau_{1}) = e^{- \frac{1}{\gamma}\int_{\tau_{1}}^{\tau_{2}} d \tau^{\prime \prime} T(\tau^{\prime
\prime})}
\eel
and
\bel{hdef}
H(s) = \frac{s^{2}}{2} + \frac{\arctan{\sqrt{\frac{1}{s^2} -1}}}{2 \, \sqrt{\frac{1}{s^2}
-1}} .
\eel
The information about initial distribution function is encoded in ${\cal E}^0(\tau)$ given by 
\begin{equation}
\label{eq.edkin0}
{\cal E}^0(\tau) = 4 \int d^4p \, \delta \left( p^2\right) \theta (p^0)
\, \frac{v^2}{\tau^2}\,  f(\tau_{0},u,\pt) =
2 \int dP \, \frac{v^2}{\tau^2}\,  f(\tau_{0},u,\pt).
\end{equation}
Note the explicit as well as implicit time-dependence in the measure and integrand in
\rf{eq.edkin}, where only the choice $\tau = \tau_{0}$ gives the initial energy density.

Equation~(\ref{eq.baym}) can be solved efficiently by fixed-point iteration~\cite{Florkowski:2013lya}. Knowing $T(\tau)$
one can find the remaining components of the energy-momentum tensor and get the quantity of interest, i.e. $\pa(w)$. It
should also be noted that one can rewrite \rf{eq.baym} directly in terms of $\pa(w)$ by, e.g., taking a derivative of
both sides with respect to $\tau$ and using \rf{fR}. This can be advantageous for some purposes, e.g. when comparing
different solutions at late times when high numerical accuracy is required to resolve subtle transient effects.

The transition to hydrodynamics in a model of QGP given by the RTA kinetic
theory has been recently studied, in an analogous way to that used for
strongly-coupled hQFTs, in Ref.~\cite{Heller:2016rtz}. This work, motivated by
earlier results obtained in Ref.~\cite{Keegan:2015avk}, discusses qualitative
similarities of the hydrodynamization process in the EKT extrapolated to
intermediate coupling, RTA kinetic theory, and strongly-coupled theories. In
particular, one can see very explicitly how generic initial conditions lead to
the late-time behaviour embodied by the universal late-time solution described
in the previous section. Below, we discuss those findings, together with
earlier results obtained in the EKT setting of Refs.~\cite{Kurkela:2014tea,
  Kurkela:2015qoa}.

\subsection{Hydrodynamization as a generic feature}
\label{sect:hydrod:generic}

As originally noticed in Ref.~\cite{Keegan:2015avk}, at the level of boost-invariant flow the patterns of thermalization
discovered in strongly coupled hCFTs and the kinetic theory models discussed in the previous section are not vastly
different. The key observation is that the leading late-time behaviour is always $1/w$, and different theories are
distinguished only by the coefficient. This means that if we consider the evolution of the anisotropy $\pa$ as a
function of a rescaled variable $\tilde{w}\equiv w/2 \pac_1$, the leading late-time behaviour of $\pa$ is universally
given by
\be
\pa_H(\tilde w) = \frac{2}{\pi \tilde w} + \mathcal{O}(\frac{1}{\tilde w^2}).
\ee
The choice of rescaling is somewhat arbitrary, but once the coefficient $\pac_1$ is scaled away, the leading
behaviour is completely universal. This means that if we compensate for the possibly very different values of
$\eta/\sd$ by using the rescaled variable $\tilde{w}$, the approach to equilibrium should be the same at
sufficiently large $\tilde{w}$. This can be checked by comparing the results of gravity calculations for
hCFTs with the results for models based on kinetic theory~\cite{Heller:2016rtz}.

\begin{figure}[t]
\center
\includegraphics[height=0.35\textheight]{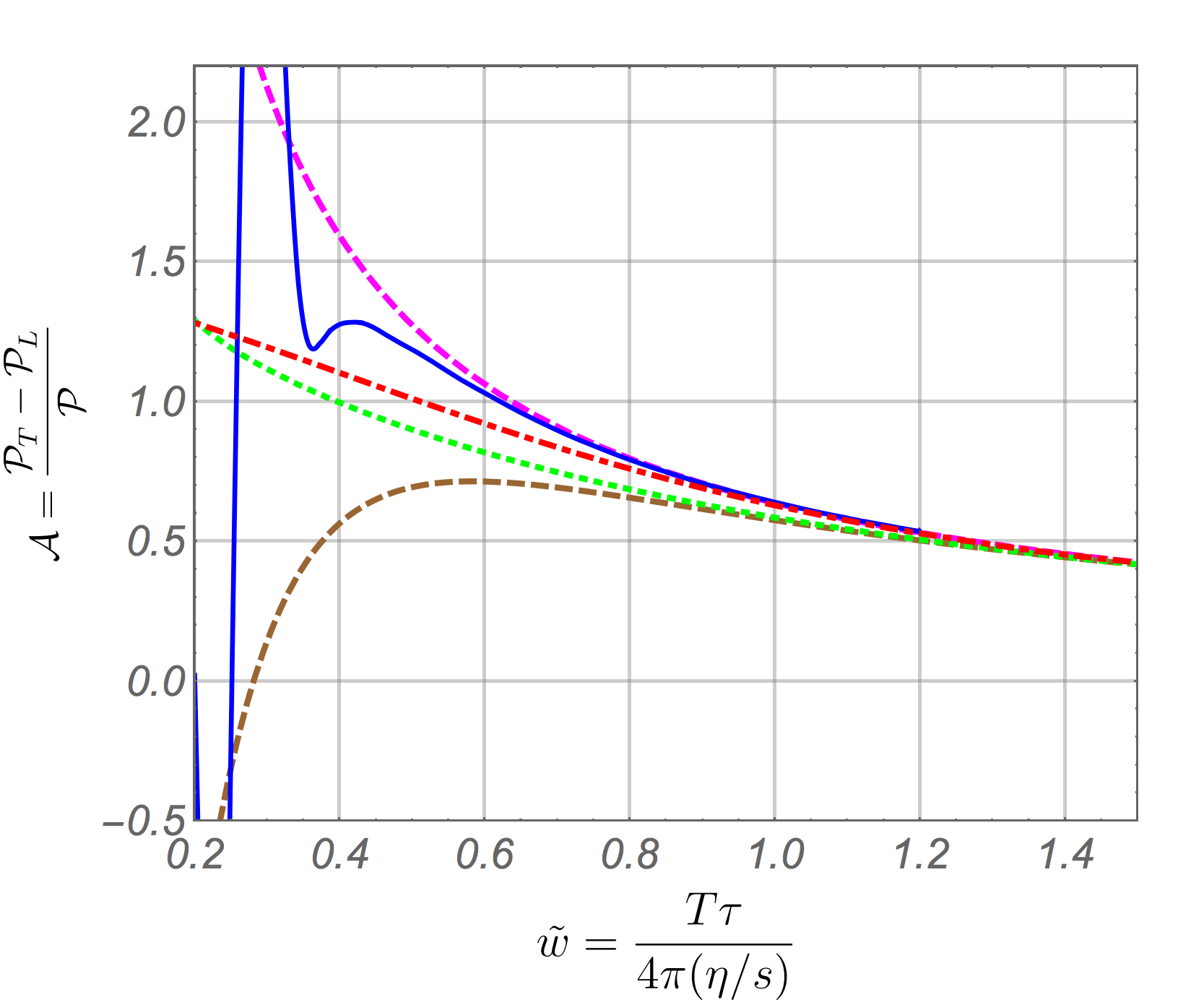}
\caption{Hydrodynamization compared: the blue curve is the result of an AdS/CFT simulation in \symm, the dashed-dotted
  red line comes from a numerical calculation using EKT, and the dashed magenta curve represents the gradient expansion
  truncated at first order.  The remaining curves come from calculations using the RTA starting from the same initial
  distribution, but with different values of $\eta/\sd$: the dashed brown curve corresponds to $\eta/\sd=1/4\pi$, while
  the dotted green curve corresponds $\eta/\sd=0.624$ (matching the data behind the red dotted line). The plot is
  adapted from Ref.~\cite{Heller:2016rtz}.}
%
\label{fig:tilde}
\end{figure}

Figure~\ref{fig:tilde} shows a comparison of the time evolution of the system
evolved according to the EKT \cite{Kurkela:2015qoa}, RTA using the methodology
of \cite{Baym:1984np,Florkowski:2013lya}, and numerical
holography~\cite{Heller:2011ju,Heller:2012je,Jankowski:2014lna}. For the EKT
and RTA simulations we adopted the initial condition used in
\cite{Kurkela:2015qoa}, whereas for the AdS/CFT simulation we took a typical
initial condition from \cite{Jankowski:2014lna}. The EKT curve describes the
evolution with 't Hooft coupling $\lambda = 10$ corresponding to $\eta/\sd
\approx 0.642$, while the holographic result has $\eta/\sd =
1/4\pi$. The~RTA~curves come from calculations with $\gamma$ fixed to
reproduce the value of $\eta/\sd$ of either model. As seen in the figure, the
evolution in all cases is similar, but distinct. Quite remarkably, in all
models, with vastly differing microphysics, the evolution converges to first
order viscous hydrodynamics roughly at the same $\tilde w \approx 1$ with a
large pressure anisotropy $\pa\approx 0.6-0.8$.
To summarize: the quantitative differences between weakly-coupled scenarios
extrapolated to intermediate couplings and genuinely strongly-coupled
scenarios arise largely from the numerical values of the hydrodynamization
time measured in units of inverse temperature, $w=\tau \, T$. In the strong
coupling scenario, $w \approx 0.7$ at the hydrodynamic threshold. In
phenomenological analysis of heavy-ion experiments hydrodynamic codes are
initialized typically at $w\approx 0.5$, which corresponds roughly to a time
$\tau = 0.5$ fm after the collision, with the temperature $T =$ 350 MeV at the
centre of the fireball at RHIC (see e.g. \cite{Broniowski:2008vp}). In the
weak coupling framework the hydrodynamization time can be much larger in
consequence of much larger shear viscosity (which, in the RTA, implies a
correspondingly longer relaxation time).

%% file: sect-fund.tex
\section{Fundamentals of relativistic hydrodynamics}
\label{sect:fund}

The microscopic calculations presented in the previous section point to universal behaviour at late times, which  we
associate with the long-lived subset of modes of equilibrium matter discussed in Secs.~\ref{sect:micro}
and~\ref{sect:adscft}. The developments reviewed in \rfs{sect:hydrod} assumed boost-invariance, but the picture which
emerges from them fits in perfectly with microscopic expectations reviewed in \rfs{sect:adscft},  which do not rely on
any symmetry assumptions. There we argued that the time evolution of non-equilibrium states in interacting systems can
be seen as comprising of two stages. The first is a transient stage during which non-hydrodynamic modes decay
exponentially fast on some scale determined by the microscopic theory. The second stage is governed by long-lived
hydrodynamic modes, which dominate the dynamics already at times when the system is still far from local thermal
equilibrium (hydrodynamization). This picture has been observed in a striking way in studies based on holography
(reviewed in \rfs{sect:sym}) as well as in models based on kinetic theory (reviewed in \rfs{sect:bifrta}). It is natural
to try to formulate an effective theory which captures the late time dynamics in terms of variables, which are
phenomenologically relevant and useful. Since the emergence of hydrodynamic behaviour is a feature seen in diverse
microscopic studies, one can formulate this effective description without assuming any specifics of the underlying
microscopic dynamics (such as, for example, the existence of quasiparticle excitations).

The universal character of late time behaviour -- the fact that no trace of initial conditions remains at late times --
comes from a significant reduction in the number of degrees of freedom characterizing the expectation value of the energy-momentum tensor. The minimum number of independent components of this matrix is four, since it needs to describe at least the fluctuations of the four conserved quantities associated with spacetime translation symmetries. These hydrodynamic fields could be taken to be the energy density $T^{00}$ and
momentum densities $T^{0i}$. Instead of these, in order to conform to standard practice in hydrodynamics, we will instead use the four-velocity local flow variable $U$ defined as the boost velocity from some fixed frame of reference to the local rest frame in which the momentum densities vanish, and the energy density $\ed$ in the local rest frame. A covariant way of introducing these quantities {\em at the microscopic level} is through the equation
\bel{udef}
\langle {\hat T}^\mu_{\,\,\nu} \rangle \, \unu = - \ed \umu
\eel
provided the vector $\umu$ is timelike. The goal of hydrodynamics is to formulate an effective  description of $\langle {\hat T}^{\mu \nu} \rangle$ in terms of classical fields which are analogous to and mimic the quantities $\ed, \umu$ defined by \rf{udef}.  The precise way in which this description, reviewed in the present section, can be compared to and reconciled with the microscopic theory is the subject of \rfs{sect:effective}.

\subsection{Dynamical variables and evolution equations: the perfect fluid}
\label{sect:hydrovars}

By a hydrodynamic description one means a theoretical framework that uses a
small set of {\it fluid variables} interpreted as the local energy density
$\ed(x)$ and local hydrodynamic flow vector $\umu(x)$, which is normalized as
$U^2=-1$.

The point of departure is the assumption that we are dealing with a system
which reaches global thermodynamic equilibrium at late times. At this stage
one can be agnostic about the fundamental physics governing this system: it
could be composed of well defined quasiparticles, but it need not be.

The equilibrium energy-momentum tensor in the rest-frame is given by
\bel{equilibrium}
\TmunuEQ = \hbox{diag}\left(\ed_{\rm EQ}, \peq(\ed_{\rm EQ}), \peq(\ed_{\rm EQ}), \peq(\ed_{\rm EQ})\right),
\eel
where we assume that the equation of state is known, so that the pressure
$\peq$ is a given function of the energy density $\ed_{\rm EQ}$. It is worth
stressing, that $\TmunuEQ$ is a classical object which we should identify with
the expectation value of the energy-momentum tensor operator in the underlying
quantum theory.

The components of the equilibrium energy-momentum tensor~\rfn{equilibrium} can be written in an arbitrary boosted frame of reference as
\begin{eqnarray}
\TmunuEQ = \ed_{\rm EQ} \umu \unu + \peq(\ed_{\rm EQ}) \Delta^{\mu \nu},
\label{TmunuEQ}
\end{eqnarray}
where $\umu$ is a constant boost velocity, and $\Dmunu$ is the operator that
projects onto the space orthogonal to $\umu$, namely
\begin{equation}
\Dmunu = \gmunu+\umu \unu.
\label{defDelta}
\end{equation}
Of course, the four-vector $\umu$ can equally well be regarded as a constant
fluid four-velocity.

The energy-momentum tensor of a perfect fluid is obtained by allowing the
variables $\ed$ and $\umu$ to depend on the spacetime point $x$.  In this way
one obtains
\begin{eqnarray}
\Tmunueq = \ed(x) \umu(x) \unu(x) + \peq(\ed(x)) \Dmunu(x).
\label{TmunuPerf}
\end{eqnarray}
In this equation, and in those which follow, the subscript ``eq'' refers to
local thermal equilibrium.

It is convenient to introduce local effective temperature $T(x)$ by the
condition that the equilibrium energy density at this temperature agrees with
the non-equilibrium value of the energy density, namely
\begin{eqnarray}
\ed_{\rm EQ}(T(x))  = \ed_{\rm eq}(x) = \ed(x).
\label{LM0}
\end{eqnarray}
One can then express the perfect fluid energy-momentum tensor in terms of the
fluid variables $T(x)$ and $U^\mu(x)$.  Note that the relativistic perfect
fluid energy-momentum tensor~\rfn{TmunuPerf} is the most general symmetric
tensor which can be expressed in terms of these variables without using
derivatives. We also note that the effective temperature $T(x)$ in this case
can be interpreted as a genuine thermodynamic quantity satisfying locally
thermodynamic identities.

The dynamics of the perfect fluid theory is provided by the conservation
equations of the energy-momentum tensor
\bel{conservation}
\p_\mu \Tmunueq = 0 .
\eel
These are four equations for the four independent hydrodynamic fields, that
form a self-consistent hydrodynamic theory.

\subsection{Viscosity and the Navier Stokes equations}
\label{sect:viscosity}

The essential physical element which is missing in the approach based
on~Eqs.~(\ref{TmunuPerf}) and (\ref{conservation}) is dissipation. The perfect
fluid evolution equations imply the covariant conservation of the 4-vector
\bel{entrocur}
s^\mu = \sd \, \umu
\eel
which is called the entropy current. This fact leads to the conclusion that
thermodynamic entropy is constant under hydrodynamic flows. This is not a
feature expected of real-world flows, except in idealized situations. Since
the conservation of the current \rf{entrocur} is in fact just a rewriting of
the projection of the conservation equation \rfn{conservation} onto $\umu$
(and using local thermodynamic relations, see e.g. \rfc{Romatschke:2009im}), a
natural step is to consider modifications of the energy-momentum tensor so as
to obtain a non-vanishing divergence of the entropy current \rfn{entrocur}
(possibly corrected by higher order terms). The requirement that this
divergence be nonnegative has been used as a guide in formulating hydrodynamic
equations (see e.g. \rfc{Romatschke:2009kr}) and has a beautiful holographic
interpretation~\cite{Bhattacharyya:2008jc,Bhattacharyya:2008xc,Booth:2009ct,Booth:2010kr,Booth:2011qy,Plewa:2013rga}.
Note that requiring that the divergence of the entropy be nonnegative
typically imposes constraints on parameters appearing in the hydrodynamic
energy-momentum tensor, see
Refs.~\cite{Romatschke:2009kr,Haehl:2014zda,Haehl:2015pja} for a very
comprehensive analysis of this issue.

To account for dissipation, we have to introduce correction terms to $\Tmunueq$ and write the complete energy-momentum tensor components $\Tmunu$ as
\begin{eqnarray}
\Tmunu = \Tmunueq + \Pimunu.
\label{Tmunu}
\end{eqnarray}
Here one can impose the condition $\Pimunu U_\nu=0$, which corresponds to the Landau definition of the
hydrodynamic flow $\umu$ \cite{LLfluid} specified by the formula
\begin{eqnarray}
T^\mu_{\,\,\,\nu} \unu = -\ed\, \umu.
\label{LandFrame}
\end{eqnarray}
This formula is the counterpart of \rf{udef} at the level of hydrodynamics.

It proves useful to further decompose $\Pimunu$ into two components,
\bel{pidecomp}
\Pimunu = \pimunu + \Pi \, \Dmunu,
\eel
which introduces the bulk viscous pressure $\Pi$ (the trace part of $\Pimunu$) and the shear stress tensor $\pimunu$ which is symmetric,
$\pimunu=\pinumu$, traceless, $\pimumu=0$, and orthogonal to $\umu$, $\pimunu \unul=0$.

Equation (\ref{Tmunu}) encodes ten independent components of $\Tmunu$ in terms
of the effective temperature, three independent components of $\umu$, five
independent components of $\pimunu$, and the bulk viscous pressure $\Pi$. We
note that the latter vanishes for conformal systems, for which the entire
energy-momentum tensor is traceless.

We still have the four conservation equations at our disposal,
\bel{fullconservation}
\p_\mu \Tmunu = 0 ,
\eel
but to obtain a closed system of  equations one needs additional information. The most
straightforward option is to express $\Pimunu$ in terms of the hydrodynamic
variables and their gradients. Since the perfect-fluid energy-momentum tensor
contains no gradients, it is natural to try to build up the theory as a series
of corrections in gradients. The simplest possibility is to include terms with
only a single gradient, which leads to the relativistic Navier-Stokes
theory~\cite{LLfluid}, in which the bulk pressure and
shear stress tensor are given by the gradients of the flow vector
\begin{eqnarray}
\Pi = -\zeta\, \p_\mu \umu, \quad  \pimunu =  -\eta \, \sigmamunu.
\label{NavierStokes}
\end{eqnarray}
Here $\zeta$ and $\eta$ are the bulk and shear viscosity coefficients,
respectively, and $\sigmamunu$ is the shear flow tensor defined as
\begin{eqnarray}
\sigmamunu = 2 \, \Dmunu_{\alpha \beta} \, \p^\alpha U^\beta,
\end{eqnarray}
where the projection operator $ \Dmunu_{\alpha \beta} $ has the form
\begin{eqnarray}
 \Dmunu_{\alpha \beta}  = \f{1}{2} \left(
 \Delta^\mu_{\,\,\,\alpha}  \Delta^\nu_{\,\,\,\beta} +  \Delta^\mu_{\,\,\,\beta}  \Delta^\nu_{\,\,\,\alpha}  \right)
 - \f{1}{3}  \Dmunu \Delta_{\alpha \beta}.
\end{eqnarray}

Equations (\ref{NavierStokes}) should be used in the conservation equations, (\ref{fullconservation}), that can be split into the parallel and orthogonal parts with respect to the flow vector $\umu$, namely
\bel{H1}
D \ed + (\ed + \peq + \Pi) \theta + \f{1}{2} \pimunu \sigma_{\mu\nu} = 0
\eel
and
\bel{H2}
(\ed + \peq + \Pi) D \umu = - \Dmunu \p_\nu (\peq+\Pi) - \Delta^\mu_{\,\,\,\nu} \p_\alpha \pi^{\alpha \nu},
\eel
where
\bel{capddef}
D\equiv  \umu \p_\mu, \quad \theta = \p_\mu \umu .
\eel
%

\subsection{The problem of causality}
\label{sect:causality}

Unlike nonrelativistic Navier-Stokes theory, its direct relativistic
generalisation \cite{LLfluid} is not consistent, because it is not
causal~\cite{Hiscock:1985zz,Lindblom:1995gp,Kostadt:2000ty}. Specifically, while the
evolution equations obtained by taking the shear stress tensor as in
\rf{NavierStokes} are covariant, they have solutions which propagate with
arbitrarily high velocities.

The easiest way to see the problem is to consider the hydrodynamic equations
linearized around the equilibrium configuration $T=\hbox{const},
\umu=\hbox{const}$. Looking for solutions of these linearized equations one
finds dispersion relations which determine the velocity of propagation $v\sim
k$, where $k$ is the modulus of the wave vector. This shows that at small
distances causality is violated. It has been argued (see e.g.
\rfc{Geroch:2001xs}) that this fact is not relevant, since one in any case
would not trust hydrodynamics on small scales. However, as stressed (for
example) in Refs.~\cite{Romatschke:2009im,Kostadt:2000ty}, this acuasality
leads to instabilities in numerical simulations for generic initial
conditions.

Adding extra terms with higher gradients on the right-hand side of (\ref{NavierStokes}) does not help to solve
the problem with causality.  The only known way to avoid it is to relax the assumption that $\Pi$ and
$\pimunu$ are expressed locally in terms of the hydrodynamic variables $T, \umu$, and (a finite number of)
their spacetime derivatives~\cite{Muller:1967zza}. This means that the shear stress tensor is treated as an
independent hydrodynamic field for which evolution equations must be provided. Thus, the conservation
equations alone are no longer enough to determine the dynamics of $\Tmunu$ and one needs to postulate
additional dynamic equations, or derive them (possibly by some heuristic means). The outcome, a closed set of
hydrodynamic equations, will clearly involve additional degrees of freedom beyond those already present in the
theory of the perfect fluid. To write down such equations additional assumptions or information will be
required, beyond what is embodied by conservation laws.

\subsection{Approaches to finding evolution equations}
\label{sect:hydro_exp}

A well-known and widely applied approach to the task of positing a set of closed equations for the
hydrodynamic fields is the M{\"u}ller-Israel-Stewart (MIS)
theory~\cite{Muller:1967zza,Israel:1976tn,Israel:1979wp}, in its modern incarnation described in
\rfc{Baier:2007ix,Baier:2006um} (BRSSS). This approach parametrises dominant contributions classified by
symmetries and the number of gradients. It does not make any special assumptions about the microscopic
dynamics, which accounts for its generality. A major advantage of this approach is that the resulting
equations are causal at least for some domain in the space of transport coefficients.

If we commit to a specific microscopic model, we gain the option of deriving (at least in a heuristic way) a
set of hydrodynamic equations which can provide a better physical picture than a generic approach such as the
MIS/BRSSS theory. An important testbed for this idea is provided by kinetic theory with an idealized collision
kernel, see Sec.~\ref{sect:ktrta}.  In this case a number of complementary approaches exist, which we now
briefly review.

One important approach developed in the context of kinetic theory makes use of the {\it hydrodynamic
expansion} of \rfc{Denicol:2012cn,Denicol:2012es,Denicol:2014loa}: this is the process  of constructing
dynamical hydrodynamic equations by making an expansion order by order in the Knudsen and inverse Reynolds
numbers.  The hydrodynamic expansion is performed around the local equilibrium state that corresponds to the
perfect-fluid limit~$\Tmunueq$. The Knudsen number is the ratio of the molecular mean free path length to a
representative physical length scale. On the other hand, the inverse Reynolds number describes deviations of
the energy-momentum components from their local equilibrium values --- they are typically expressed by the
ratios $\sqrt{\pimunu \pimunul}/\peq$ and $\Pi/\peq$.  The hydrodynamic expansion serves as a tool to systematically
derive hydrodynamic equations from kinetic theory.

Another approach to the task of formulating a closed set of hydrodynamic equations for models of kinetic
theory is known under the name of anisotropic hydrodynamics~\cite{Florkowski:2010cf,Martinez:2010sc} (for a
recent review see \cite{Strickland:2014pga}).  This name originated in the desire of finding hydrodynamic
equations suited to describing early stages of evolution of QGP produced in heavy-ion collisions. Equations of
anisotropic hydrodynamics were formulated in such a way as to capture some features of highly anisotropic
initial states, but also to ensure that at late times their predictions should be consistent with MIS/BRSSS.
In modern formulations, the equations of anisotropic hydrodynamics are suitable for studying arbitrary flows.

\subsection{M\"uller-Israel-Stewart (MIS) theory}
\label{sect:mis}

The simplest example of a theory of relativistic hydrodynamics where no
causality violations appear (at least at the  linearized level) is MIS theory,
which adds a single purely damped non-hydrodynamic degree of freedom. This is
done by replacing the
Navier-Stokes form of the shear-stress tensor (see \rf{NavierStokes}) by an
independent field satifying a relaxation equation of the
form~\cite{Muller:1967zza,Israel:1976tn,Israel:1979wp}
\bel{relax}
\left(\tpi U^\alpha \p_\alpha + 1 \right) \pimunu = - \eta \, \sigmamunu,
\eel
where $\tpi$ is a new parameter called the relaxation time.  Equation \rfn{relax} guarantees that at late
times the shear stress will tend to the Navier-Stokes form \rfn{NavierStokes} -- this point will be made much
more explicit below (in \rfs{sect:effective}). At the same time, it introduces new modes of propagation and
alters the dispersion relations in a way which renders the propagation velocity finite.

The relaxation equation described in \rfc{Israel:1979wp} includes some additional terms with new transport coefficients.
In fact, a number of variants of the MIS equations have been written down and applied to QGP evolution -- some of these
will be reviewed in \rfs{sect:nonconf}. For simplicity of presentation we will focus on the case of conformal
hydrodynamics, which is also interesting for comparison with the results of holographic calculations.  For a conformal
theory in Minkowski spacetime the energy-momentum tensor must be traceless, which requires a conformal equation of
state, vanishing of the bulk viscous pressure, and tracelessness of the shear-stress tensor. These features must be
preserved under time evolution, which requires augmenting \rf{relax} by additional terms. The simplest way to present
the resulting equation is to realize that the additional terms complete the spacetime derivative in \rf{relax} to a
Weyl-covariant derivative denoted by $\D$:
\bel{crelax}
\left(\tpi U^\alpha \D_\alpha + 1 \right) \pimunu = - \eta \sigmamunu.
\eel
The explicit form of the derivative $\D$ is given in \rfs{app:weyl}. Its main feature is that it preserves the
tracelessness and transversality of the shear stress tensor under time evolution. We will refer to equation
\rfn{crelax} as the conformal MIS equation.

Modes of this theory have been worked out in \rfc{Baier:2007ix} (see also \rfc{Romatschke:2009im}). Note that this analysis is not sensitive to the presence of the additional terms which appear in BRSSS theory (discussed below in \rfs{sect:brsss}). As discussed in \rfs{sect:linresp} and assuming momentum aligned with the $x^{3}$-direction, one obtains three channels (we just list independent components):
\begin{itemize}
\item scalar channel: non-vanishing $\delta \pi^{12}$;
\item shear channel: non-vanishing $\delta u^{a}$ and $\delta \pi^{3 a}$ with $a = 1, 2$;
\item sound channel: non-vanishing $\delta T$, $\delta u^{3}$, $\delta \pi^{33}$.
\end{itemize}
For example, in the sound channel one obtains the following dispersion relation:
\bel{disprel}
\omega^3 + \f{i}{\tpi} \omega^2 - \f{k^2}{3} \left(1 + 4 \, \f{\eta/\sd}{T \, \tpi} \right) \,\omega - \f{i k^2}{3 \tpi} = 0 .
\eel
For small $k$ one finds a pair of hydrodynamic modes (whose frequency tends to zero with $k$)
\bel{hydrom}
\omega_{\rm H}^{(\pm)} = \pm \f{k}{\sqrt{3}} - \f{2 i}{3 \, T} \, \f{\eta}{\sd} \,k^2 + \ldots
\eel
and a single transient mode
\bel{nonhydrom}
\omega_{\rm NH} = - i \left(\f{1}{\tpi} - \f{4}{3 \, T} \f{\eta}{\sd} \, k^2\right) +
\ldots
\eel
The dominant mode at long wavelengths is the one whose imaginary part is the least negative, so at large distances the hydrodynamic modes dominate.

Using the dispersion relation \rfn{disprel} we can calculate the speed of propagation of linear perturbations. The result is
\bel{velo}
v = \f{1}{\sqrt{3}} \sqrt{1 + 4 \,\f{\eta/\sd}{T \tpi}} .
\eel
This formula implies that as long as the relaxation time is sufficiently large,
\bel{causal}
T \, \tpi > 2 \eta/\sd ,
\eel
there is no transluminal signal propagation. This is clearly not the case if one tries to eliminate the
relaxation time by taking it to vanish.

\subsection{The non-hydrodynamic sector as a regulator}
\label{sect:regulator}

The presence of transient modes is essential for the consistency of the hydrodynamic description in the
relativistic case. The success of relativistic hydrodynamics in describing the dynamics of QGP can be ascribed
to the exponential decay of these modes, which leads to the fast emergence of quasiuniversal, attractor
behaviour of this system \cite{Heller:2015dha,Romatschke:2017vte}. This will be discussed in detail in \rfs{sect:brsss}; here we wish to focus on another aspect: the inequality \rfn{causal} suggests that the non-hydrodynamic sector
of relativistic hydrodynamics may be thought of as a regulator, ensuring that the speed of propagation does not exceed the speed of light.

This regulator cannot be removed, but its effects may or may not be practically significant in the regime of
interest.  It is important to understand when these effects may be ignored, otherwise one may be studying the
physics of the regulator rather than universal hydrodynamic behaviour. This is certainly the case at very
early times. It will also be an issue in the case of small systems~\cite{Habich:2015rtj,Spalinski:2016fnj},
where it may happen that the non-hydrodynamic modes do not have time to decay and hydrodynamic simulations
become sensitive to the choice of the non-hydrodynamic sector -- that is, to the choice of
regulator.\footnote{One can think of the regulator sector as an analogue of the notion of a
``UV-completion'', which arose in the context of effective field theories.}

One can try to make this a little more quantitative by writing down explicitly the condition for the
non-hydrodynamic mode to be subdominant using \rf{disprel}. This leads to the
condition~\cite{Spalinski:2016fnj}
\bel{bound}
R T > 2\pi \sqrt{2 (T\tpi) (\eta/\sd)} \ .
\eel
where $R$ is the spatial extent of the system. The violation of this bound, crude as it is, is an indication that the regulator sector cannot be ignored. The interesting point is that for reasonable values of the parameters this bound translates into a rather weak condition on the final charged particle multiplicity
\bel{mulbound}
\left(\f{dN}{dY}\right)_{\rm MIN} \approx 3 \ .
\eel
which is consistent with another (but related) line of reasoning developed in \rfc{Habich:2015rtj}. This conclusion makes it less surprising that relativistic hydrodynamics is successful not only in describing heavy ion collisions, but also for the case of pA, or even pp collisions.

If the bound \rfn{bound} is violated, it may be necessary to compare different regulators. Examples of
hydrodynamic theories with a qualitatively different non-hydrodynamic sector were discussed in
Ref.~\cite{Heller:2014wfa} and will be reviewed in \rfs{sect:beyond} below.

%% file: sect-eff.tex
\section{Hydrodynamics as an effective theory -- insights from holography}
\label{sect:effective}

As stressed in \rfs{sect:hydrovars}, once dissipative effects are incorporated within a hydrodynamic
framework, one loses the universality of perfect fluid theory, and many different sets of hydrodynamic
equations are possible. This raises the question of how they are to be compared, and how they can be
reconciled with computations carried out directly in the microscopic theory. In this section we argue that a useful way to proceed is to compare the gradient expansion of a microscopic theory with the gradient expansion generated from hydrodynamic models. This way one can try to mimic the hydrodynamic behaviour of the microscopic theory in a systematic, quantitative way. Throughout this section we assume conformal invariance, apart from \rfs{sect:nonconf}. The reasons for this are twofold. From the point of view of holography this allows us to focus on its most complete and best understood instance. The second reason is simplicity: the assumption of conformality restricts the number of terms appearing in hydrodynamic equations in a significant~way.

\subsection{The gradient expansion as an infinite series}
\label{sect:gradexp}

We have seen is \rfs{sect:fund} that causal theories of relativistic hydrodynamics are constructed in such a way as to reproduce at late times the approach to equilibrium captured by the Navier-Stokes terms. Indeed, solving \rf{crelax} by iteration, starting from the Navier-Stokes term, one obtains a formal solution in the form of an infinite series graded by the numer of spacetime gradients
\bel{pigradex}
\pimunu = - \eta \sigmamunu + \tpi U^\alpha\p_\alpha \left(\eta \sigmamunu \right) + \dots
\eel
For the case of BRSSS theory this expansion will be discussed at length in \rfs{sect:brsss}. The point we wish
to make here is that the gradient expansion of the energy-momentum tensor in causal theories of hydrodynamics
contains an infinite number of terms. These  expansions arise as a generic late-time solution. Given the set
of evolution equations for the shear stress tensor, we can always find such a solution explicitly by writing
down the most general gradient series consistent with Lorentz symmetry (and any other constraints, such as
perhaps conformal invariance), and determine the scalar coefficient functions by using the evolution equation
order by order in gradients. This leads to a formal, infinite series expansion in powers of gradients of the
fluid variables $T(x)$ and $\umu(x)$:
\bel{gradexp}
\Tmunu &=& \Tmunueq + \hbox{powers of gradients of } T \hbox{ and } \umu .
\label{TmunuGrad}
\eel
By comparing such formal solutions, one may quantify differences between different hydrodynamic
theories~\cite{Florkowski:2016zsi}. It is important to note at this point, that at each order in the number of
gradients there is a finite number of terms which can appear in \rf{gradexp}. These terms are of course
subject to the constraints of Lorentz covariance, and possibly other symmetries, such as conformal symmetry.
The terms which can appear in \rf{gradexp} have been classified up to
second~\cite{Baier:2007ix,Romatschke:2009kr} and third~\cite{Grozdanov:2015kqa} orders.

Crucially, an infinite expansion of the form of \rf{gradexp} also arises in microscopic theories. We have
already seen this implicitly in \rfs{sect:hydrod} for the special case of Bjorken flow, where we in fact
referred to the large-$w$ expansion as the gradient expansion. We can now fully justify using this term: once we recognize that for Bjorken flow, see e.g. \rf{U},
\be
\sqrt{\sigma_{\mu \nu} \, \sigma^{\mu \nu}} \sim \f{1}{\tau}
\ee
we can see that $w$ is the only dimensionless quantity of order one in the gradient expansion which is consistent with boost invariance:
%
\be
\f{1}{T} \, \sqrt{\sigma_{\mu \nu} \, \sigma^{\mu \nu}}  \sim \f{1}{w} .
\ee
This shows that the large-$w$ expansion is in fact precisely the gradient expansion for the special case when boost-invariance is imposed. Furthermore, we also see that the gradient expansion is a partial resummation of the large proper-time expansion of Bjorken flow.

This example demonstrates explicitly that the gradient expansion of the energy-momentum tensor expectation value in a
microscopic theory contains an infinite series of terms. Comparing such a microscopic calculation to the
gradient solution of a given hydrodynamic model provides a definite and unique way to relate the parameters of
the hydrodynamic description to those of an underlying theory. From this perspective we might say that
phenomenological theories of hydrodynamics can be viewed as theoretical devices constructed in such a way as
to reproduce the gradient expansion of a given microscopic theory to some order in gradients.

\subsection{Matching MIS to holography and BRSSS theory}
\label{sect:matching}

A beautiful illustration of these ideas is fluid-gravity duality~\cite{Bhattacharyya:2008jc}: the relationship
between the near-equilibrium behaviour of hCFT plasma, or more generally hQFT plasma, and a generalisation of
MIS theory. This is in fact a direct extension of the large proper-time expansion to general flows. The basic
insight appears already in \rfc{Janik:2005zt}, where it was observed that the leading hydrodynamic effect
corresponds to a ``slow'' dependence of the black-brane horizon on the proper time see also
\rfs{sect:resu:holo}. In a general setting this suggests looking for solutions in which the horizon position
in the bulk of spacetime is allowed to depend on the boundary spacetime position $x$. The key insight
of~Ref.~\cite{Bhattacharyya:2008jc} was to consider a boost and dilation of the black brane solution given by
\rf{eq.AdS-Schw-EF} and to allow the boost and dilation parameters to depend on $x$ in a smooth way. This
naturally leads to a solution of the Einstein equations which takes the form of an expansion in the gradients
of the $x$-dependent parameters $T, U^\mu$, up to first order in gradients given by
\small
\beal{fgdmetric}
ds^{2} = \frac{L^{2}}{u^{2}} \left\{2\, U_\mu dx^\mu \, [du + \WA_\nu dx^\nu]
+ [P_{\mu\nu} - \left( 1 - \pi^{4} T^{4} u^{4} \right) U_\mu U_\nu ] dx^\mu dx^\nu 
+  \f{F(\pi T u)}{\pi T}  \sigma_{\mu\nu} dx^\mu dx^\nu \right\},
\eea
\normalsize
where $\WA_\nu$ is defined in \rf{weylconn} and
\bel{fordone}
F(x) =  \arctan(x) + \log(1+x) + \half \log(1+x^2).
\eel
Remarkably, at each order of this gradient expansion, the constraint equations among Einstein's
equations~(\ref{eq.Einstein}) take the form of conservation laws for the boundary energy-momentum tensor, which has the
form of a gradient expansion in the boost and dilation parameters.  The holographic dictionary, see
Eq~(\ref{eq.nearbdry}), therefore leads to a form of the expectation value of the energy-momentum tensor in a hCFT which
has the form of a gradient expansion in local temperature (the dilation parameter) and 4-velocity (the boost
parameters).

The result of this holographic calculation extended to second order in gradients~\cite{Bhattacharyya:2008jc} is
consistent with what one a priori could have expected based only on the symmetries of the problem. Indeed, all the terms
which can appear in such calculations in relativistic conformal hydrodynamics have been
classified~\cite{Baier:2007ix}. Allowing for a fixed, but not necessarily flat, background metric one has
\bel{emt2}
\pimunu &=& -\eta \sigmamunu + \eta \tpi  \D\sigmamunu +
\kappa\left(R^{\langle\mu\nu\rangle}-  2 U_\alpha
R^{\alpha\langle\mu\nu\rangle\beta}  U_\beta\right) \nonumber \\
  &+& \lambda_1 {\sigma^{\langle\mu}}_\lambda \sigma^{\nu\rangle\lambda}
  + \lambda_2 {\sigma^{\langle\mu}}_\lambda \Omega^{\nu\rangle\lambda}
  + \lambda_3 {\Omega^{\langle\mu}}_\lambda \Omega^{\nu\rangle\lambda} + \ldots \, ,
\eel
where
\bel{weyld}
\D\sigmamunu = {}^\langle D\sigmamunu{}^\rangle +
\frac1{3} \sigmamunu (\partial_{\alpha} \, U^{\alpha})
\eel
and
\bel{omdef}
\Omega^{\mu\nu} = \f{1}{2} \Delta^{\mu\alpha} \Delta^{\nu \beta}
\left(\nabla_\alpha U_\beta - \nabla_\beta U_\alpha \right),
\eel
while $\nabla_\alpha$ is the covariant derivative and  $R^{\mu\nu\rho\sigma}$ and $R^{\mu\nu} (R)$ denote the Riemann
tensor and Ricci tensor (scalar). The coefficients $\eta$, $\tpi$, $\kappa$, $\lambda_{1}$, $\lambda_{2}$ and
$\lambda_{3}$ are scalar functions of the effective temperature. The ellipsis in \rfn{emt2} denotes terms with 3 and
more derivatives of fluid velocity, temperature and background metric. The angular brackets denote taking the symmetric,
orthogonal and traceless part of a tensor, namely
\bel{Amunu}
{}^{\langle}   A^{ \mu\nu \rangle}   \equiv A^{\langle \mu\nu \rangle} = \f{1}{2}   \Delta^{\mu\alpha} \Delta^{\nu \beta} \left( A_{\alpha \beta} + A_{\beta\alpha} \right)
- \f{1}{3}  \Delta^{\mu\nu} \Delta^{\alpha \beta} A_{\alpha \beta} .
\eel
The form of \rf{emt2} is a consequence of Lorentz and conformal symmetries, as well as a choice of integration constants which is equivalent to a choice of frame in hydrodynamics (see e.g. \rfc{Bhattacharya:2011tra}).

In conformal theories, the only equilibrium scale is set by the temperature and the non-trivial theory-dependent content of thermodynamic quantities and transport coefficients $\eta$, $\tpi$, $\kappa$, $\lambda_{1}$, $\lambda_{2}$ and $\lambda_{3}$ sits in their dimensionless combinations with appropriate powers of the local temperature (or, equivalently, the entropy density $s$):
\beal{CsInBRSSS}
\eta &=& C_\eta\ \sd,\qquad
\tpi = \frac{ C_{\tpi}}{T}, \qquad
\kappa = C_\kappa \frac{\eta}{T}\nn\\
\lambda_1 &=&  C_{\lambda_1} \frac{\eta}{T},\qquad
\lambda_2 =  C_{\lambda_2} \frac{\eta}{T},\qquad
\lambda_3 =  C_{\lambda_3} \frac{\eta}{T}.
\eel
The relevant dimensionless constants for \symm\ in the holographic regime read~\cite{Bhattacharyya:2008jc}:
\beal{symvalues}
C_\eta &=& \frac{1}{4 \pi},\qquad
C_\kappa = \frac{1}{\pi},\qquad
C_{\tpi} = \frac{2-\log (2)}{2 \pi} , \nn\\
C_{\lambda_1} &=& \frac{1}{2\pi}, \qquad
C_{\lambda_2} = -\frac{\ln 2}{\pi}, \qquad
C_{\lambda_3} = 0.
\eel
Assuming that one wants to model the late-time dynamics of, say~\symm, using the conformal MIS equation, it would be natural to match the gradient expansion of the latter with Eq.~(\ref{emt2}) to a desired order in derivatives, e.g. to second order. However, the gradient expansion of the conformal MIS equation \rfn{relax} reads (up to second order in gradients)
\bel{mis2}
\pimunu = -\eta \sigmamunu + \eta \, \tpi \, \D\sigmamunu.
\eel
Comparing the above to \rf{emt2}, we thus see that the expression\rfn{mis2} is not complete at second order,
since it does not allow to account for all possible gradient structures which can appear. Indeed, as seen in
\rf{emt2}, one cannot reproduce the result of the  holographic calculation of \rfc{Bhattacharyya:2008jc}
at second order in gradients by using MIS theory. The solution is to include more terms on the RHS of
\rf{relax}. The point made in \rfc{Baier:2007ix} was that one needs to modify the MIS relaxation equation
\rfn{relax} in such a way that it generates all second order terms admitted by symmetries with tunable
coefficients. If that is ensured, then one is guaranteed to match any microscopic calculation up to second
order.

Specifically, Baier et al. noted~\cite{Baier:2007ix} that if one replaces $\sigma^{\mu \nu}$ by $-\pi^{\mu \nu}/\eta$ in some terms of \rf{emt2}, one can rewrite that relation in a form which resembles \rf{relax}:
%
\beal{brsss}
\left(\tpi U^\alpha \D_\alpha + 1 \right) \pimunu
&=& -\eta \sigmamunu + \kappa\left(R^{\langle\mu\nu\rangle}- 2 U_\alpha R^{\alpha\langle\mu\nu\rangle\beta} U_\beta\right) + \nn\\
&+&  \lambda_1 {\pi^{\langle\mu}}_\lambda \pi^{\nu\rangle\lambda} +
\lambda_2 {\pi^{\langle\mu}}_\lambda \Omega^{\nu\rangle\lambda} +
\lambda_3 {\Omega^{\langle\mu}}_\lambda \Omega^{\nu\rangle\lambda}
\eel
The replacement of $\sigma^{\mu \nu}$ by $-\pi^{\mu \nu}/\eta$ which leads to \rf{brsss} is allowed at first order in gradients due to \rf{NavierStokes}. The resulting equation clearly generates the desired expansion \rfn{emt2} plus an infinite number of higher order terms with coefficients expressed in terms of the scalar functions already present in \rf{emt2} and their spacetime derivatives. This particular way of turning the gradient expansion result of a microscopic calculation into an effective hydrodynamic evolution equation is not unique, but it guarantees matching up to second order in the gradient expansion. Other possibilities exist which coincide with the above up to second order, but differ at higher orders. At the level of the gradient expansion this may not be a significant issue, but the properties of the differential equations will be different and some may be more suitable than others for numerical evaluation and phenomenological applications.

Finally, let us comment briefly on the physics of shear viscosity in hQFTs. It has been understood very early on, see Ref.~\cite{Buchel:2003tz}, that whenever the gravitational action does not contain terms of second or higher order in curvature, the ratio of shear viscosity to entropy density takes the universal value given in \rf{symvalues}. This occurs for conformal and non-conformal hQFTs in any spacetime dimension, also when on the gravity side fields other than the bulk metric are nonzero. This observation, together with other indications, has led to the famous conjecture that $1/4\pi$ is the lowest value of the shear viscosity to entropy density ratio permitted in physics~\cite{Kovtun:2004de}. We know now due to Refs.~\cite{Kats:2007mq,Buchel:2008vz} that this conjectured KSS bound is certainly violated in certain hCFTs whose gravity dual contains the Gauss-Bonnet term~(\ref{eq.SGB}). In these cases, the shear viscosity is given by the exact\footnote{Since viscosity cannot get negative (otherwise one has an instability in the system), \rf{etasGB} explains why one should certainly not consider $\lambda_{GB} > \frac{1}{4}$, see also the discussion below \rf{eq.SGB}.} expression~\cite{Brigante:2007nu}
\begin{equation}
\label{etasGB}
\frac{\eta}{\sd} = \frac{1}{4\pi} \left( 1- 4 \, \lambda_{GB} \right),
\end{equation}
and Ref.~\cite{Buchel:2008vz} found a very specific class of hCFTs for which gravity duals are consistent and characterized by a very small positive $\lambda_{GB}$. The emerging picture is that viscosity in consistent holographic QFTs is very close to the famous result, but can be a tiny bit lower. How much lower than $1/4\pi$ $\eta/\sd$ can get in a controllable setting and if any viscosity to entropy density bound actually exists do not really seem to be clear at the moment of writing this review.

\subsection{Hydrodynamics without conformal symmetry}
\label{sect:nonconf}

The point of departure for the arguments leading to the BRSSS evolution equation \rfn{brsss} was the
observation expressed in \rf{emt2}, which encapsulates the information about the symmetries of the underlying
microscopic theory, specifically Lorentz and conformal invariance. Conformal invariance offers significant
constraints which greatly reduce the number of terms which can appear in \rf{emt2} if only Lorentz covariance
is imposed. Furthermore, for QGP the assumption of conformal symmetry is a reasonable one at temperatures
above the chiral transition. However, this symmetry is an approximation which clearly has to be abandoned at
lower temperatures, and \rf{emt2} has to be replaced by a more general expansion. The terms which can appear
after imposing Lorentz covariance alone have been classified \rfc{Romatschke:2009kr}. Since there is a
significant number of allowed terms (2 at first order and 15 at second order -- compared to 1 and 5
respectively in the conformal case), it is hard, and probably impractical to try to include them all in a
generalized MIS relaxation equation. The approach which has been adopted in practice has been to include only
a subset of the terms allowed by symmetries. Indeed, in various practical applications of MIS theory to
relativistic heavy-ion collisions no single form of the relaxation equations has universally been adopted,
with different authors using different sets of terms. The choice of this subset is sometimes motivated by
consistency with microscopic models, especially in the framework of kinetic theory. Indeed, as reviewed in
some detail in \rfs{sect:kint}, there are rather sophisticated methods of motivating particular forms of the
relaxation equations.

The hydrodynamic evolution equations may be viewed as a parametrisation of a stage of non-equilibrium dynamics which
encapsulates conservation laws and the constraints of Lorentz symmetry. They are valid under very general conditions,
regardless of whether quasiparticle excitations are present. In their general form they guarantee matching the gradient
expansion of any microscopic theory. At the same time, they can be used in a purely phenomenological manner, allowing a
description of a rich spectrum of behaviours in terms of a finite number of transport coefficients. Since calculations
of the transport coefficients in the regimes of interest directly from QCD are not available, practitioners often resort
to intuitive choices of the possible terms on the RHS of the shear stress tensor relaxation equation. One of the most
popular versions of the MIS approach, which has been used frequently in phenomenological applications and is sometimes
identified with the Israel-Stewart approach, is a system of two equations for the shear stress tensor and the bulk
pressure~\cite{Muronga:2001zk,Muronga:2003ta,Heinz:2005bw,Bozek:2009dw,Shen:2014vra},
\beq
D{\pi}^{\langle\mu\nu\rangle} +\frac{\pimunu}{\tau_{\pi}} &=&
- \beta_{\pi} \sigmamunu
- \frac{\beta_\pi T}{2} \pimunu \,\partial_\lambda \left(  \frac{U^\lambda}{\beta_\pi T}  \right),  \label{MISshear} \\
D{\Pi} +\frac{\Pi}{\tau_{\Pi}}&=& - \beta_{\Pi} \theta
-\frac{\beta_\Pi}{2}  \Pi \,\partial_\lambda \left( \frac{U^\lambda}{\beta_\Pi T} \right),
\label{MISbulk}
\eeq
%
The shear viscosity of the system, $\eta$, is given by the product of the shear relaxation time, $\tpi$, and the coefficient $\beta_\pi$, $\eta = \tpi \beta_\pi$. Similarly, for the bulk viscosity we have $\zeta = \tpi \beta_\Pi$.

%
%

Even though one may justify the specific form of hydrodynamic equations such as \mbox{(\ref{MISshear})--(\ref{MISbulk})}
on the basis of a microscopic model, they may be (and often are) used on their own terms, with values of the transport
coefficients determined by other considerations. For example, a popular choice at the moment is to use the value of
$\eta/\sd$ suggested by holography, $\eta/\sd = 1/(4\pi)$, or its multiple. Similarly, the equation of state is not
necessarily  taken as that of an ideal gas as in all CFTs but may follow from other calculations. In this case, one
usually uses the QCD equation of state obtained from the lattice simulations~\cite{Borsanyi:2010cj} (for systems at zero
baryon chemical potential). Recently, several attempts have been made to consistently include the QCD equation of state
in kinetic theory by using an effective, temperature dependent mass and introducing mean
fields~\cite{Tinti:2016bav,Alqahtani:2017jwl}. Needless to say, Eqs.~\rfn{MISshear}--\rfn{MISbulk}  with some
parameterisation of the transport coefficients can be used in order to determine them from the
experiment~\cite{Pratt:2015zsa,Bass:2017zyn}.

\subsection{Bjorken flow in BRSSS theory}
\label{sect:brsss}

It is very instructive to consider a restriction of the BRSSS equation \rf{brsss} to Bjorken flow. The reason for this is that while the resulting equations are relatively simple, they provide a very explicit model where higher orders of the gradient expansion may be easily calculated and their significance can be directly explored. It can also usefully be regarded as a model of full, nonlinear dynamics where the interplay between hydrodynamic and non-hydrodynamic modes is in full view~\cite{Heller:2015dha,Aniceto:2015mto}. This attitude  is natural if one views theories of MIS type as ``UV completions'' of a purely hydrodynamic theory such as Navier-Stokes.

In the case of Bjorken flow described in \rfs{sect:bif}, the BRSSS equations \rfn{brsss} reduce to the following, simple form
\beal{miseqn}
\tau  \dot{{\cal E}} &=& - \frac{4}{3} \, {\cal E} + \phi\nonumber\, , \\
\tpi \dot{\phi} &=&
\frac{4 \, \eta}{3 \, \tau }
- \frac{\lambda_1\, \phi^2}{2 \, \eta^2}
- \frac{4 \, \tpi \, \phi}{3 \tau }
- \phi \, ,
\eea
where the dot denotes a proper time derivative and $\phi\equiv-\pi^{y}_{y}$, is the single independent component of the shear stress tensor~\cite{Baier:2007ix}.

Combining \rf{miseqn} with \rf{CsInBRSSS} one can derive a single second order equation for the energy
density, or equivalently the temperature~\cite{Heller:2015dha}:
%
\beal{misevol}
C_{\tpi} \tau \ddot{T}
+\frac{3}{2} \tau \left(\frac{C_{\lambda_1}  \tau}{C_\eta}+\frac{2  C_{\tpi}}{T}\right) \dot{T}^2
&+& \left(\frac{\tau (C_\eta+C_{\lambda_1})   T}{C_\eta}+\frac{11 C_{\tpi}}{3}\right) \dot{T}  \nn\\
&+& \frac{(2  C_\eta +C_{\lambda_1}) T^2}{6   C_\eta}
-\frac{4  (C_\eta -C_{\tpi})  T}{9}
= 0.
\eea
%
To proceed further it is convenient to rewrite \rf{misevol} in first order form. Introducing the dimensionless clock
variable $w$ as in \rf{wdef}, and using \rf{fR}, the evolution equation \rf{misevol} can be expressed as an evolution
equation for the pressure anisotropy $\pa$: 
%
\beal{feqn}
C_{\tpi} \, w \, (1 + \f{1}{12} \pa) \, \pa'
+ \left(\f{1}{3} C_{\tpi} + \f{1}{8} \f{ C_{\lambda_{1}}}{ C_{\eta}} w \right) \pa^2
+ \f{3}{2} \, w \, \pa
- 12 \, C_\eta = 0,
\eea
where the prime denotes a derivative with respect to $w$. Equations \rfn{wdef}, \rfn{fR}  and
\rfn{feqn} are equivalent to \rf{misevol} as long as the function $w(\tau)$ is invertible.

At large values of $w$ (late times) we expect universal hydrodynamic behaviour. \rf{feqn}
indeed possesses a unique stable solution which can be presented as a series in powers of $1/w$
\bel{rhydro}
\pa(w) = 8 \, C_\eta \,  \f{1}{w}
+\frac{16}{3} \, C_\eta \, (C_{\tpi} - C_{\lambda_{1}}) \, \f{1}{w^2}+ O\left(\frac{1}{w^3}\right).
\eel
This is precisely of the form of the gradient expansion \rf{rgradex} discussed earlier at the microscopic
level in \rfs{sect:hydrod}.

It is easy to see that linear perturbations around this formal solution decay at late times
exponentially on a time scale set by $\tpi$:
\bel{linpert}
\delta \pa(w) \sim e^{-\frac{3}{2 C_{\tpi}} w}
w^{\frac{C_\eta- 2  C_{\lambda_1} }{C_{\tpi}}}
\left\{1 + O\left(\frac{1}{w}\right)\right\}.
\eel
This is in fact the short-lived (nonhydrodynamic) mode introduced by the MIS prescription. In the language of the gravity dual to ${\cal N} = 4$ SYM, see \rfs{sect:qnm}, this would be an analogue of a transient QNM whose frequency is purely imaginary~\cite{Heller:2013fn,Heller:2014wfa}. The additional factor $3/2$ in the frequency (decay rate) when the mode is considered on expanding background is attributed to the fact that the frequency at every instant of proper time $\tau$ depends on the local temperature at that instant, see Refs.~\cite{Janik:2006gp,Heller:2013fn,Heller:2014wfa} for an extensive discussion of this point. The integral $\int_{\tau_{0}}^{\tau} d\tau' \, T(\tau')$ evaluated for the leading late time term from \rf{thydro} gives $3/2 \, w$, which explains the exponent in \rf{linpert}. Extending this analysis to the first subleading term explains the possibility of power-law modification. The logic presented here applies to all conformal models and will become particularly important in \rfs{sect:resu}.

These observations show that the term $\tpi \, \D\pimunu$ from which the $C_{\tpi}$ coefficient originates plays simultaneously two a priori \emph{different} roles in the BRSSS and related frameworks:
\begin{itemize}
\item it captures the purely hydrodynamic effect described by the second-order term $\tpi \,\D\sigmamunu$;
\item it controls the decay rate of a purely imaginary transient mode.
\end{itemize}
If the microscopic theory contains such a purely decaying excitation as the least damped non-hydrodynamic effect, this
provides us with two (in general incompatible) possibilities: matching $C_{\tpi}$ to the second order hydrodynamic
transport coefficient or relating it to the decay rate of a transient mode. If the least damped non-hydrodynamic mode in
the microscopic theory is not purely decaying, then MIS theory can only be useful once the effects of the transient mode
dies away -- its role is purely to act as a regulator ensuring causality.

The presence of the exponentially decaying mode \rfn{linpert} suggests that at least for sufficiently large
values of $w$ \rf{feqn} possesses an attractor solution which the gradient expansion is trying to approximate. As we will see in \rfs{sect:resu} the gradient expansion has to be interpreted appropriately for this to be successful.

\begin{figure}
\begin{center}
\includegraphics[height=0.3\textheight]{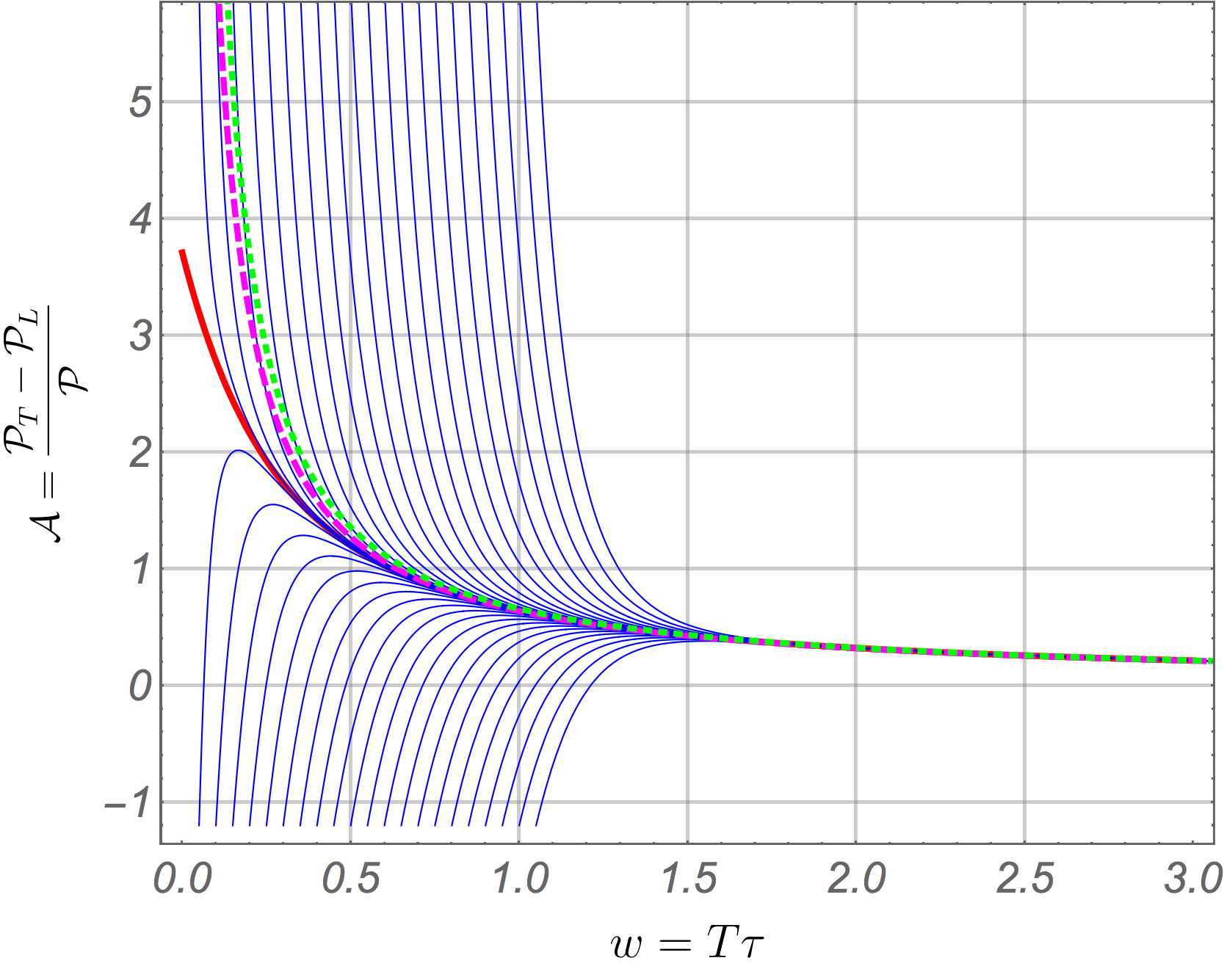}
\includegraphics[height=0.3\textheight]{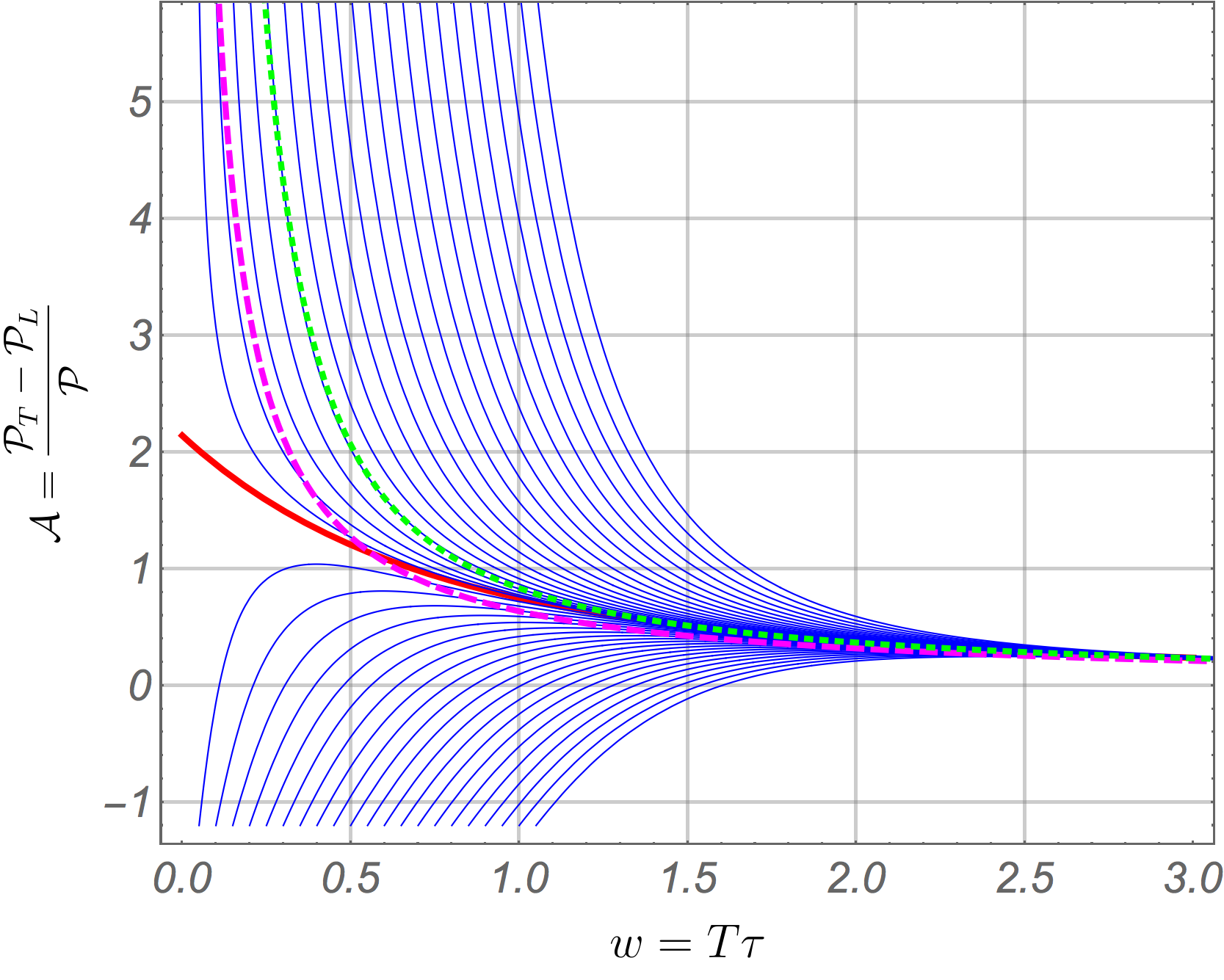}
\end{center}
\caption{Numerical solutions of \rf{feqn} for various initial conditions are represented by the blue curves. They asymptote
to the numerically determined attractor solution shown as the red line. The magenta, dashed and green, dotted curves are
approximations to the attractor given by the hydrodynamic gradient expansion truncated at first and second order in
derivatives respectively. The upper plot was made using parameter values appropriate for \symm, while the lower plot has
the relaxation time increased by a factor of $3$.} 
\label{fig:numa}
\end{figure}
The existence of the hydrodynamic attractor in \rf{feqn} is supported by examining the behaviour of generic
numerical solutions of \rf{fdef}, with initial conditions set at various values of~$w$.  Examining the
behaviour of $\pa$ for $w$ close to zero analytically, one finds two solutions, one of which is stable
\bel{smallw}
\pa(w) = 6 \sqrt{\f{C_\eta }{C_{\tpi}}} + O(w) .
\eel
By setting the initial value of $\pa$ at small $w$ close to the value in \rf{smallw}, the attractor can
be determined numerically with the result shown in Fig.~\ref{fig:numa}.

Note also, that one can determine the attractor by using a scheme similar to the slow-roll expansion used in  inflationary cosmology, see~e.g.~Ref.~\cite{Liddle:1994dx}, in which case one at first solves \rf{feqn} by neglecting the $\pa'$ term and then one accounts for it iteratively~\cite{Heller:2015dha}. At leading order this gives
\bel{slow}
 R\left(w\right) =
 \frac{6 C_\eta \left(\sqrt{64 C_\eta C_{\tpi} + 24 C_{\lambda_1} w+9 w^2}-3 w\right)}{8
   C_\eta C_{\tpi} +3 C_{\lambda_1} w} \, .
\eel
The resulting series provides then an alternative to the standard hydrodynamic gradient expansion, in particular it does not break down at $w = 0$.

As seen in Fig.~\ref{fig:numa}, a generic solution rapidly decays to the attractor. Furthermore, the attractor
appears to persist even at very small values of~$w$, where the truncated gradient expansion cannot be reliable
(but the slow-roll approach works quite well). Unsurprisingly, truncating \rf{rhydro} at first or second order
gives results distinctly different from the attractor at very small~$w$, but the difference becomes negligible
at larger values of $w$. The magnitude of the difference depends on the values of the transport coefficients.
Assuming \symm\ parameter values, we see that adopting just the viscous hydrodynamics constitutive relations
provides a remarkably good approximation to the attractor for a wide range of $w$. In particular, this holds
with an error smaller than $10 \%$ for $w > 0.5$.

It is tempting to propose that the hydrodynamic attractor constitutes the definition of hydrodynamic behaviour: it is
the universal solution to which the system tends as transient non-hydrodynamic modes decay. Recent results indicate the
presence of an attractor solution also at the microscopic level: for boost-invariant expansion in the RTA kinetic
theory~\cite{Romatschke:2017vte} and in \symm~\cite{Romatschke:2017vte,Spalinski:2017mel}, which significantly
strengthens the case for this idea. At the moment of writing this review, it remains to be seen how this putative
attractor would manifest itself when the conformal symmetry is broken or for more complicated flows (but
see \rfs{sect:outlook}). .

\subsection{Beyond BRSSS: HJSW and transient oscillatory behaviour}
\label{sect:beyond}

BRSSS theory includes the minimal regulator sector necessary to ensure causality.
However, given some information about the non-hydrodynamic sector of a specific microscopic theory or model
one may try to mimic it at the level of an effective hydrodynamic description.  Heuristic arguments given
in~Ref.~\cite{Heller:2014wfa}, and earlier in~Ref.~\cite{Denicol:2011fa,Noronha:2011fi}, suggest modifying the
dynamics of the shear-stress tensor, which in MIS theory and similar frameworks is governed by relaxation
equations such as \rf{relax}. This is of interest if one aims to account for phenomena where the non-hydrodynamic
sector is not subdominant, such as for example early time dynamics or the behaviour of small systems. One may
hope that by including a more faithful representation of the non-hydrodynamic sector at least some transient
effects of non-hydrodynamic modes can be captured. For some observables (such as the final multiplicities)
this may not be quantitatively important. However for observables sensitive to the pre-equilibrium stages of
evolution (such as photon \cite{Schenke:2006yp,McLerran:2014hza} or dilepton emission
\cite{Martinez:2008di,Shuryak:2012nf}) capturing the early time dynamics may be valuable.

The goal of this section is to review an attempt to capture the dynamics of the least damped non-hydrodynamic modes of
\symm\ plasma based on studies of AdS Schwarzschild black-brane QNMs and to incorporate them in an extension of the
hydrodynamic framework. The basic observation is that in  \symm\ and most other holographic strongly-coupled QFTs the
least damped transient modes are oscillatory. The idea is then to generalize the BRSSS framework to try to capture not
only the effects of viscosity, but also some of the effects of the least-damped oscillatory transient excitation. It is
not obvious that this idea can be successful: for instance, in theories where the least damped non-hydrodynamic modes
are not clearly separated,  one has to accept that the hydrodynamic description can only be useful after the
non-hydrodynamic sector has decayed.

The simplest way to incorporate additional non-equilibrium degrees of freedom into a causal (and hyperbolic)
theory is to set
\be
\label{newtmunu}
\pi^{\mu \nu} =  \pi_{MIS}^{\mu \nu} + \tilde{\pi}^{\mu \nu}
\ee
with $\pi_{MIS}^{\mu \nu}$ satisfying \rf{crelax} and $\tilde{\pi}^{\mu \nu}$ obeying a second order equation
designed to reproduce the damped-oscillatory behaviour of transient QNMs:
\bel{dampedosc}
(\f{1}{T} \, \D)^2 \,  \tilde{\pi}_{\mu\nu} +2\,\Omega_I\, \f{1}{T} \, \D \, \tilde{\pi}_{\mu\nu}+
|\Omega|^2 \, \tilde{\pi}_{\mu\nu}=0,
\eel
where $|\Omega|^2\equiv \Omega_I^2+\Omega_R^2$. This is formally the equation of motion of a damped harmonic
oscillator. To match the least-damped QNMs of \symm\ plasma one needs to choose the parameter $\Omega$ appropriately. For example, in the case of interest, i.e. the energy-momentum tensor operator, at vanishing momentum $\vk$ one has in all channels the following leading transient QNMs frequency~\cite{Kovtun:2005ev,Nunez:2003eq}:
\bel{omegavalue}
\frac{1}{T} \, \omega\Big|_{k = 0}  = \pm \Omega_{R} - i \, \Omega_{I} \approx \pm 9.800 - i \, 8.629.
\eel
Although the frequency of this mode (as that of any other mode) depends on the momentum~$\vk$, as explicit results
discussed in \rfs{sect:qnm} and displayed in Fig.~\ref{fig:qnmsk}, this dependence is weak for a whole range of momenta
around zero. Neglecting this dependence is thus a justified approximation. Furthermore, the use of Weyl-covariant
derivatives here ensures that the evolved shear-stress tensor remains transverse and traceless. These two traceless and
transverse quantities, $\Pi_{MIS}^{\mu \nu}$~and~$\Pit^{\mu \nu}$, are coupled together by the conservation law of the
energy-momentum tensor.

The resulting theory satisfies the same causality and stability properties as the MIS theory.  At the
linearized level, in addition to the standard hydrodynamic modes it contains damped modes which resemble
the transient QNMs seen in holography. However, in addition we still have the purely decaying MIS mode, which is
spurious from the perspective of reproducing the pattern seen in holographic plasma.

An alternative which disposes of the MIS mode altogether is to abandon the relaxation equation such as \rf{relax} entirely and adopt the following second order equation for the shear-stress tensor:
\bel{eqpi2s}
 \left\{\left(\f{1}{T} \, \D\right)^2 + 2 \, \Omega_I \,  \f{1}{T} \, \D + |\Omega|^2\right\}
\pi^{\mu \nu} = - \eta \,  |\Omega|^2 \, \sigma^{\mu\nu}  - c_\sigma \, \f{1}{T} \, \D\left(\eta
\, \sigma^{\mu\nu}\right) + \ldots
\eel
where the ellipsis denotes contributions of second and higher order in gradients. Of all possible second order terms
only one term has been kept, with a coefficient $c_\sigma$. We will refer to the model based on \rf{eqpi2s} as the HJSW
model. The appearance of the second derivative in \rf{eqpi2s} leads to non-hydrodynamic modes
which are not purely decaying, in contrast with what we saw in MIS (or BRSSS) theory.  Indeed, the
linearization of equations \rfn{fullconservation} and \rfn{eqpi2s} around flat space reveals a pair of
non-hydrodynamic modes with complex frequencies $\Omega$ and $\bar{\Omega}$.

The key property of~\rf{eqpi2s} is that linearization around an equilibrium background leads to a system of partial
differential equations which is hyperbolic and causal in a region of parameter space. Choosing values of $\eta/\sd$ and
$\Omega_{R,I}$ characteristic of \symm\ we find that causality requires $-\pi \leq c_\sigma \leq 2\pi$. Requiring
stability at the linearized level imposes further conditions which cannot be met. While these unstable modes appear far
outside the range of applicability of the long wavelength description, for numerical calculations this is an issue.
This means that one cannot insist on \symm\ parameter values, unless a symmetry (such as boost invariance) is imposed
which eliminates unstable modes -- this is discussed further below.

Both schemes discussed above make no attempt to match the gradient expansion to second order: they only ensure that the first order matches. The rationale for this is that while viscosity is a critically important physical effect seen in QGP dynamics, the second order terms appearing in MIS theory and its generalizations such as BRSSS are more important for consistency and less for modeling actual physical processes.

Regardless of which of the two alternative dynamical descriptions one chooses, for practical applications to
QGP dynamics one needs to develop an effective heuristic for setting initial conditions for the
non-hydrodynamic modes. One of the possible approaches might be to extract these initial conditions from the
early post-collision state following from the numerical simulations of \cite{Gelis:2013rba} or
\cite{Berges:2013fga}.

\begin{figure}
\center
\includegraphics[height=0.4\textheight]{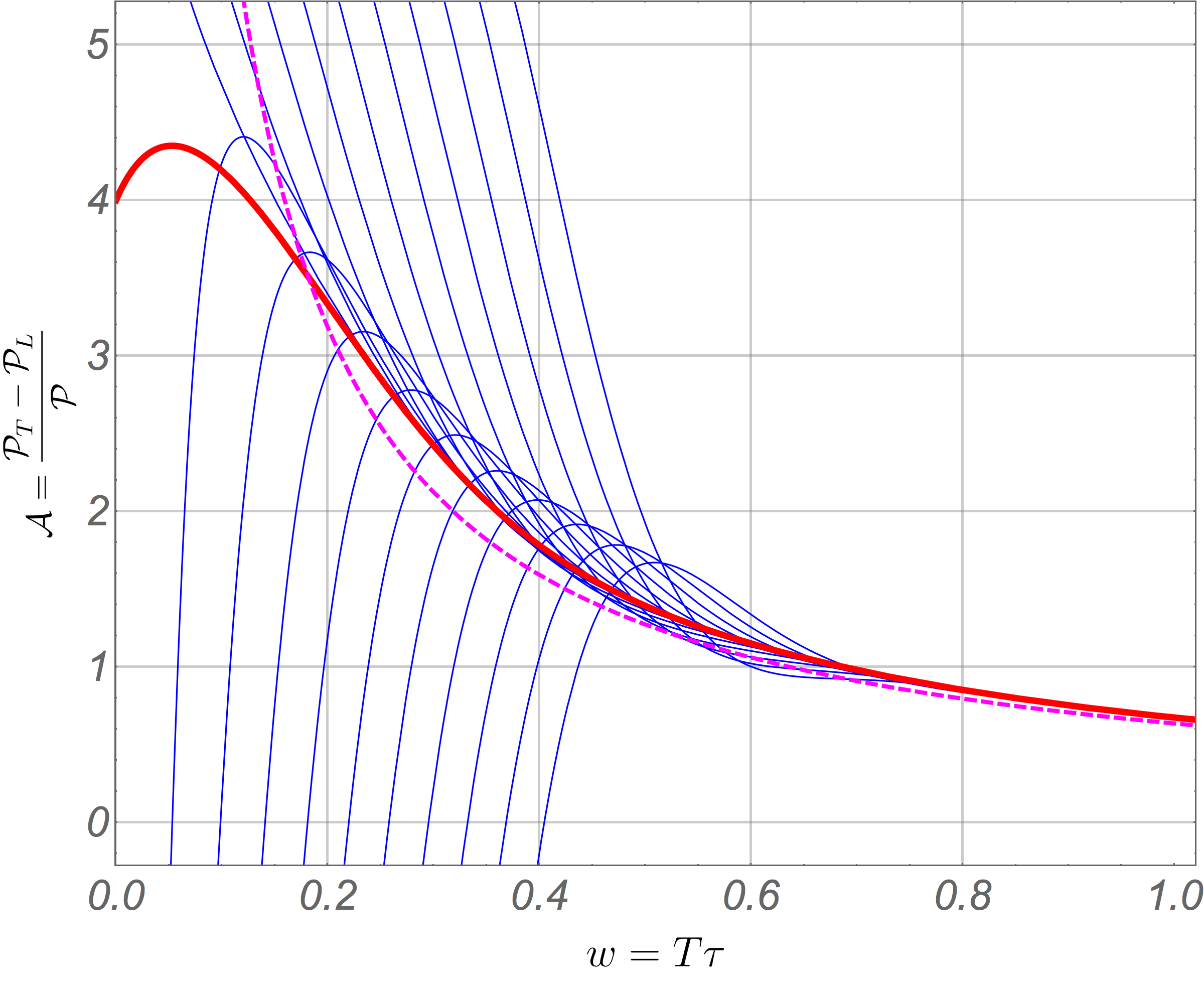}
\caption{Numerical solutions of \rf{vcp} are shown in blue; they converge (non-monotonically) to the numerically
determined attractor (red). The dashed magenta curve is the first order truncation of the gradient expansion.} 
\label{fig:attractvcp}
\end{figure}

As in the case of BRSSS it is very interesting to study Bjorken flow in this theory.
Imposing boost-invariance, the hydrodynamic equations reduce to a third order ordinary differential equation for the
temperature. Introducing new variables as in Eqs.~\rfn{wdef}~and~\rfn{fR} one can rewrite it as a second
order differential equation for the function $\pa(w)$:
\bel{vcp}
\alpha_1 \pa''+ \alpha_2 \, \pa'^2+\alpha_3 \, \pa'+12 \, \pa^3+\alpha_4 \, \pa^2+\alpha_5 \, \pa+\alpha_6 = 0,
\eel
where
\beal{vcp.coeffs}
\alpha_1 &=&  w^2 \, (\pa+12)^2,\nn\\
\alpha_2 &=& w^2 \, (\pa+12),\nn\\
\alpha_3 &=& 12 \,  w \,  (\pa+12) \, (\pa+3 \, w \,  \Omega_I),\nn\\
\alpha_4 &=& 48 \,  (3 \,  w\, \Omega_I - 1),\nn\\
\alpha_5 &=& 108 \, \left(- 4\,  C_\eta \, C_{\sigma} + 3 \, w^2 \, \Omega ^2\right),\nn\\
\alpha_6 &=& -864 \, C_\eta \, \left(- 2 \, C_{\sigma}+3\, w\, \Omega^2\right).
\eel
This is the analog of \rf{relax} of MIS theory. At large $w$, the corresponding hydrodynamic gradient expansion takes the form
\bel{hjsw.largew}
\pa(w) =  8 \, C_\eta \, \f{1}{w} +
\f{16 \, C_\eta \, (- C_{\sigma}  + 2 \,  \Omega_I)}{3 \, |\Omega|^2} \, \f{1}{w^2} + \dots
\eel
As expected, the first term, which encodes the shear viscosity, coincides with what one obtains in BRSSS theory, see \rf{rhydro}, whereas the higher order terms are necessarily different.

At early times
one finds a unique real power series solution of the form
\bel{hjsw.smallw}
\pa(w) =  4 + \f{54 \, C_\eta \, |\Omega|^2- 48 \, \Omega_I}{20 - 9  \, C_\eta \,
  C_{\sigma}} \, w + \dots
\eel
By examining numerical solutions of \rf{vcp} it is clear that (similarly to the case of MIS theory) this
is the small $w$ behaviour of an attractor solution valid in the entire range of $w$.

Since \rf{vcp} is of second order, one must specify both $\pa$ and $\pa'$ at the initial value of~$w$. As seen in
Fig.~\ref{fig:attractvcp}, setting initial conditions at various values of $w$ shows that the numerical
solutions converge to the attractor. However, unlike in the MIS case, the numerical solutions do not decay
monotonically, but oscillate around the attractor. Note that because the phase space is three-dimensional, a
plot like \rff{fig:attractvcp} is a two-dimensional section of a proper phase-space picture.


\begin{figure}
\includegraphics[height=0.28\textheight]{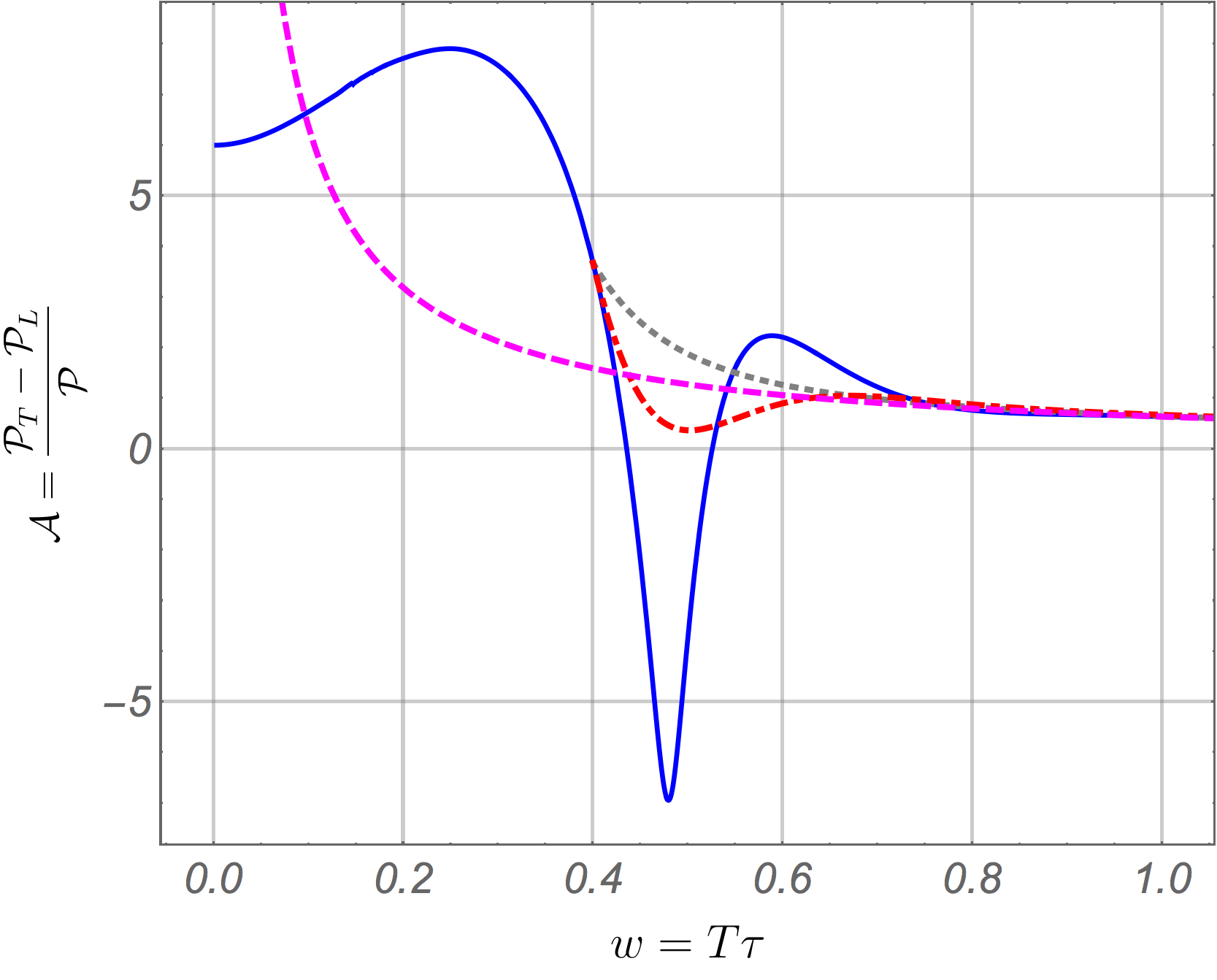}
\quad
\includegraphics[height=0.28\textheight]{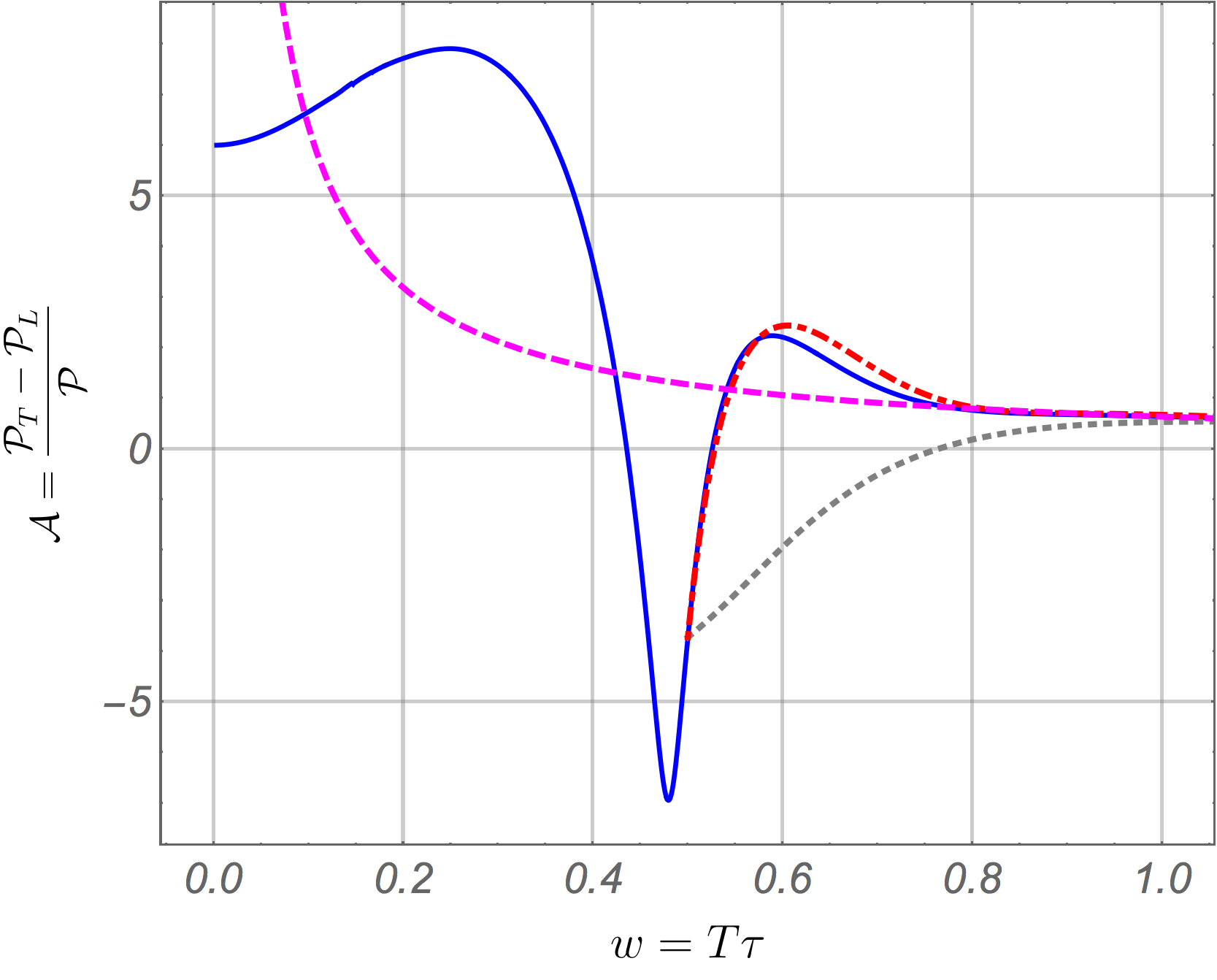}
\caption{Comparing BRSSS ans HJSW: the dotted grey curves are solutions of BRSSS theory while the dot-dashed red curves
are solutions of the HJSW equations. The blue curves represent numerical solutions obtained using holography. For
reference, the prediction of first order hydrodynamics is displayed as the dashed magenta curve. The plots show the
results of setting initial conditions at $\tau \, T = 0.4$ (left), $0.5$ (right). We see that the HJSW curve captures
the oscillation missed by BRSSS.}
\label{figbif}
\end{figure}

An important check of the theories reviewed in this section is to compare the evolution of $\pa$ as governed by them
with the microscopic computation of Bjorken flow using AdS/CFT. This requires setting the parameters to values
appropriate for \symm, i.e. $\eta/\sd = 1/4\pi$ and $\Omega_{R,I}$ as in \rf{omegavalue}.  These calculations also
assumed $c_\sigma=2\pi$ (the largest value allowed by causality).~\footnote{Since boost invariance does not allow for
instabilities, this causes no problems, and the results are in fact essentially indistinguishable from simulations
using \rf{dampedosc}.}  A sample comparison is presented in \rff{figbif}, which shows that the MIS approach captures the
late time tail very well, as do the second order equations reviewed here. However, at earlier times evolution based
on \rf{vcp} provides a much more accurate picture.

%% file: sect-kint.tex
\section{Hydrodynamics as an effective theory  -- insights \\  from kinetic theory}
\label{sect:kint}

The approach embodied in the BRSSS formulation of relativistic hydrodynamics is elegant from the theoretical
perspective and completely general. Given a specific microscopic model it may however be possible to find a
non-generic hydrodynamic description, which is crafted to reproduce that particular theory. A significant body
of work exists whose goal was to find such hydrodynamic descriptions of kinetic theory models. From the
perspective of heavy-ion collisions such an approach is limited by our insufficient knowledge of an
appropriate collision kernel for QCD and non-perturbative phenomena, but nevertheless this approach is an
invaluable theoretical laboratory for studying relativistic hydrodynamics.

\subsection{The gradient expansion in kinetic theory}
\label{sect:kint:gradex}

In full analogy with the cases discussed in Secs.~\ref{sect:matching} and \ref{sect:nonconf}, the connection
between kinetic theory models and effective hydrodynamic descriptions can be made by comparing gradient
expansions. Indeed, the Boltzmann equation can be solved iteratively in a gradient expansion in a way similar
to what was discussed in \rfs{sect:bifrta:grad}. The infinite series obtained this way is called the
Chapman-Enskog expansion.~\footnote{Strictly speaking,  the Chapman-Enskog method is introduced formally as an
expansion in the Knudsen number,  which is the ratio of the characteristic microscopic and macroscopic scales
describing the system --- see Ref.~\cite{Denicol:2014loa} where the case of a general collision term is
discussed in this context. For the RTA kinetic equation, a Chapman-Enskog-like expansion has been introduced
in Refs.~\cite{Jaiswal:2013npa,Jaiswal:2013vta}, which is based directly on the gradient expansion.}
Depending on the collision kernel, the generated series will contain a subset (not necessarily proper) of all
possible Lorentz covariant terms with specific coefficients. This can then be used to match generalised MIS
equations such as \rf{MISshear} or \rf{MISbulk}. This matching determines the hydrodynamic transport
coefficients much in the same way as in the case of holography in \rfs{sect:matching}. If this is done for
\rf{rta}, one finds at first order that the shear viscosity coefficient $\eta$ is connected with the
relaxation time and entropy density~$\sd$. For classical statistics, one
finds~\cite{Anderson:1974a,Czyz:1986mr}
\begin{equation}
\trel = \frac{5 \eta}{T \sd}.
\label{teq-etabar}
\end{equation}
Equation~\rfn{teq-etabar} connects the relaxation time with the ratio $\eta/\sd$. Thus, in this case, larger
values of $\eta/\sd$ lead to larger values of the relaxation time and, consequently, longer timescales for the
decay of  non-hydrodynamic modes (delayed hydrodynamization). This relation is a consequence of there being
only a single parameter in the RTA collision kernel. We note however that results quantitatively consistent
with \rf{teq-etabar}  have been recently obtained in \rfc{Czajka:2017bod} in the context of scalar $\phi^4$
theory.

Thus, starting from the gradient expansion of kinetic theory we can determine the transport coefficients appearing in the hydrodynamic equations, i.e., their dependence on temperature, chemical potential and the particle mass. In addition, the equation of state which follows from kinetic theory is known and corresponds typically to a gas of weakly interacting particles.

It is worth stressing -- and is the subject of this section -- that the approach outlined above is not the one most often used in practice in the context of kinetic theory. The standard approach is to write down a set of moments of the distribution function and impose an additional constraint in such a way as to obtain a closed differential equation of the MIS form. Various ways of carrying this out in practice are described below.

\subsection{DNMR}
\label{sect:dnmr}

The approach by Denicol, Niemi, Molnar and Rischke (DNMR) represents a general
derivation of relativistic fluid dynamics from the underlying Boltzmann
equation, with an arbitrary collision kernel, using the method of
moments~\cite{Denicol:2012cn,Denicol:2014loa,Denicol:2010xn}. The authors
derive first a general expansion of the distribution function $\delta f = f-
f_{\rm eq}$ in terms of its irreducible moments. Then, the exact equations of
motion for these moments are derived.  It turns out, that there is an infinite
number of such equations and, in general, one has to solve this infinite set
of coupled differential equations in order to determine the time evolution of
the system. The reduction of the number of equations is possible, however, if
the terms are classified according to a systematic power-counting scheme in
the Knudsen and inverse Reynolds numbers. As long as one keeps terms of second
order (in both parameters) the equations of motion can be closed and expressed
in terms of only 14 dynamical variables.

\begin{figure}[t]
\begin{center}
\includegraphics[angle=0,width=0.75\textwidth]{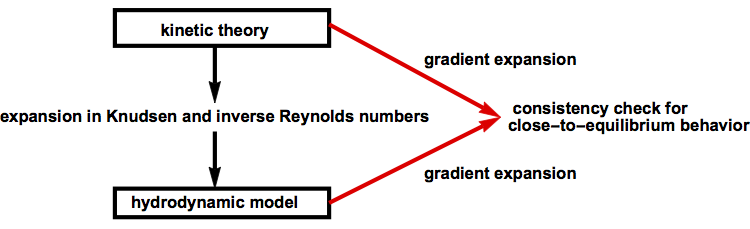}
\end{center}
\caption{Schematic view of the hydrodynamics expansion that serves to obtain the hydrodynamic equations from the underlying kinetic theory by simultaneous expansion in the Knudsen and inverse Reynolds numbers. At the same time the gradient expansion can be used to check the agreement of the two approaches for close-to-equilibrium dynamics. }
\label{fig:hyd_exp}
\end{figure}

The simultaneous expansion in the Knudsen and inverse Reynolds numbers can be treated as a kind of {\it hydrodynamic expansion} that allows for a systematic derivation of hydrodynamic equations from the underlying kinetic theory. We wish to stress, however,  the need to make a clear distinction between the gradient expansion discussed in this review and the hydrodynamic expansions treated as a means of deriving hydrodynamic  equations.  As we have showed so far, various methods exist for writing down hydrodynamic equations; once some set of hydrodynamic equations is found, one can look for their formal solutions in the form of a gradient expansion of the energy-momentum tensor as in \rf{TmunuGrad}. In particular, the gradient expansion can be applied to both the underlying microscopic theory and the hydrodynamic model that serves as its approximation, in order to check the overall consistency between the two approaches, see~Fig.~\ref{fig:hyd_exp}.

For the case of the RTA kinetic equation, the DNMR formalism yields the following two equations for the bulk pressure and shear stress tensor
\bel{dnmrpi1}
D{\pi}^{<\mu\nu>}+\f{\pimunu}{\tau_\pi} = -\beta_\pi  \sigmamunu - 2 \pi_\gamma^{< \mu} \omega^{\nu > \gamma}
 - \delta_{\pi\pi} \pimunu \theta +\f{1}{2} \tau_{\pi\pi} \pi_\gamma^{< \mu} \sigma^{\nu > \gamma}
 +  \f{1}{2} \lambda_{\pi \Pi} \Pi \sigma^{\mu\nu}
\eel
and
\bel{dnmrPi1}
D{\Pi} + \f{\Pi}{\tau_\Pi}  = -\beta_\Pi \theta -\delta_{\Pi\Pi} \Pi \theta -\f{1}{2} \lambda_{\Pi\pi} \pi^{\mu\nu} \sigma_{\mu\nu},
\eel
where $\omega^{\mu\nu} = (1/2) (\nabla_\perp^\mu U^\nu- \nabla_\perp^\nu U^\mu)$ is the vorticity tensor
($\nabla_\perp^\mu = \Delta^{\mu\nu} \p_\nu$). In the conformal limit the second equation is automatically
fulfilled, while the coefficients appearing in the first equation are: $\delta_{\pi\pi} = 4/3$ and
$\tau_{\pi\pi} = 10/7 $. In the general, non-conformal case, it is interesting to observe that
Eqs.~\rfn{dnmrpi1} and \rfn{dnmrPi1} are directly coupled through the last terms appearing on their right-hand
sides. The importance of such a shear-bulk coupling has been emphasised in~\cite{Denicol:2014mca}.

\subsection{Jaiswal's third order theory}
\label{sect:jaiswal}

In the case where the collision term in the kinetic equation is given by the relaxation-time form, see Eq.~\rfn{rta}, the second and third order theory of viscous hydrodynamics has been worked recently out by Jaiswal, using the Chapman-Enskog-like expansion for the distribution function close to equilibrium~\cite{Jaiswal:2013vta,Jaiswal:2014raa}. Separating the space-time gradient  appearing in \rfn{rta} into the longitudinal and transverse part
\bel{dmu}
\p_\mu = g_{\mu\nu} \p^\nu = \left(\Delta_{\mu\nu} - U_\mu U_\nu \right)  \p^\nu =  \nabla^\perp_\mu -U_\mu D,
\eel
we can write
\bel{deltafdot}
D \delta f = -\f{\delta f}{\trel} - D{f_{\rm eq}} + \f{1}{\pdotU} \, p^\mu \nabla^\perp_\mu \, f,
\eel
where $\delta f = f - f_{\rm eq}$ is the difference between the non-equilibrium distribution function and the reference equilibrium one. Similarly as in \rf{iter} we write $\delta f = f_0 + f_1 + f_2 + \cdots$, where the leading term $f_0$ is identified with the equilibrium distribution, $f_0 = f_{\rm eq}$, and the next terms are obtained by consecutive differentiations of the previous terms, namely
\vspace{-0.1cm}
\bel{f1f2}
f_1 = \f{\trel}{\pdotU} p^\mu \p_\mu f_0, \quad f_2 = \f{\trel}{\pdotU} p^\mu \p_\mu f_1, \quad \ldots \,\,\,\,.
\eel

As first shown in Ref.~\cite{Denicol:2010xn}, the formula (\ref{deltafdot}) can be used to find the exact equations satisfied by the shear and bulk pressures. For the shear sector in RTA one finds
\bel{pimunudot1}
D{\pi}^{<\mu\nu>}+\f{\pimunu}{\trel} = - \Delta^{\mu\nu}_{\alpha\beta} \int dP \, p^\alpha p^\beta \left(
D{f_0} - \f{1}{\pdotU} \, p^\mu \nabla^\perp_\mu \, f \right).
\eel
In order that~\rf{pimunudot1} is second-order in gradients, the distribution function on the right-hand side of \rfn{pimunudot1} should be computed only up to the first-order in gradients, $f = f_0 + f_1$. In this way, for massless particles one derives the equation
\bel{pimunudot2}
D{\pi}^{<\mu\nu>}+\f{\pimunu}{\trel} = -\beta_\pi  \sigmamunu - 2 \pi_\gamma^{< \mu} \omega^{\nu > \gamma}
+ \f{5}{7} \pi_\gamma^{< \mu} \sigma^{\nu > \gamma} - \frac{4}{3} \pimunu \theta.
\eel
In the transition from \rfn{pimunudot1} to \rfn{pimunudot2} one uses the first order relation $\pi^{\mu\nu} = -\eta
\sigma^{\mu\nu}$ in order to remove the relaxation times on the right-hand side of \rfn{pimunudot2}.
It is interesting to compare this to the equation which follows from \rf{MISshear} in the case of conformal systems,
where 
the bulk pressure and the bulk viscosity vanish, and $\beta_\pi 
= \frac{4}{5} \, \peq$ and $\ed = 3 \, \peq$.
This leads to
\beq
D{\pi}^{\langle\mu\nu\rangle} +\frac{\pimunu}{\tau_{\pi}} &=&
- \beta_{\pi} \sigmamunu - \frac{4}{3} \pimunu \theta
\label{MISshear1}.
\eeq
which misses two terms present in \rf{pimunudot2}. 


The strategy to find the third order equation is similar. In this case,  the right-hand side of \rfn{pimunudot1} is computed including the second-order terms, $f = f_0 + f_1 + f_2$, and \rf{pimunudot2} is used to replace $\sigma^{\mu\nu}$. For details of this procedure we refer to \cite{Jaiswal:2013vta,Chattopadhyay:2014lya}. It is worth noting, that comparing the gradient expansion generated by these third-order equations~\cite{Florkowski:2016zsi} with the gradient expansion obtained directly from kinetic theory in the RTA~\cite{Heller:2016rtz} one finds, as expected, that the third order contributions indeed match (see also \rfs{subsect:consistency}).

\subsection{Anisotropic hydrodynamics}
\label{sect:ahydro}

As we have discussed above, in the context of kinetic theory standard viscous
hydrodynamics can be constructed as an  expansion in the  Knudsen and  inverse
Reynolds numbers around the local equilibrium
state~\cite{Denicol:2014loa}. This type of expansion may be questioned in the
situation where space-time gradients and/or deviations from the local
equilibrium are large.  The goal of the anisotropic hydrodynamics (AHYDRO) 
program~\cite{Florkowski:2010cf, Martinez:2010sc, Strickland:2014pga} is to
create a  framework of dissipative hydrodynamics that is better suited to deal
with such cases and accurately describes several features such as:  the early
time dynamics of the QGP created in heavy-ion collisions, dynamics near the
transverse edges of the nuclear overlap region, and temperature-dependent and
possibly large shear viscosity to entropy density ratio.

\subsubsection{Reorganized hydrodynamic expansion}

In the standard approach to viscous hydrodynamics, the energy-momentum tensor components $T^{\mu\nu}$ may be
treated as functions of $T$, $\umu$, $\pimunu$, and $\Pi$, which we write schematically as
\beq
\Tmunu &=&  \Tmunu \left(T,\umu,\pimunu,\Pi \right) \nn \\
&=& \Tmunueq (T,U) + \pimunu + \Pi  \Delta^{\mu \nu} \nn \\
&\equiv& \Tmunueq + \delta \Tmunu.
\label{TmunuDecEq}
\eeq
The hydrodynamic equations that determine the dynamics of~$\Tmunu$ contain various terms that may be characterised by
the power of space-time gradients and/or the power of the dissipative terms (strictly speaking one considers the powers
of the ratios $\sqrt{\pi^{\mu\nu} \pi_{\mu\nu}}/\peq$ and $\Pi/\peq$ which are known as the inverse Reynolds numbers).
For example, at first order (viscous hydrodynamics) one deals with $\pi^{\mu\nu}$ and $\Pi$ and also with the gradients
of $T$ and $U^\mu$, see Eqs.~(\ref{NavierStokes}). At second order, the products of $\pi^{\mu\nu}$ and $\Pi$ appear, as
well as the gradients of $\pi^{\mu\nu}$ and $\Pi$. Such approach may be continued to higher orders but, in practical
applications, one stops at the third order (see our discussion in Sec.~\ref{sect:jaiswal}).

Anisotropic hydrodynamics can be treated as a method to reorganize this kind
of expansion within the framework of the kinetic theory. One uses the classical concept
of the phase space distribution function $f(x,p)$ and expresses different
physical quantities as the moments of $f(x,p)$ (in the three-momentum space).

Within \AH one separates the description of dissipative effects into two parts.
The first part is characterised by new fluid variables $\ximunu$ and $\phi$, see Refs.~\cite{Martinez:2012tu,Tinti:2013vba,Nopoush:2014pfa,Tinti:2014yya,Tinti:2015xra,Tinti:2015xwa}.
They may account for possibly large values of the shear stress tensor and bulk viscosity and should be treated in a non-perturbative manner, similarly to $T$ and $\umu$. The second part is characterised by the tensors $\pimunut$ and $\Pit$ that are treated similarly as $\pimunu$ and $\Pi$ in the standard
case~\cite{Bazow:2013ifa,Bazow:2015cha,Molnar:2016vvu,Molnar:2016gwq}. Thus, we write
\beq
\Tmunu &=&  \Tmunu \left(T, \umu, \ximunu, \phi, \pimunut, \Pit \right) \nn \\
&=& \Tmunua(T,\umu,\ximunu,\phi) + \pimunut + \Pit  \Dmunu \nn \\
&\equiv& \Tmunua + \delta {\tilde T}^{\mu\nu}.
\label{TmunuDecA}
\eeq
At this point it is important to emphasize several issues:

\begin{itemize}
\item{}
Introducing the fluid variables $\ximunu$ and $\phi$ simultaneously with $\pimunut$ and $\Pit$ implies that the ``dissipative'' degrees of freedom are ``doubled'' and we need more than the typical ten hydrodynamic equations to determine the dynamics of $\Tmunu$. This can easily be accomplished within kinetic theory approach by including, for example, a corresponding number of moments of the kinetic equation. The choice of moments is, however, not well defined. One takes into account usually the lowest possible moments~\cite{Bazow:2013ifa}, since they are most sensitive to the low momentum sector which is expected to be well described by in the language of hydrodynamics. We come back to the discussion of this ambiguity below.
\item{}
The anisotropy tensor  $\ximunu$ has analogous geometric properties as the shear stress tensor $\pimunu$, i.e., it is symmetric, transverse to $\umu$ and traceless~\cite{Tinti:2014yya,Tinti:2015xra,Tinti:2015xwa}. This means that it has in general five independent components. However, in practical applications one often uses simplified versions of $\ximunu$ that contain one or two independent fluid variables. In such cases only these degrees of freedom may be ``doubled''. We note that the use of simplified forms of  $\ximunu$ is very often a consequence of symmetries such as boost invariance, homogeneity in the transverse plane or cylindrical symmetry.  The tensor   $\pimunut$ is also symmetric, transverse to $\umu$ and traceless. Consequently, using the Landau frame, one can determine the effective temperature of the system $T$ and the flow $\umu$ by the equation
%
\begin{eqnarray}
\Tmunua(x) \unul(x) = - \ed_{\rm EQ} (T(x))\, \umu(x).
\label{LandFrameA}
\end{eqnarray}
Since a substantial part of the viscous effects is included through the use of $\ximunu$ and $\phi$, one expects that the  terms $\pimunut$ and ${\tilde \Pi}$ are small compared to the equilibrium pressure $\peq$. The expansion in the ratios  $\sqrt{\pimunut {\tilde  \pi}_{\mu\nu}}/\peq$ and $\Pit/\peq$ is discussed in this context as an expansion in the modified inverse Reynolds  numbers~\cite{Bazow:2013ifa,Bazow:2015cha,Molnar:2016vvu,Molnar:2016gwq}.
\item{}
Using the kinetic theory approach, Eq.~(\ref{TmunuDecA}) is reproduced with the distribution function that has a structure
\begin{eqnarray}
f(x,p) = \fa(x,p)  + \delta {\tilde f}(x,p).
\label{deltafa}
\end{eqnarray}
Here $\fa(x,p)$ is the anisotropic distribution function in momentum space. It can be regarded as an
extension of the equilibrium distribution $f_{\rm eq}(x,p)$, which depends not only on $T$ and $U^\mu$ but
also on $\ximunu$ and $\phi$. In the limit $\ximunu \to 0, \phi \to 0$ one finds that $\fa(x,p) \to f_{\rm
eq}(x,p)$. Several special forms of $\fa(x,p)$ will be discussed below. In the {\it leading order of
anisotropic hydrodynamics}, we neglect the corrections $ \delta {\tilde T}^{\mu\nu}$ in (\ref{TmunuDecA}) and
$\delta {\tilde f}(x,p)$ in (\ref{deltafa}). Then, the complete energy-momentum tensor is given by the formula
\begin{eqnarray}
\Tmunu = \Tmunua = k \int dP \,\,p^\mu p^\nu \fa(x,p),
\label{TmunuA}
\end{eqnarray}
where we have added a degeneracy factor $k$ to account for internal degrees of freedom. In this case, the
components of $\Tmunua$ depend, in general, on ten independent fluid variables contained in the set: $T$, $\umu$,
$\ximunu$ and $\phi$. The equations of anisotropic hydrodynamics specify the dynamics of $\Tmunua$. They
include four equations that follow directly from the energy-momentum conservation law and additional six
equations that should be derived from kinetic theory.
\item{}
The doubling of dissipative degrees of freedom can be avoided if the use of a certain fluid variables in the set
($\ximunu, \phi$) is accompanied with the elimination of some variables in the set ($\pimunut, \Pit$). For
example, using the bulk field $\phi$ we may assume that $\Pit=0$. The extreme strategy in this context is
to assume that the variables $\ximunu$ and $\phi$ are chosen in such a way that
\begin{eqnarray}
\Tmunu = \Tmunua .
\label{TmunuAP}
\end{eqnarray}
This formula represents the {\it anisotropic matching principle} introduced by Tinti~\cite{Tinti:2015xwa}. We note that (\ref{TmunuAP}) looks exactly like (\ref{TmunuA}), however, these two equations are obtained with different assumptions: instead of neglecting the term $ \delta {\tilde f}(x,p)$ in (\ref{deltafa}) one assumes that $ \delta {\tilde f}(x,p)$ might be finite but it does not contribute to $\Tmunu$.

\end{itemize}

We can now define the gradient expansion for the leading-order anisotropic hydrodynamics. Given $T(x)$ and $\umu(x)$ we construct $f_{\rm eq} (x,p) $ and $\Tmunueq$ and write
\begin{eqnarray}
\Tmunu  = \Tmunueq + \delta \Tmunu = \Tmunueq + \left(\Tmunua - \Tmunueq \right).
\label{TmunuGrada}
\end{eqnarray}
This formula suggests to use the gradient expansion of anisotropic hydrodynamics in the form
\begin{eqnarray}
\Tmunu  = \Tmunueq + \hbox{powers of gradients of } T, \umu, \ximunu \hbox{and  } \phi .
\label{TmunuGrada2}
\end{eqnarray}
Compared to (\ref{TmunuGrad}), the expansion  (\ref{TmunuGrada2}) includes in addition the gradients of $\ximunu$ and $\phi$. On the other hand, similarly to (\ref{TmunuGrad}),  the expansion  (\ref{TmunuGrada2})  should be done around the perfect-fluid solution that is determined solely by the conservation law.

\subsubsection{Phenomenological vs. kinetic-theory formulation}
\label{subsect:anisohist}

The original concepts of anisotropic hydrodynamics were introduced in Refs.~\cite{Florkowski:2010cf,Martinez:2010sc}, see also \cite{Barz:1987pq,Kampfer:1990qg}. The approach of Ref.~\cite{Florkowski:2010cf} was based on the energy-momentum conservation law and used an ansatz for the entropy source term which defined the off-equilibrium dynamics. On the other hand, the approach of Ref.~\cite{Martinez:2010sc} was based on the kinetic theory, and employed the zeroth and first moments of the RTA Boltzmann kinetic equation~\cite{Bhatnagar:1954zz}. It was demonstrated in \cite{Ryblewski:2010ch}  that these two approaches are equivalent as the first moment of the Boltzmann equation yields the energy-momentum conservation law, while the zeroth moment can be interpreted as a special form of the entropy source.

The two original approaches describe boost invariant, transversally homogeneous systems and refer to the (quasi)particle picture, where the phase-space distribution function is given by the Romatschke-Strickland (RS) form~\cite{Romatschke:2003ms}. In the covariant formulation, the latter takes the form
\beq
\label{eq:RS}
\fRS = \f{1}{(2\pi)^3}\exp \left( - \f{1}{\Lambda}  \sqrt{ (\pdotU)^2 + \xi \,\,(\pdotZ)^2 } \,  \right),
\eeq
where $\Lambda$ is the transverse-momentum scale variable, $\xi$ is the anisotropy variable, while $U$ and $Z$ are the  two four-vectors that define a simple boost-invariant geometry, $U=(t/\tau,0,0,z/\tau)$ and $Z=(z/\tau,0,0,t/\tau)$ with $\tau = \sqrt{t^2-z^2}$ being the longitudinal proper time. The distribution function (\ref{eq:RS}) leads to the following form of the energy-momentum tensor
\beq
\label{eq:Tmunu}
\Tmunu =  \ed  \umu \unu  + \pT \, \Dmunu + (\pL - \pT) \zmu \znu.
\eeq

Within the kinetic-theory formulation, if we restrict ourselves to conformal, boost-invariant and cylindrically symmetric systems, the form of the hydrodynamic flow is fixed and the \AH\, equations are reduced to two coupled ordinary differential equations for the anisotropy fluid variable $\xi$ and the transverse-momentum scale variable  $\Lambda$~\cite{Martinez:2010sc}. They read
\beq
\frac{1}{1+\xi} \f{d\xi}{d \tau} &=&
 \frac{2}{\tau} - 4 \, \Gamma \, \cR(\xi) \, \frac{\cR^{3/4}(\xi)\sqrt{1+\xi}-1}{2 \cR(\xi) + 3 (1+\xi) \cR'(\xi)} \, ,
 \label{ah1} \\
\frac{1}{1+\xi}\frac{1}{\Lambda} \f{d\Lambda}{d\tau} &=& \Gamma \, \cR'(\xi) \,
\frac{\cR^{3/4}(\xi)\sqrt{1+\xi}-1}{2 \cR(\xi) + 3 (1+\xi) \cR'(\xi)} \, .
\label{ah2}
\eeq
Here the parameter $\Gamma$ is proportional to the inverse relaxation time and $ \cR'(\xi) = d\cR(\xi)/d\xi$. If we demand that Eqs.~(\ref{ah1}) and (\ref{ah2})  describe the system with the same shear viscosity as that obtained directly from the RTA kinetic theory based on \rf{bebif} we should use the relation~\cite{Florkowski:2013lya}
\beq
\f{1}{\Gamma} = \f{\trel}{2}.
\eeq
The function $\cR(\xi)$, appearing in Eqs.~(\ref{ah1}) and (\ref{ah2}), is defined as~\cite{Martinez:2010sc}
\beq
\cR(\xi) = \frac{1}{2}\left(\frac{1}{1+\xi}
+\frac{\arctan\sqrt{\xi}}{\sqrt{\xi}} \right).
\label{Rofxi}
\eeq
%

In the analogous way we introduce
\bel{RTofxi}
\cR_T(\xi)
= \frac{3}{2 \xi}
\left( \frac{1+(\xi^2-1){\cal R}(\xi)}{\xi + 1}\right)
\eeq
and
\bel{RLofxi}
\cR_L(\xi)
= \frac{3}{\xi}
\left( \frac{(\xi+1){\cal R}(\xi)-1}{\xi+1}\right)  .
\eeq
The functions $\cR(\xi)$, $\cR_T(\xi)$, and $\cR_L(\xi)$ are used to relate the non-equilibrium energy density and the two pressures, depending on $\xi$ and $\Lambda$, to the equilibrium (isotropic) quantities computed at the temperature $\Lambda$, namely
\beq
\ed(\xi,\Lambda) &=&  \cR(\xi) \ed_{\rm iso}(\Lambda), \label{Eiso} \\
\pT(\xi,\Lambda) &=&  \cR_T(\xi) \peq_{\rm iso}(\Lambda), \label{PTiso} \\
\pL(\xi,\Lambda) &=&  \cR_L(\xi) \peq_{\rm iso}(\Lambda) \label{PLiso}.
\eeq
If the anisotropy variable $\xi$ vanishes, the scale variable $\Lambda$ coincides with the effective temperature $T$, while the two pressures become identical and equal to one third of the energy density. Since $3 \cR - 2 \cR_T - \cR_L = 0$ and $\ed_{\rm iso}=3\peq_{\rm iso}$, we also find $\ed-2\pT-\pL=0$, as required for conformal systems.

\subsubsection{Perturbative vs. non-perturbative approach}
\label{subsect:anisoleadorder}

Subsequent developments of anisotropic hydrodynamics were based exclusively on the kinetic theory and they may be classified either as perturbative or as non-perturbative schemes.
\begin{itemize}
\item{}
In the perturbative approach~\cite{Bazow:2013ifa,Molnar:2016vvu,Molnar:2016gwq} one assumes that the distribution function has the form  $f = \fRS + \delta f $, where $\fRS$ is the leading order described by the RS form, which accounts for the difference between  the longitudinal and transverse pressures, while $\delta f$  describes a correction.  In this case, advanced methods of traditional viscous hydrodynamics are used to restrict the form of $\delta f$ and to derive hydrodynamic equations. In this way non-trivial dynamics may be included in the transverse plane and, more generally, in the full (3+1)D case.
\item{}
In the non-perturbative approach one starts with the decomposition  $f = \fa +  \delta f $, where $\fa$ is the leading order distribution function given by the generalised RS form. In this case, all effects due to anisotropy are included in the leading order, while the correction term $ \delta f$ is typically neglected. The generalised RS form includes more variables than the original RS ansatz, namely one uses the expression
\beq
\fa  = f_{\rm iso}\left(\frac{\sqrt{p^\mu \Ximunul p^\nu}}{\lambda}\right),
\eeq
where $\lambda$ is the transverse momentum scale, $\Ximunul$ is the anisotropy
tensor defined below and $f_{\rm iso}$ denotes the isotropic distribution (in
practice the equilibrium Boltzmann, Bose-Einstein or Fermi-Dirac
distribution). 
\end{itemize}

The structure of the generalised RS distribution together with the corresponding hydrodynamic equations have been gradually worked out for: (1+1)D conformal case~\cite{Tinti:2013vba} (with two independent anisotropy fluid variables), (1+1)D non-conformal case~\cite{Nopoush:2014pfa} (with two anisotropy variables and one bulk variable), and full (3+1)D case~\cite{Tinti:2014yya,Tinti:2015xra} (five anisotropy variables included in the tensor $\ximunu$ and one bulk variable $\phi$). In the latter case,  one uses the following parameterisation
\beq
\label{eq:par}
&& \Ximunu  = \umu \unu + \ximunu + \Dmunu \phi,  \nn \\
&&  \umul \ximunu = 0, \quad  \ximumu = 0.
\eeq
The second line in (\ref{eq:par}) indicates that the symmetric tensor
$\ximunu$ is orthogonal to $\umu$ and traceless, thus has indeed five
independent variables. These properties are similar to those characterising
the shear stress tensor $\pimunu$ commonly used in the formalism of the
standard dissipative hydrodynamics. As a matter of fact, $\pimunu$ becomes
proportional to $\ximunu$ for systems approaching local
equilibrium. Similarly, in this case the variable $\phi$ becomes proportional
to the bulk viscous pressure~$\Pi$. 

\subsubsection{Anisotropic matching principle}
\label{subsect:amp}

To derive the hydrodynamic equations obeyed by the fluid variables $\ximunu$ and $\phi$ one most often uses the moments of the RTA kinetic equation. The number of included moments is equal to the number of variables to be determined. An alternative to this approach is the procedure where one first derives, directly from the RTA Boltzmann equation,  the equations for the pressure corrections  $\pimunu$  and~$\Pi$, and expresses them in terms of the function~$\fa $. This is the latest development for the leading order, known as the anisotropic matching principle~\cite{Tinti:2015xwa}, that may be supplemented by next-to-leading terms in a perturbative approach~\cite{Bazow:2013ifa,Molnar:2016vvu}.

For conformal, boost-invariant, and transversally homogeneous systems  the bulk variable $\phi$ can be set equal to zero, while the tensor $\ximunu$ has the structure
\beq
\ximunu = \xi_T \left( X^\mu X^\nu +  Y^\mu Y^\nu \right) + \xi_L  Z^\mu Z^\nu,
\label{ximunu}
\eeq
where $X^\mu = (0,1,0,0)$ and $Z^\mu=(0,0,1,0)$. Then, the distribution function can be written as
\bel{TF1}
 f_{\rm a} =   \exp\left(- \frac{E}{\lambda}\, \right),
\eeq
where
\bel{E2}
E^2 = (1+\xi_X) (p\cdot X)^2 + (1+\xi_Y) (p\cdot Y)^2 + (1+\xi_Z) (p\cdot Z)^2
\eeq
with $\xi_X=\xi_Y=\xi_T$ and $\xi_Z=\xi_L$.  We note that Eqs.~(\ref{TF1}) and (\ref{E2}) with $\xi_X \ne \xi_Z$ represent a generalized ellipsoidal parameterization of the anisotropic distribution function proposed first in~\cite{Tinti:2013vba}. The symmetries of $\ximunu$ imply that the anisotropy variables $\xi_I$ introduced in  (\ref{E2}) satisfy the condition \cite{Tinti:2013vba}
\begin{eqnarray}
\sum_I \xi_I = \xi_X + \xi_Y + \xi_Z = 0 \, .
\label{sumofxis}
\end{eqnarray}
Consequently, the parameterizations (\ref{eq:RS}) and (\ref{TF1}) are connected through the following set of simple transformations
\begin{eqnarray}
\xi_X &=& \xi_Y = \xi_T = -\frac{\xi/3}{1+\xi/3}, \nonumber \\
\xi_Z &=& \xi_L = \frac{2\,\xi/3}{1+\xi/3}, \nonumber \\
\lambda &=& \Lambda (1+\xi/3)^{-1/2}.
\label{RS-TF}
\end{eqnarray}

Using the anisotropic matching principle introduced above, one obtains two coupled ordinary differential equations for the effective temperature $T$ and the anisotropy variable $\xi$,
\beq
4\,\frac{ \cR(\xi)}{T} \frac{dT}{d\tau} = -\frac{1}{\tau}
\left(\cR(\xi) + \frac{\cR_L(\xi)}{3}  \right)
\label{enecon1}
\eeq
and
\begin{eqnarray}
\frac{d\,\Delta P}{d\tau} = - \frac{T\,\Delta P }{c } - \frac{F}{\tau},
\label{DeltaPeq}
\end{eqnarray}
Here $\Delta P$ is the difference of the longitudinal and transverse pressures. Using definitions given in~\cite{Tinti:2015xwa} one finds that  $\Delta P$  can be expressed as
\begin{eqnarray}
\Delta P = -\frac{6 k \pi   \Lambda^4}{\xi}
 \left(\frac{\xi+3}{\xi+1} + \frac{(\xi-3) \arctan\sqrt{\xi}}{\sqrt{\xi}} \right).
 \nonumber \\ \label{DeltaPofy}
\end{eqnarray}
Similarly, one finds the form of the function $F$ appearing on the right-hand
side of (\ref{DeltaPeq}), namely
\begin{eqnarray}
F = -2 (1+\xi) \frac{\partial \Delta P}{\partial \xi}.
\nonumber \\ \label{F}
\end{eqnarray}

\subsection{Comparisons with exact solutions of the RTA kinetic equation}
\label{subsect:comparisons}

Anisotropic hydrodynamics and viscous hydrodynamics predictions have been checked against exact solutions available for the RTA kinetic equation~\cite{Baym:1984np,Baym:1985tna}. Such studies have been done for the one-dimensional Bjorken geometry~\cite{Florkowski:2013lza,Florkowski:2013lya,Florkowski:2014sfa} and for the Gubser flow which includes transverse expansion~\cite{Denicol:2014xca,Denicol:2014tha}. The results of those studies show that \AH in general better reproduces the results of the underlying kinetic theory than the standard viscous hydrodynamics.

\begin{figure}[t]
\begin{center}
\includegraphics[angle=0,width=0.95\textwidth]{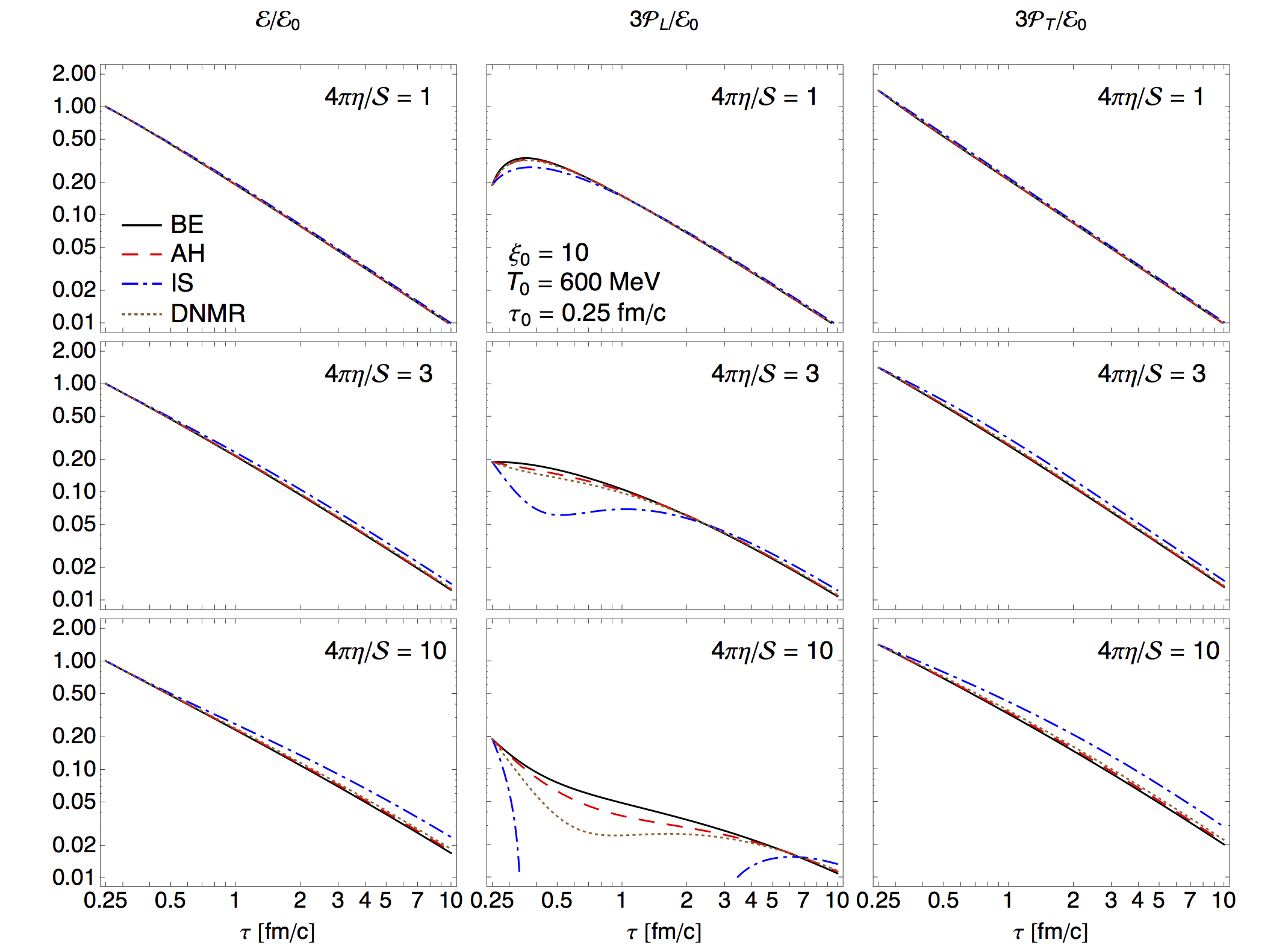}
\end{center}
\caption{Time dependence of the energy density (first column), longitudinal pressure (second column) and the transverse pressure (third column) for three values of the shear viscosity: $4\pi \eta/\sd=1$ (first row), $4\pi \eta/\sd=3$ (second row), and $4\pi \eta/\sd=10$ (third row). The solid black, dashed red, dashed-dotted blue and dotted brown curves describe the results obtained with: RTA kinetic theory, AHYDRO, MIS, and DNMR.}
\label{fig:PLPTE_600_10}
\end{figure}

In Fig.~\ref{fig:PLPTE_600_10} we show the results for the proper-time dependence of the energy density (first column), longitudinal pressure (second column) and the transverse pressure (third column). All these quantities are normalised in such a way as to have the same late-time asymptotics. The results presented in the three rows (from up to down) correspond to three values of the shear viscosity: $4\pi \eta/\sd=1$, $4\pi \eta/\sd=3$, and $4\pi \eta/\sd=10$. The solid black curves show the results obtained by solving the RTA kinetic equations. They are compared with the results obtained within the framework of \AH (dashed red curves), the MIS hydrodynamics (dashed-dotted blue lines), and the DNMR hydrodynamics (dotted brown curves). The \AH approach is defined in this case by Eqs.~(\ref{ah1}) and~(\ref{ah2}).

All these results describe a one-dimensional, boost-invariant expansion with the initialization time $\tau_0~=~0.25$~fm/c, the initial anisotropy variable $\xi_0=10$, and the initial effective temperature $T_0=$~600~MeV. One observes that the \AH results  follow most closely the exact kinetic-theory results. This is most clearly seen in the case of the longitudinal pressure and large viscosity. For $4\pi \eta/\sd=10$ the longitudinal pressure calculated in the MIS approach becomes negative, which is not allowed in the kinetic-theory framework. Although such large values of the QGP shear viscosity are excluded by the recent phenomenological models of heavy-ion collisions, one may find such values at the edges of the system where hadronic gas is present.  Therefore, the use of the MIS approach should be avoided in such regions. Interestingly, the more complete DNMR approach reproduces the exact result much better.

\subsection{Consistency with the gradient expansion}
\label{subsect:consistency}

Recently, the gradient expansion has been studied for anisotropic hydrodynamics (in the formulation based on the anisotropic matching principle), its underlying kinetic theory in the relaxation time approximation, and for different formulation of standard viscous hydrodynamics~\cite{Florkowski:2016zsi}. The first four coefficients of the gradients expansion generated for these theories are shown in Table~\ref{tab:grad_coeff}. One finds that the formulation of anisotropic hydrodynamics based on the anisotropic matching principle~\cite{Tinti:2015xwa} yields the first three terms of the gradient expansion in agreement with those obtained for the RTA kinetic theory~\cite{Heller:2016rtz}, see the second and the last columns. This finding gives further support for this particular hydrodynamic model as a good approximation of the kinetic-theory framework.

\begin{table}[t]
  \begin{center}
  \begin{tabular}{|c|c|c|c|c|c|}
    \hline
    $n$ & RTA & BRSSS & DNMR & MIS & AHYDRO\\
    \hline
    $0$ & $2/3$ & $2/3$ & $2/3$ & $2/3$ & $2/3$\\
    \hline
    $1$ &  $4/45$ &  $4/45$ & $4/45$ & $4/45$ & $4/45$ \\
    \hline
    $2$ & $16/945 $ & $16/945$ & $16/945$ & $8/135 $ & $16/945$ \\
    \hline
    $3$ & $- 208/4725$ & $-1712/99225$  & -$304/33075$ & $112/2025$ & -$176/6615$ \\
    \hline
  \end{tabular}
  \end{center}
    \caption{Leading coefficients of gradient expansions for RTA,
      BRSSS, DNMR, MIS, and AHYDRO \cite{Florkowski:2016zsi}.}
       \label{tab:grad_coeff}
\end{table}

Leading coefficients of the gradient expansion have also been calculated for other versions of the
hydrodynamic evolution equations. The results for BRSSS, DNMR and MIS are shown in the third, fourth and fifth
column of Table~\ref{tab:grad_coeff}, respectively. We find that both BRSSS and DNMR agree up to terms of
second order in gradients ($n=2$). It is important to stress, however, that there are different reasons for
this agreement in the two cases. For BRSSS one fixes the free parameters (transport coefficients) in such a
way as to get the agreement up to second order (and since MIS has fewer transport coefficients to adjust, it
does not reproduce the second order exactly). In contrast, in the DNMR and AHYDRO constructions the values of
the transport coefficients are obtained directly from the RTA kinetic equation (without any adjustment).
Recall finally, that the third order theory due to Jaiswal~\cite{Jaiswal:2014raa} (see \rfs{sect:jaiswal}) matches the first three orders, as guaranteed by its construction.

We note at this point that in a recent work~\cite{Florkowski:2016kjj}, the off-equilibrium behaviour
described by different hydrodynamic models has been analysed and compared in numerical simulations of
non-boost invariant expansion. It was found that the results of anisotropic hydrodynamics and viscous
hydrodynamics agree for the central close-to-equilibrium part of the system, however they differ at the edges
where the approach of anisotropic hydrodynamics helps to control the undesirable growth of viscous corrections
observed in standard frameworks.

%% file: sect-resu.tex


\section{Asymptotic nature of the late proper time expansion}
\label{sect:resu}

Much of this review has been devoted to showing how the gradient expansion
truncated at low orders provides an effective way to quantify the approach to
equilibrium and at the same time provides a way of matching calculations in
microscopic models with effective descriptions in terms of hydrodynamics. In
the present section we review recent progress in understanding large order
behaviour of gradient expansions, which has led to the discovery of their
asymptotic character both at the microscopic level~\cite{Heller:2013fn} and in
hydrodynamics~\cite{Heller:2015dha}. It is important to realize that the
divergence of the gradient expansion is connected to the phenomenon of
hydrodynamization, reviewed in \rfs{sect:hydrod}, which is the statement that
hydrodynamics works even when leading order corrections to the perfect fluid
limit are very large.

These ideas were partly motivated by the early papers of Lublinsky and
Shuryak~\cite{Lublinsky:2007mm,Lublinsky:2009kv}, which were later developed
in the context of linearized solutions of gravitational equations in
Refs.~\cite{Bu:2014sia,Bu:2014ena,Bu:2015bwa}. The main difference with the
works extensively discussed in this section is the linearized character of
problems studied in these papers. As advocated in Ref.~\cite{Heller:2016gbp}
there is a way to understand the asymptotic character of the hydrodynamic
gradient expansion at the level of linear response theory and we refer
interested readers to this reference.

It has been observed long ago that divergent series can be meaningful, see
e.g.~Ref.~\cite{Boyd1999}. Indeed, a useful way to think about such series is
not as prescriptions to add up all the contributions in a naive manner, but
rather as a formal method of presenting an ordered sequence of numbers which
encodes some information. We cannot attempt to review this subject here, but
we will summarize and briefly explain the basic methods which have been used
in the literature to deal with the divergent series arising in the context of
boost-invariant conformal hydrodynamics.

The main message is that the pattern of divergence conveys information about
the transient dynamics of effective hydrodynamic models as well as of
microscopic theories. In the former case, as reviewed in \rfs{sect:fund},
hydrodynamic modes are augmented with a non-hydrodynamic sector which acts as
a regulator to maintain causality. It is most straightforward to begin with
this setting, because here one has much better access to the full dynamics
which generates the gradient expansion, so the connection between the pattern
of divergence and the non-hydrodynamic sector can be made completely
explicit. This will be the subject of \rfs{sect:resu:hydro}, where we will
address BRSSS theory as well as anisotropic hydrodynamics and the HJSW theory
reviewed in Secs. \ref{sect:ahydro} and \ref{sect:beyond}. Building intuition
on these examples, we will review microscopic calculations in holography,
which initiated this line of research, and the most recent analysis of these
issues performed within RTA kinetic theory.

\subsection{Large order gradient expansion in hydrodynamics}
\label{sect:resu:hydro}

A very fruitful point of view is to think of the gradient expansion calculated
in a hydrodynamic theory, such as MIS or any of the other models reviewed
here, precisely in the same way as if it had been obtained from a fundamental
theory.  This is useful, because both at the microscopic level and the level
of hydrodynamics the hydrodynamic modes dominate at late times, the difference
lying only in the spectrum of transient modes. We can then study it in
hydrodynamic theories where it is simple and where we have reliable
information about it (from linearized analysis, for example). This leads to a
very detailed understanding of the interplay between the structure of the
non-hydrodynamic sector and large-order behaviour of the gradient expansion.

\subsubsection{BRSSS theory}
\label{sect:resu:mis}

The simplest hydrodynamic theory to consider is MIS theory, but in this section, following
Refs.~\cite{Heller:2015dha,Basar:2015ava,Aniceto:2015mto}, we will instead deal with BRSSS theory since it
captures more physics than MIS, yet is not much more difficult to analyse.

The gradient expansion of the pressure anisotropy~$\pa(w)$,
\bel{divser}
\pa(w) = \sum_{n=1}^\infty \f{\pac_n}{w^n},
\eel
can in this case be calculated analytically directly from the differential equation \rf{feqn}. The leading terms of this expansion were already presented  in \rf{rhydro}. It is straightforward to calculate the expansion coefficients to very high order. In our presentation,
following Ref.~\cite{Heller:2015dha}, we will adopt the \symm\ values of the transport coefficients. The basic
feature of the resulting series is that at large order the expansion coefficients exhibit factorial growth.
More precisely, one can check that at large $n$ these coefficients grow with $n$ in a way consistent with the
Lipatov form~\cite{Lipatov:1976ny}
\bel{lipatov}
\pac_n \sim \f{\Gamma[n+\beta]}{A^n},
\eel
where $A$ and $\beta$ are real parameters. Given the series coefficients ${\pac_n}$, one can easily check whether \rf{lipatov} applies by examining the ratios of neighbouring coefficients. Specifically, if \rf{lipatov} describes the large order behaviour, then one should find asymptotically linear behaviour:
\bel{rat}
\f{\pac_{n+1}}{\pac_n} \sim \f{1}{A} \, n + \f{\beta}{A}.
\eel
For the case of the gradient series in BRSSS hydrodynamics, this linear behaviour can be seen in \rff{fig:diverge}. It is worth noting that by fitting a straight line to the ratio \rfn{rat} one can determine the parameters $A\approx 7.21$ and $\beta\approx -1.15$, which have a clear physical interpretation, described later on in this section.
\begin{figure}[ht]
\begin{center}
\includegraphics[height=0.3\textheight]{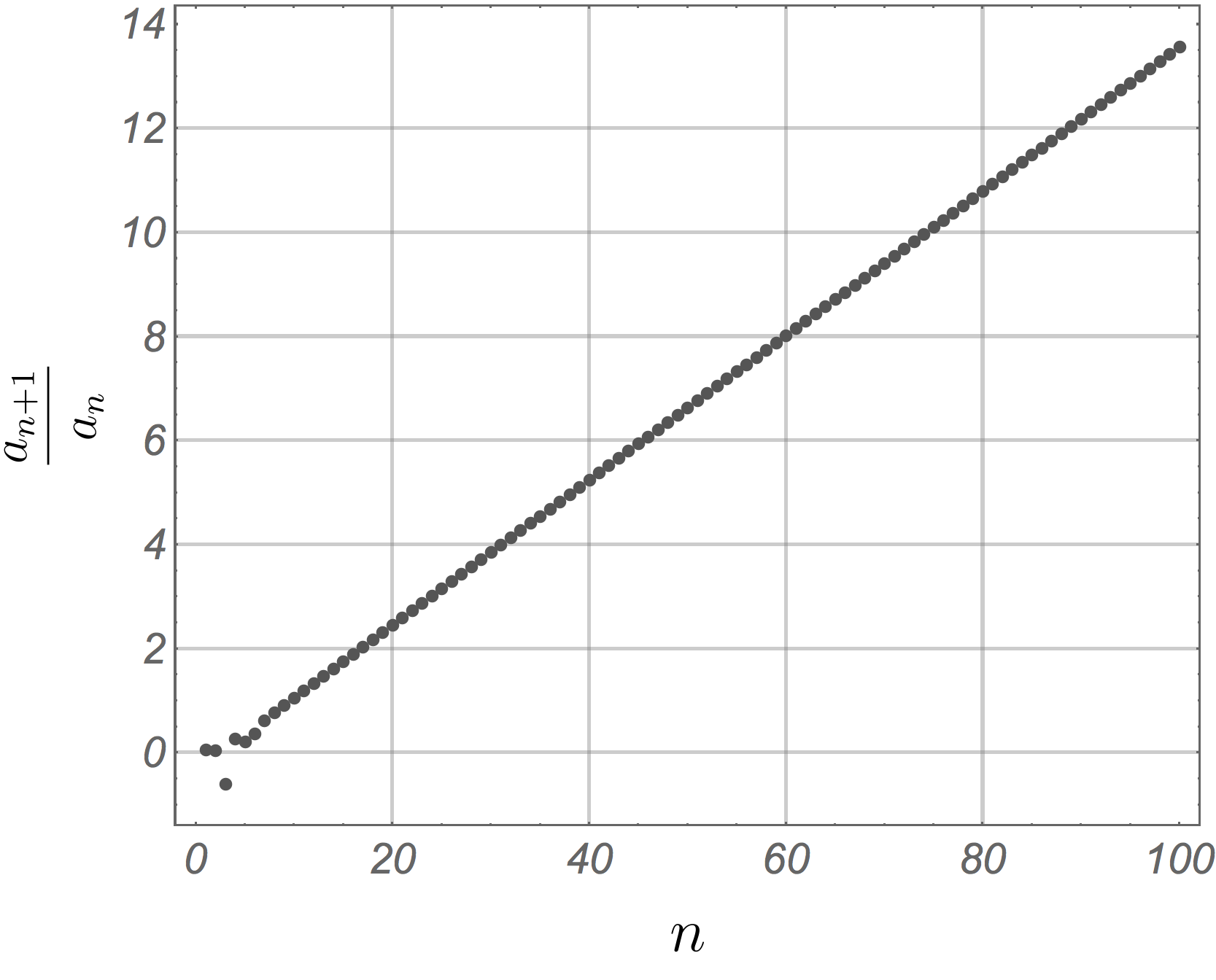}
\caption{The large order behaviour of the hydrodynamic gradient series for BRSSS
hydrodynamics. The linear behaviour of $\pac_{n+1}/\pac_{n}$ at large $n$ determines $A$ and
$\beta$ through \rf{rat}.} 
\label{fig:diverge}
\end{center}
\end{figure}

The standard tool in dealing with factorially divergent series is the (generalized) Borel transform,
see~e.g.~Ref.~\cite{Boyd1999}, which amounts to removing the leading order factorial growth of the series
coefficients:
\bel{borel}
\bort \pa(\xi) = \sum_{n=1}^\infty \frac{\pac_n}{\Gamma(n+\beta)} \, \xi^{n}.
\eel
This series will possess a non-vanishing radius of convergence, and will define an analytic function within the disc of radius $A$ around the origin in the complex $\xi$ plane. The inverse transform is given by the (generalized) Borel summation formula
\bel{iborel}
\pa_{\mathrm{resummed}}(w) = w^\beta \int_{\cal C} d\xi \, e^{- w \xi}\, \xi^{\beta-1}\, \borta \pa(\xi),
\eel
where $\borta R$ is the analytic continuation of the Borel transform \rf{borel} and $\cal C$ is the contour
connecting~$0$~and~$\infty$. The analytic continuation of $\bort \pa(\xi)$ necessarily contains singularities that are
responsible for the vanishing radius of convergence of the original series.

In practice, the analytic continuation of the Borel transform is typically performed using Pad\'e approximants. A Pad\'e approximant is a rational function
\bel{pade}
\borta \pa(\xi) = \f{P_m(\xi)}{Q_n(\xi)},
\eel
where $P_m(\xi)$ and $Q_n(\xi)$ are polynomials of order $m$ and $n$ respectively, with the constant term in $Q_n$ scaled to unity and the remaining coefficients chosen so as to match a given polynomial of order $m+n+1$ -- in our case the truncated Borel transform of the gradient series.

\begin{figure}[ht]
\begin{center}
\includegraphics[height=0.3\textheight]{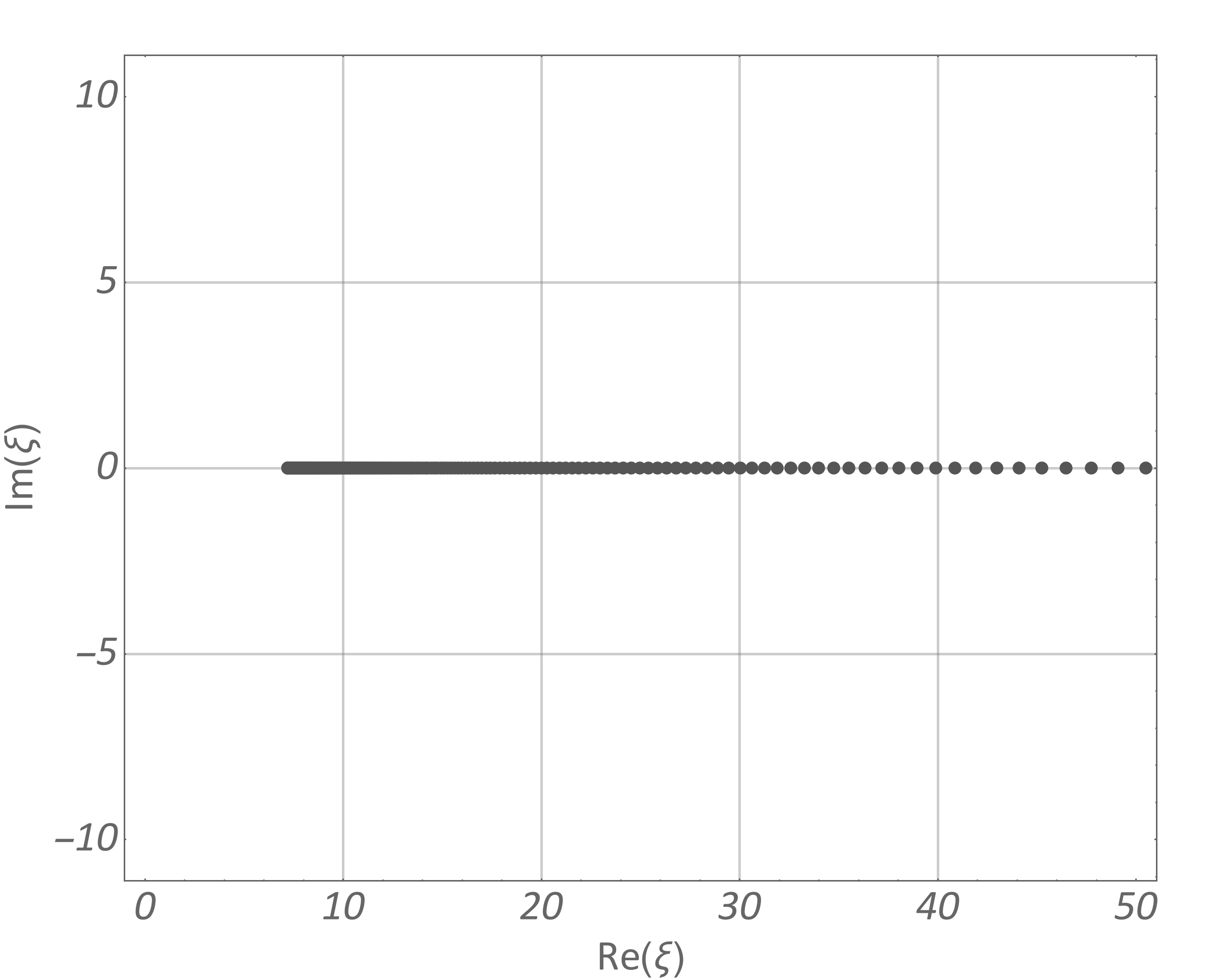}
\caption{Poles of the symmetric Pad\'e approximant to the Borel transform of
  the gradient series for BRSSS hydrodynamics truncated at 600
  derivatives. The leading singularity is a branch point at $\xi=A$ with a cut
  starting there and running to infinity along the real axis. An in-depth
  analysis in Refs.~\cite{Heller:2015dha,Basar:2015ava,Aniceto:2015mto}
  revealed that this singularity hides further branch points located at
  positive integer multiples of $A$ as expected from the theory of resurgence
  (see also Fig.~\ref{fig.BorelPolesHJSW}).}
\label{fig:padecut}
\end{center}
\end{figure}

The singularities of the Pad\'e approximants are poles, but it is known that functions possessing branch point singularities result in Pad\'e approximants featuring dense sequences of poles originating at the locations of the branch points, see e.g. Ref.~\cite{2013arXiv1308.4453Y}. For the case of the gradient expansion of BRSSS theory one finds a branch cut starting at $\xi \approx 7.2118$; this is shown in \rff{fig:padecut}. The location of the branch point is numerically very close to the value of the parameter $A\approx 7.2117$ fitted using \rf{lipatov}.

In fact, the origin of the cut is consistent with the supposition that the analytically continued Borel transform contains a singular piece of the form
\bel{nonanalytic}
\borta \pa(\xi) \sim \log{\left( A-\xi \right)} 
\eel
where the parameter $A$ describing the location of the branch point is the same as the one which appears in
\rf{rat}. One can convince oneself of this claim by calculating Pad\'e approximants for \rf{nonanalytic} and
studying the pattern of singularities.

The key observation is that the presence of a branch point singularity on the real axis introduces complex
ambiguities in the Borel summation, given by the difference in the values obtained for \rf{iborel} by integrating
above and below the cut. For large values of $w$ the ambiguity is proportional to
\bel{amb}
AMB = e^{-w \, A} w^{\beta},
\eel
where we have neglected subleading corrections in the large-$w$ expansion. This ambiguity is a critically
important feature of the hydrodynamic series in BRSSS theory and its presence is an indication of physics
encoded in its large-order behaviour.

The form \rf{amb} implies that to cancel the ambiguity one needs exponential corrections to the gradient
series.  One can easily see that such corrections are in fact required by the differential equation which we
are solving, \rf{feqn}. Indeed, we have already seen in \rfs{sect:effective} that there are non-analytic,
exponentially suppressed corrections to the hydrodynamic series following from the presence of the
non-hydrodynamic MIS mode, given in \rf{linpert}, which we display here again for convenience:
\bel{linpert2}
\delta \pa(w) \sim e^{-\frac{3}{2 C_{\tau_{\pi}}} w}
w^{\frac{C_\eta- 2  C_{\lambda_1} }{C_{\tau_{\pi}}}}
\left\{1 + O\left(\frac{1}{w}\right)\right\}.
\eel
These have precisely the correct structure to eliminate the ambiguity in inverting the Borel transform~(\ref{amb}) and suggests that one should make the following identifications:
\bel{pertparams}
A = \frac{3}{2 \, C_{\tau_{\pi}}}\quad \mathrm{and} \quad \beta = \frac{C_\eta- 2 \,C_{\lambda_1} }{C_{\tau_{\pi}}}.
\eel
Evaluating these combinations with parameter values appropriate for
\symm\ (which were used to generate the gradient expansion under discussion)
gives excellent agreement with the values obtained by fitting $A$ and $\beta$
using \rf{lipatov}. Both Eq.~(\ref{amb})~and~(\ref{linpert2}) receive
corrections in $1/w$ which one also expects to match. One therefore sees that
the branch-cut in the Borel plane represents the transient excitation of BRSSS
theory evaluated at vanishing momentum, see \rf{nonhydrom}. The fact that the
exponential decay receives an infinite series of corrections in powers $1/w$
can be interpreted by saying that the transient excitation seen earlier in the 
linearized analysis described in \rfs{sect:mis} becomes
``dressed'' by the hydrodynamic flow. This issue was addressed for the first
time in the AdS/CFT-based calculations of Ref.~\cite{Janik:2006gp}.

The exponential correction \rf{linpert2} is in fact just the leading term in
an infinite set of exponentially suppressed contributions which need to be
added to the original asymptotic series \rf{rgradex}. This naturally follows
from the nonlinear nature of the BRSSS evolution equation, which in our
variables is given by \rf{feqn}. The resulting expression for the pressure
anisotropy takes the form of a transseries:
\bel{trans}
\pa(w) = \sum_{m=0}^\infty \sigma^m \, e^{-m A w} \, \Phi_m(w),
\eel
where $m$ labels the different transseries sectors and $\sigma$ is a so-called transseries parameter, which is complex, see~e.g.~Refs.~\cite{Dunne,Aniceto:2013fka}. In each sector we have a divergent series of the form
\bel{sector}
\Phi_m(w) = w^{m \, \beta} \, \sum_{n=0}^\infty \f{\pac^{(m)}_n}{w^n},
\eel
whose resummation involves an exponentially-suppressed ambiguity at large values of $w$, as in \rf{amb}. The sector with $m=0$ is the original gradient expansion \rf{divser}, so that $\pac^{(0)}_n\equiv \pac_n$. The $m=1$ sector corresponds to the only transient excitation that is present in the BRSSS theory, and the higher order sectors follow from the presence of nonlinear effects in \rf{feqn}.

This transseries structure is well known in physics due to its appearance in studies of large-order behaviour of perturbation theory in quantum mechanics. For this reason one sometimes speaks of $\Phi_0$ as being the ``perturbative'' sector, and the $\Phi_m$ with $m>0$ as the ``instanton'' sectors. In the same spirit, the parameter $A$ in \rf{lipatov} is sometimes referred to as an ``action''. As we shall see below, one often has many such~``actions''.

To make sense of the transseries expansion \rf{trans} one proceeds by performing a Borel summation in each sector. The basic idea is that the complex parameter $\sigma$ carries physical information about the initial state. \rf{feqn} requires specifying an initial condition to determine its solution: one real number, the value of $\pa$ at some value of $w$. This provides one constraint on $\sigma$. To fully determine $\sigma$ one has to impose the cancellation of the ambiguity in the choice of contour $\cal C$ when performing the integral in the inverse Borel transform for the perturbative series $\Phi_0$. The resurgent structure of the transseries then guarantees that there is a choice of integration contours when resumming the $m = 1, 2, \ldots$ sectors in \rf{trans} such that the answer, $\pa(w)$, is real and unambiguous up to the physical choice of the initial condition.

The ideas reviewed above in the context of BRSSS theory carry over to more complicated cases in spirit, but need to be suitably adapted and generalized. Before we move on to this subject, let us remark that it is possible to explicitly carry out the Borel summation procedure, as outlined in this section. This was
done for the first three transseries sectors ($m = 0, 1, 2$) in \rfc{Heller:2015dha}. It is interesting to note that in order to match this expression with the special attractor solution discussed in \rfs{sect:brsss}, one is required to set
$\Re(\sigma)\approx 0.875$, while naively one might expect that the attractor should correspond to omitting the exponential terms. In consequence, instead of thinking about non-hydrodynamic modes in terms of perturbations about the gradient expansion one should perhaps think of them as perturbations around the hydrodynamic attractor. The presence of such attractor solutions in other models, see Ref.~\cite{Romatschke:2017vte}, certainly strengthens this idea.
\begin{figure}
\begin{center}
\includegraphics[height=0.32\textheight]{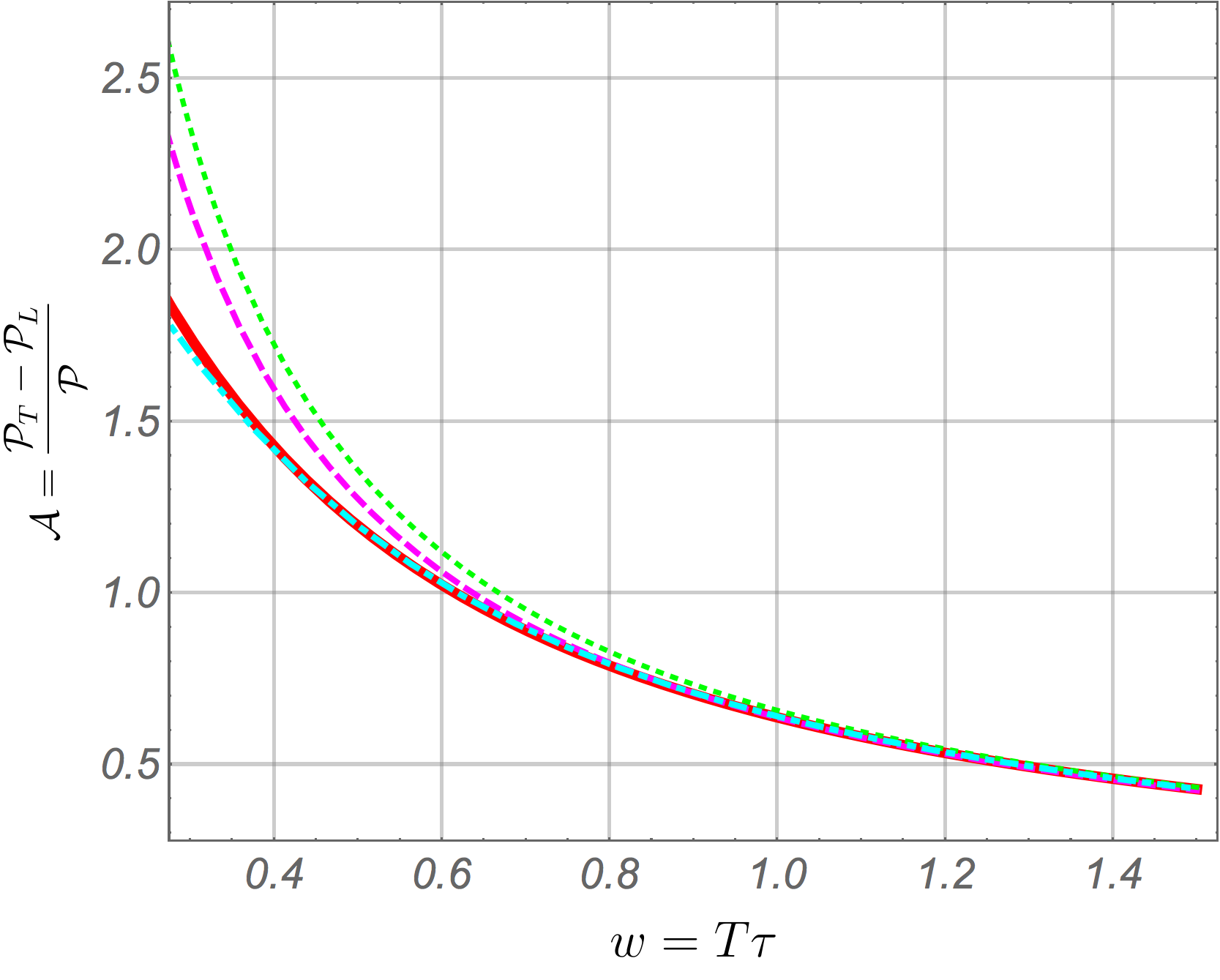}
\caption{The hydrodynamic attractor (solid red), compared with the
  resummation result (cyan, dot-dashed) and the gradient expansion truncated at first
  (magenta, dashed) and second (green, dotted) orders. Quite curiously, the comparison required fitting
  \mbox{$\Re(\sigma)\approx 0.875$}~in~\rf{trans}.} 
\label{fig.all}
\end{center}
\end{figure}

\subsubsection{Anisotropic hydrodynamics}
\label{sect:resu:aniso}

The large order behaviour of the gradient expansion of the pressure anisotropy in boost-invariant anisotropic hydrodynamics for a conformal fluid was calculated in \rfc{Florkowski:2016zsi}. The calculation proceeds a little differently there, since it does not seem to be possible to write down a single differential equation for $\pa(w)$. Instead, one has to solve equations \rfn{enecon1} and \rfn{DeltaPeq} as series in the proper-time $\tau$ and then use the result to find the coefficients $\pac_n$ appearing in the expansion of $\pa(w)$. The~analysis of this series then proceeds as in the case of BRSSS theory and leads to the same conclusion, namely that anisotropic hydrodynamics contains a single purely decaying non-hydrodynamic mode with the decay rate set by the relaxation time. This is the outcome expected on the grounds that the order of the boost-invariant evolution equations is the same as in the case of BRSSS (unlike the HJSW theory discussed in the following section).

\subsubsection{HJSW theory}
\label{sect:resu:hjsw}

It is interesting to consider the large order behaviour of the gradient expansion of the simplest hydrodynamic
theory which avoids acausality by extending Navier-Stokes theory not by a single, purely decaying mode (as
MIS, BRSSS and AHYDRO do), but by a pair of conjugate modes. The second of the models described in
\rfs{sect:beyond} is an example of such a theory. When restricted to boost-invariant flows one arrives at
\rf{vcp}. It is easy to calculate higher order terms in the gradient series \rfn{hjsw.largew} numerically and
as one would expect, this series diverges. If we use a transseries Ansatz of the type~\rfn{trans} we find two complex conjugate values for the ``action'':
\be
\label{eq.AvsOMEGA}
A_{\pm}=\frac{3}{2}\left(\Omega_{I}\pm\mathrm{i}\Omega_{R}\right).
\ee
We then have two types of ``non-perturbative'' contributions and thus, following \rfc{Aniceto:2011nu}, we
find that we need a two-parameter transseries to fully describe solutions of this equation:
\beq
\label{transtwopar}
\pa\left(w\right)=\sum_{n_{\pm}=0}^{\infty}\sigma_{+}^{n_{+}}\,\sigma_{-}^{n_{-}}
\,\mathrm{e}^{-(n_{+}\,A_{+}+n_{-}\,A_{-})\,w}\,w^{n_{+} \, \beta_{+} + n_{-} \beta_{-}}\,\Phi_{(n_{+}|n_{-})}\left(w\right).
\eeq
In the expression above $\Phi_{(n_{+}|n_{-})}(w)$ are series expansions in $w^{-1}$ in each sector; the
``perturbative'' sector is given by taking $n_{+}=n_{-}=0$. The coefficients of these expansions were studied
in \rfc{Aniceto:2015mto} and found to satisfy very nontrivial relations following from the theory of
resurgence, similarly to the BRSSS case.
\begin{figure}
\begin{center}
\includegraphics[height=0.32\textheight]{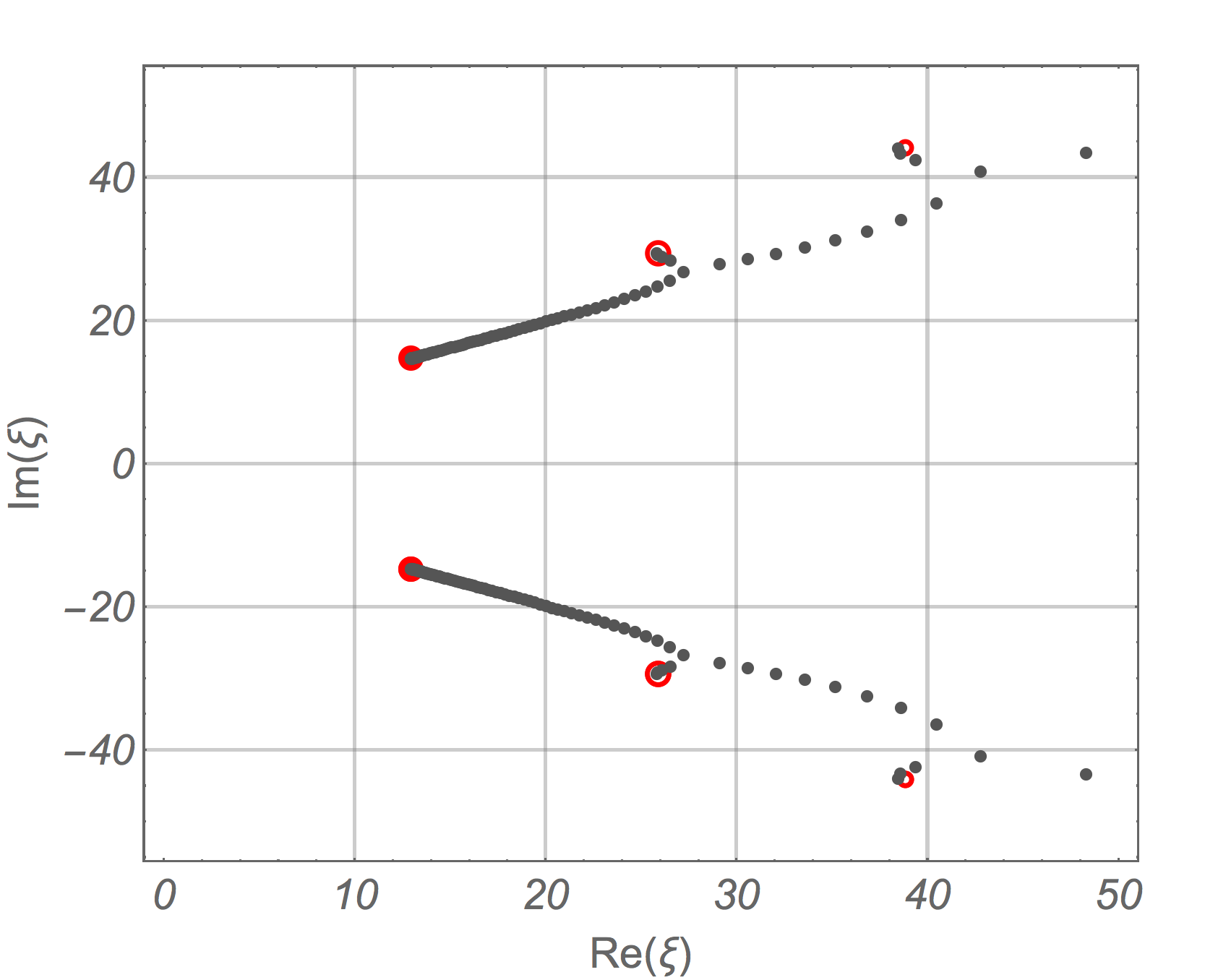}
\caption{The plot shows singularities of the approximate analytic continuation
  of the Borel transform of hydrodynamic gradient expansion in the HJSW theory
  truncated at 600 derivatives. One can clearly see that the two leading
  singularities match $3/2 \, i$ frequency of the lowest non-hydrodynamic
  excitation in the model at vanishing momentum $\vk$ and unit temperature $T$
  (solid red dots). The new clearly visible feature as compared to the BRSSS
  case depicted in Fig.~\ref{fig:padecut} is the presence of additional
  singularities starting at positive integer multiples of the leading branch
  points (empty red dots denoting the first two multiples). They necessarily
  follow from the nonlinear nature of the underlying equation and are
  reflected in the structure of transseries ansatz. Following this logic, they
  also exist in the BRSSS theory, but are impossible to locate in
  Fig.~\ref{fig:padecut} since the relevant branch-cuts partially~overlap.}
\label{fig.BorelPolesHJSW}
\end{center}
\end{figure}

The presence of two transseries parameters is, of course, related to the fact that now the evolution equation for $\pa(w)$ is second order and requires specifying two initial conditions. This structure can be viewed as a model for the nonhydrodynamic sector found previously in the holographic study of hydrodynamic gradient expansion at strong coupling~\cite{Heller:2013fn}, which is the subject of \rfs{sect:resu:holo}.

\subsection{Large order gradient expansion from microscopic models}
\label{sect:resu:micro}

\subsubsection{Holographic CFTs}
\label{sect:resu:holo}

The holographic dual to the hydrodynamic gradient expansion for Bjorken flow is a special case of fluid-gravity
duality~\cite{Bhattacharyya:2008jc} in which $T$ becomes a function of $\tau$ only and symmetry dictates $u^{\mu} \,
\partial_{\mu} = \partial_{\tau}$. This finding leads to the following metric ansatz for its gravity
dual~\cite{Heller:2008mb,Kinoshita:2008dq,Kinoshita:2009dx}
\be
\label{eq.ds2bif}
ds^{2} = \frac{L^{2}}{u^{2}} \left\{ - 2 d\tau du - q \, d\tau^{2} + (\tau + u)^{2} e^{\frac{2}{3} d - 2 b}\, dY^{2} +
e^{\frac{2}{3} d - b} \,\left( \left(dx^{1}\right)^{2} + \left(dx^{2}\right)^{2} \right) \right\}, 
\ee
where the off-diagonal bulk metric element implies the choice of the so-called generalized ingoing
Eddington-Finkelstein coordinates\footnote{As a result, the bulk coordinates are distinct from the ones used
in~\rf{eq.AdS} despite being denoted by the same symbols.} and the metric components $q$, $b$ and $d$ are
functions of $\tau$ and $u$. As opposed to the original work~\cite{Heller:2013fn}, and in the spirit of the
fluid-gravity duality, we will repackage this dependence as a dependence on $\tau \, T(\tau)$ and $u \,
T(\tau)$ with $T(\tau)$ being the effective temperature from~\rf{efftemp}. The former is the familiar
$w$-variable encountered in Sect.~\ref{sect:hydrod} which casts the hydrodynamic gradient expansion as the
Taylor series around $w = \infty$. The latter quantity, $u \, T(\tau) \equiv \rho$, roughly measures the
radial  energy scale $u$ in the units of local temperature $T(\tau)$. One can superficially understand this
parametrisation as a statement that in the hydrodynamic regime we measure all dimensionful parameters with
respect to the energy density or, equivalently, effective temperature.

The key idea is to seek for the bulk metric in the gradient-expanded form
\be
q (w, \rho)= q_{0} (\rho) + \frac{1}{w} q_{1} (\rho) + \frac{1}{w^2} q_{2} (\rho) + \ldots
\ee
where the ellipsis denotes further terms with three or more derivatives. Analogous expressions hold also for $b$ and
$d$. As shown originally in Ref.~\cite{Heller:2013fn}, it is possible to solve Einstein's equations (\ref{eq.Einstein})
in a semi-analytic manner up to a very high order in this large-$w$ expansion and infer from it the large order
behaviour of the hydrodynamic derivative expansion of the normalised pressure anisotropy $\pa(w)$ of strongly coupled
expanding plasma\footnote{Ref.~\cite{Heller:2013fn} parametrizes the problem differently, which, at least superficially,
  appears to lead to a more efficient algorithm for solving the relevant equations. The implementation presented here,
  which we introduce especially for the needs of the present review, is conceptually nicer, but most likely
  computationally slower.}. We will not discuss here the detailed procedure and its implementation, but, instead, we
want to indicate the following crucial ideas behind it. The starting point for the whole analysis is the metric
describing the Bjorken perfect fluid solution:
\bel{eq.holobifLO}
q_{0} (\rho) = 1 - \pi^{4} \, \rho^4 \quad \mathrm{and} \quad b_{0}(\rho) = d_{0}(\rho) = 0.
\eel
This solution has a horizon at $\rho = \frac{1}{\pi}$. Higher order corrections can now be calculated by
solving at each order in the large-$w$ expansion three linear second-order ordinary differential equations
outside the horizon. The remaining two Einstein's equations turn out to be trivially obeyed. Lower order
solutions (e.g. zeroth order when solving for the first order solution) appear in the source terms and as a result the solution at a given order is determined by boundary conditions at $\rho = 0$ and $\rho =
\frac{1}{\pi}$. As discussed in \rfs{sect:adscft}, the physical solution should not blow up outside the
horizon, which can be taken care of by representing functions by sums of orthonormal polynomials bounded on
this domain, see e.g. Refs.~\cite{Grandclement:2007sb,Chesler:2013lia} for reviews of these methods in the context of general relativity.

All the \emph{physical} boundary conditions can be specified at $\rho = 0$ using \rf{eq.nearbdry}. First, we
wish to impose that the plasma evolves in flat Minkowski space. Second, the leading order expression given by
\rf{eq.holobifLO} through the definition of $\rho$ and \rf{eq.nearbdry} already accounts for the full
dependence of the energy density on the effective temperature $T(\tau)$. As a result, when calculating higher
order corrections, we need to ensure, through the use of \rf{eq.nearbdry}, that this result remains intact.
This corresponds to the Landau matching condition in relativistic hydrodynamics. Let us point out, as it will
become apparent to the reader after reading the next section, that the above way of phrasing the gradient
expansion on the gravity side can be very closely mimicked within kinetic theory. This allows us to build up a
very close parallel between these two a priori very distinct languages of describing collective systems.

\begin{figure}
\begin{center}
\includegraphics[height=0.32\textheight]{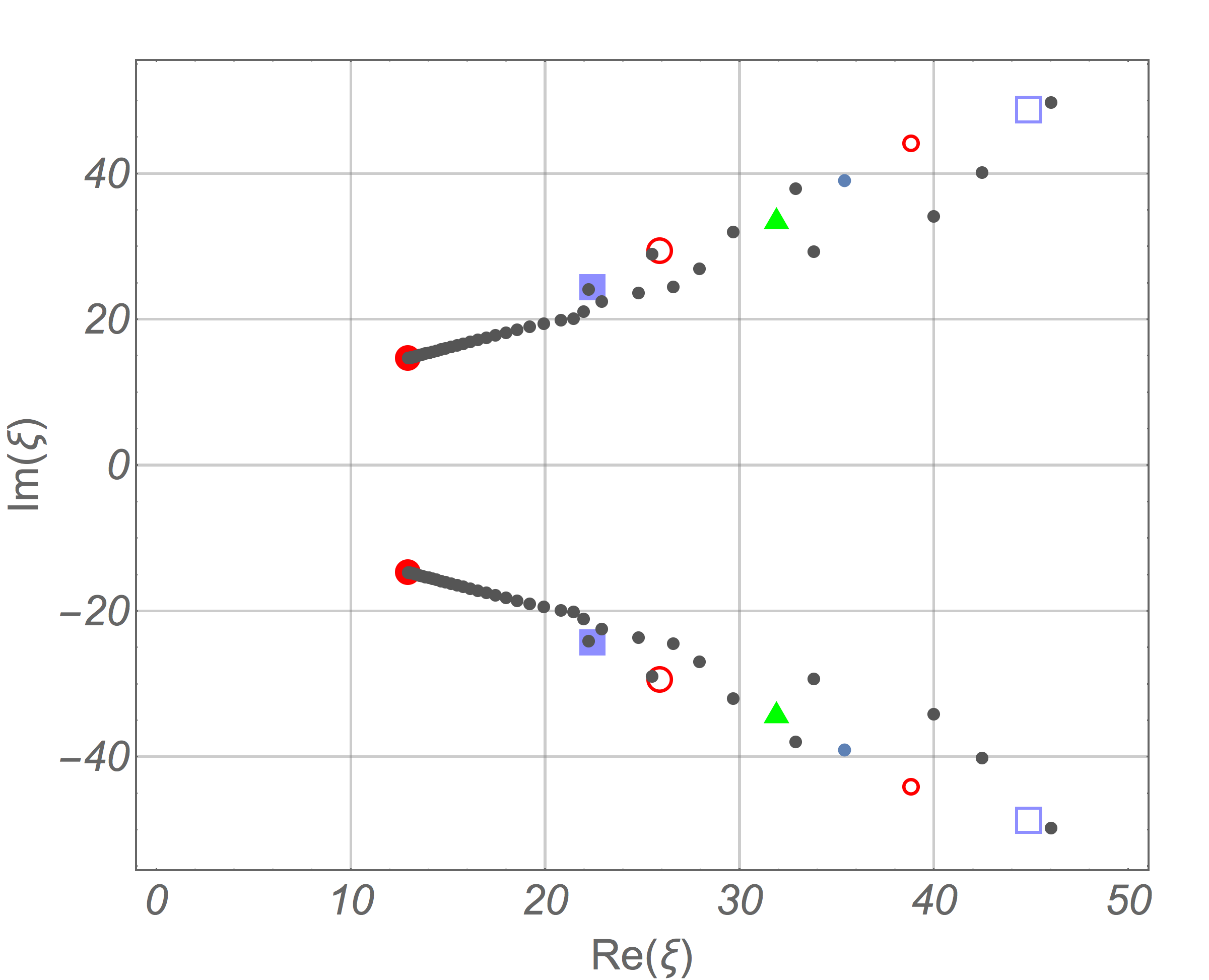}
\caption{In this plot, grey dots denote singularities of the approximate
  analytic continuation of the Borel transform, given by \rf{borel} with
  $\beta = 1$, of the hydrodynamic gradient expansion in a hCFT such as ${\cal
    N} = 4$~SYM truncated at 240 derivatives obtained in
  Ref.~\cite{Heller:2013fn}. Big red dot, pale blue square and green triangle
  correspond to $3/2 \, i$ times $\omega/T$ for the three least damped
  transient QNMs at vanishing momentum $\vk$, see
  \rfs{sect:qnm}. Corresponding empty shapes denote integer multiples of
  relevant frequencies. One can clearly see that the position of the leading
  singularity -- a pair of branch points -- coincides with what would be given
  by the lowest non-hydrodynamic mode frequency, as in the studies of
  hydrodynamic theories. Quite remarkably, one also sees singularities lying
  very close to the positions corresponding to frequencies of the second
  transient mode and the doubled frequency of the first transient mode. One
  expects that upon increasing the number of orders and accuracy of the
  calculation, one would see the associated branch cuts, as well as signatures
  of higher QNMs. This points, quite emphatically, to a transseries
  structure. Note that some of the poles in the plot do not correspond to
  physical singularities and are artefacts of the approximate analytic
  continuation. Note also that one should see interference effects between
  different modes, a feature absent in the HJSW theory, but we believe that
  the accuracy of the analytic continuation is not high enough for them to
  appear.}
\label{fig.borelNeq4}
\end{center}
\end{figure}

The outcome of analogous calculations in \rfc{Heller:2013fn}, using a faster numerical implementation, is the form of the normalized pressure anisotropy $\pa(w)$ for strongly-coupled plasma up to terms having 240 derivatives  of the fluid variables. The coefficients, again, exhibit factorial growth and, as can be seen in \rff{fig.borelNeq4}, they give rise to very intricate singularity structure in the Borel plane. The new feature, as compared with hydrodynamic theories reviewed to date, is the presence of multiple non-hydrodynamic excitations (see also Ref.~\cite{Buchel:2016cbj} for a sharp manifestation of their presence). By inverting the Borel transform along two inequivalent contours one obtains the following large-$w$ contribution associated with the two leading singularities:
\be
\delta \pa \sim \mathrm{e}^{- A^{(1)}_{\pm} \,w}\,w^{\beta^{(1)}_{\pm}},
\ee
where $A^{(1)} = -14.7003 \pm 12.9435 \, i$ and $\beta^{(1)} = 0.6867 \pm 0.7799 \, i$. In the original \rfc{Heller:2013fn} these values were matched to the behaviour of the lowest transient QNM evaluated in the Bjorken background, in complete agreement with the overall picture advocated in this review. The presence of other transient QNMs suggests the following form of the transseries~\cite{Heller:2016gbp}
\begin{eqnarray}
\pa\left(w\right)=\sum_{n^{(1)}_{\pm},n^{(2)}_{\pm}, \ldots=0}^{\infty} &&\, \Phi_{\left(n^{(1)}_{+}|n^{(1)}_{-} | n^{(2)}_{+}|n^{(2)}_{-} | \ldots \right)}\left(w\right) \times \nonumber \\
&& \times  \prod_{j =1}^{\infty} \left(\sigma^{(j)}_{+}\right)^{n^{(j)}_{+}}\,\left(\sigma^{(j)}_{-}\right)^{n^{(j)}_{-}} \,
\mathrm{e}^{-\left(n^{(j)}_{+}\,A^{(j)}_{+}+n^{(j)}_{-}\,A^{(j)}_{-}\right)\,w}\,w^{n^{(j)}_{+} \, \beta^{(j)}_{+} + n^{(j)}_{-} \beta^{(j)}_{-}}, \quad \quad
\label{eq:trans-series-infty-param}
\end{eqnarray}
which generalizes \rf{transtwopar} to a case in which infinitely many independent modes are present. Again, the ``perturbative'' series considered in Ref.~\cite{Heller:2013fn} is represented by $\Phi_{\left(0|0|0|0| \ldots \right)}$ and ``actions'' $A^{(j)}_{\pm}$ represent frequencies of subsequent QNMs at vanishing momentum $\vk$, as in \rf{eq.AvsOMEGA}. Different products present in \rf{eq:trans-series-infty-param} correspond to interactions between modes triggered by nonlinearities of Einstein's equations~(\ref{eq.Einstein}).

One can intuitively understand the proposal in \rf{eq:trans-series-infty-param} in the following way. BRSSS
and AHYDRO theories give rise to first order ODEs for $\pa(w)$, which require providing one real number as an
initial condition. They also have one transient mode and, through the transseries, there is a relation between
the transseries parameter $\sigma$ and a relevant initial condition. This is illustrated in Fig.~\ref{fig.all}
in the context of the attractor solution of BRSSS theory. The HJSW theory gives rise to a second order ODE for
$\pa(w)$, which requires providing two real numbers as an initial condition, and there is one pair of
oscillatory transient modes. This leads to a resurgent transseries which has two parameters. In holography,
Einstein's  equations~(\ref{eq.Einstein}) for the bulk metric Ansatz~(\ref{eq.ds2bif}) are second order PDEs
in the  $\rho$ and $w$ variables and require specifying a function of $\rho$ (or, in practical calculations, a
function of $u$) in order to solve the initial value
problem~\cite{Beuf:2009cx,Chesler:2009cy,Heller:2012je,Jankowski:2014lna}. Such a function contains infinitely
many real parameters. At the level of the energy-momentum tensor of the hCFT this freedom manifests itself in
its early time dynamics through \rf{epstau}, since the ${\cal E}_{n}$ from this equation are in one-to-one
correspondence with the form of the near-boundary expansion of the initial condition~\cite{Beuf:2009cx}. In
the late-time dynamics, the same feature manifests itself through the presence of infinitely many modes in
equilibrium, see \rfs{sect:qnm}. At least superficially, one can therefore think of Einstein's equations as
equivalent to some tentative ODE of infinite order for~$\pa(w)$. It remains to be seen if this perspective will
bring further insights.

Despite the complex situation created by the infinite sequence of QNMs it is interesting to explicitly perform the Borel
summation of the hydrodynamic gradient series of \symm\ and compare it to numerical solutions of actual flows obtained
using AdS/CFT~\cite{Spalinski:2017mel}.
\begin{figure}
  \begin{center}
    \includegraphics[height=0.27\textheight]{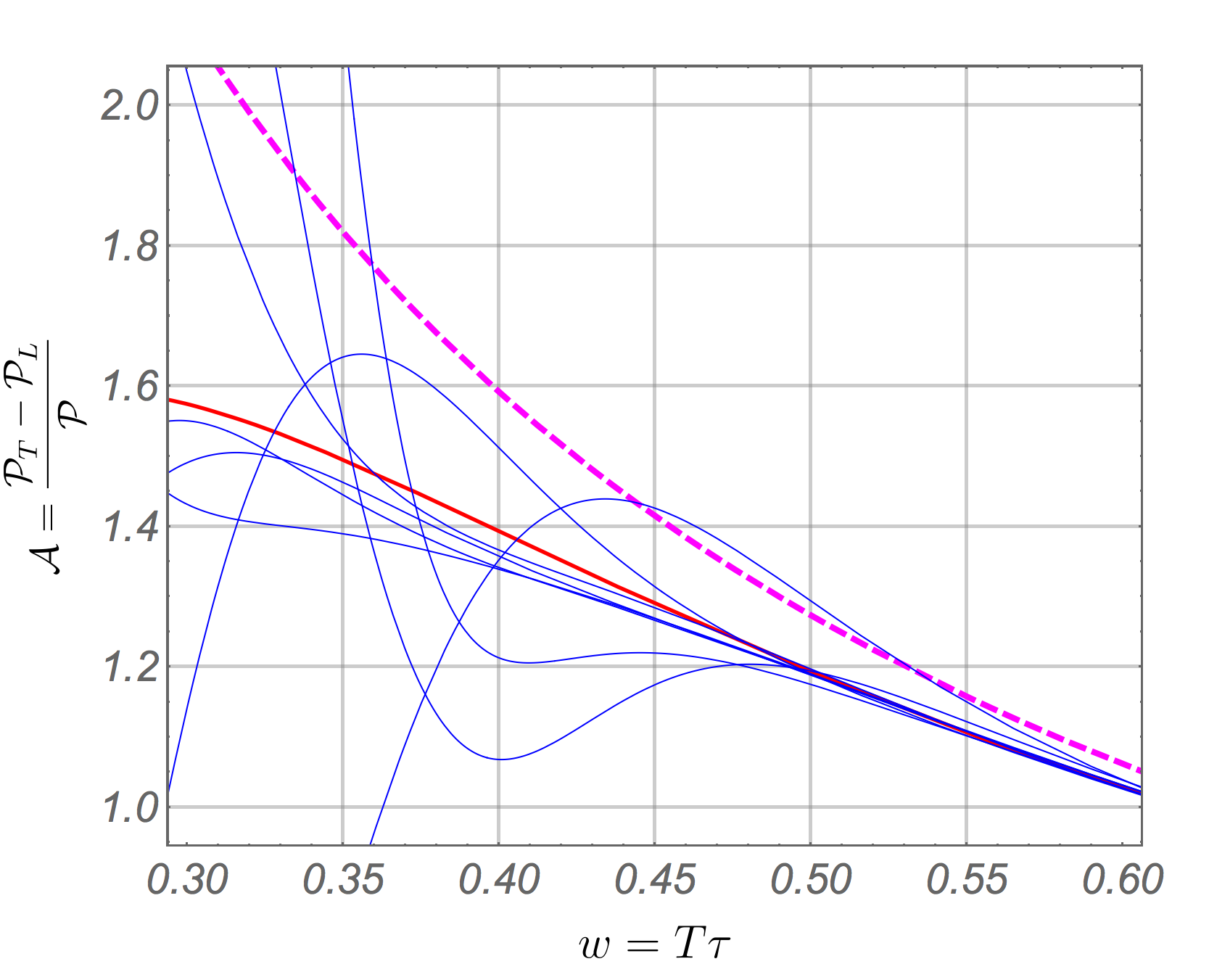}
    \includegraphics[height=0.27\textheight]{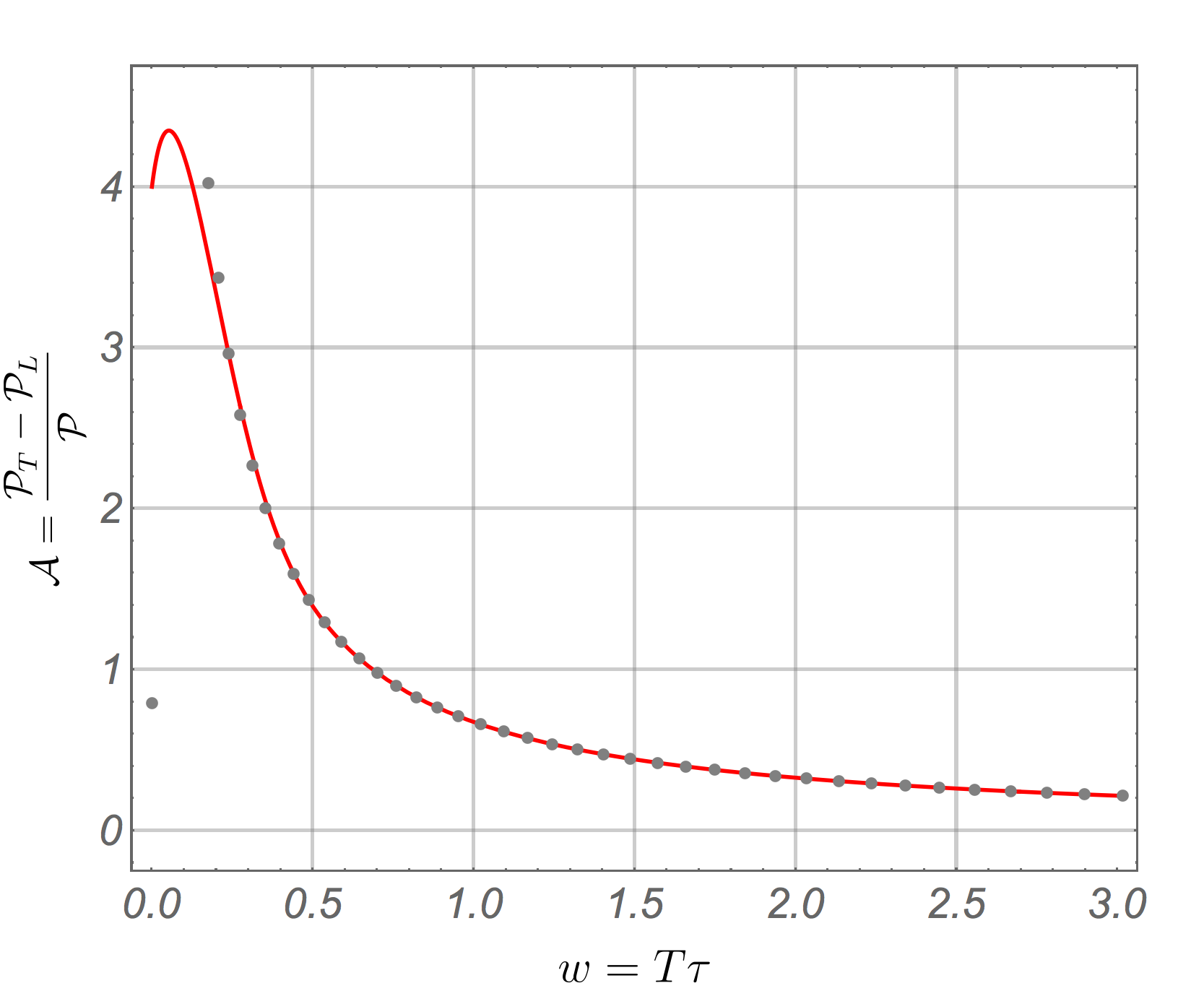}    
\caption{The left plot shows the attractor of \symm\ plasma obtained by Borel summation (the red curve), along with
  results of numerical simulations using AdS/CFT (blue curves); the dashed magenta curve represents the hydrodynamic gradient
  expansion truncated at first order. The right plot shows the attractor in HJSW theory, calculated numerically from the
  hydrodynamic equations (red curve), compared with the results of Borel summation (gray dots).  }
\label{fig:attrSYM}
\end{center}
\end{figure}
The result of the summation is shown in \rff{fig:attrSYM} (left). Note that it acts as an attractor
for the numerically calculated histories, which clearly decay to it, rather than to the truncated gradient
expansion. This behavior is very similar to the attractor found in BRSSS hydrodynamics (see \rfs{sect:brsss}).
One's confidence in this procedure is strengthened by the analogous exercise in HJSW theory, where (as shown on
the right plot in \rff{fig:attrSYM}) the result of the Borel summation matches the numerically determined attractor
rather well for $w>0.3$~\cite{Spalinski:2017mel}. This is encouraging, because the leading singularities on the Borel
plane are the same in both cases.

\subsubsection{RTA kinetic theory}
\label{sect:resu:rta}

At the moment of writing this review, the latest developments on the large-order behaviour of the hydrodynamic gradient expansion concern the RTA kinetic theory. We will review here the conformally-invariant case analysed in Ref.~\cite{Heller:2016rtz} with the aim to draw parallels with holography and present the Borel plane analysis of the gradient expansion as a novel tool in diagnosing excitations of time-dependent systems. It should be stressed that the asymptotic character of the gradient expansion in the RTA kinetic theory has been also observed ~in~Ref.~\cite{Denicol:2016bjh} in a non-conformal model with a constant relaxation time~$\trelm$. See also Ref.~\cite{Blaizot:2017lht} in this context.

To obtain the solution of the RTA Boltzmann equation~\rfn{bebif} at high orders of the gradient expansion it is far more efficient to proceed with the wisdom of hindsight and calculate the anisotropy directly as a function of the dimensionless evolution parameter~\rfn{wdef}, rather than as it was presented in \rfs{sect:bifrta:grad} or in the original article~\cite{Heller:2016rtz}. It is convenient to view the distribution function as dependent on the boost-invariant dimensionless variable $w$, as well as on
\be
\upsilon \equiv \frac{U\cdot p}{T} \quad \mathrm{and} \quad \psi \equiv \frac{\left(p^0\right)^{2}-\left(p^{3}\right)^2}{T^2},
\ee
i.e. to consider~$f(w,\upsilon,\psi)$. To derive the requisite form of the
Boltzmann equation we start with \rf{rta} with the relaxation time set as in
\rf{conf-teq}. One has to view the new variables
$w$, $\upsilon$ and $\psi$ as functions of the boost-invariant combination
$\sqrt{\left(x^{0}\right)^2-\left(x^{3}\right)^2}$ of Cartesian
coordinates~$x^{0}$~and~$x^{3}$. After the differentiation is carried out we
can, appealing to boost-invariance \cite{Baym:1984np}, set $x^{3}=0$ and
$t=\tau$ to obtain
\bel{rtanewvars}
 \left(\f{2}{3} +\f{1}{18} \pa\right)\p_w f +
\left(\f{1}{w}\f{\psi}{\upsilon} - \left(\f{2}{3} +\f{1}{18} \pa\right) \f{\upsilon}{w}\right) \p_\upsilon f +
\left(\f{2}{3} - \f{1}{9} \pa\right) \f{\psi}{w} \p_\psi f = \f{e^{-\upsilon} - f}{\gamma} \,,
\eel
where derivatives of the temperature $T$ have been eliminated in favour of the
pressure anisotropy $\pa$ defined in~\rf{rdef}. The key idea now is to look
for $\pa(w)$ and $f(w, \upsilon, \psi)$ in the gradient-expanded form,
i.e. use the ansatz for $\pa(w)$ given by \rf{rgradex} and the following one
for $f(w, \upsilon, \psi)$:
\begin{equation}
\label{eq.gradexpfRTA}
f(w, \upsilon, \psi) = f_{0} (\upsilon,\psi) + \frac{1}{w} f_{1} (\upsilon,\psi) + \frac{1}{w^2} f_{2} (\upsilon,\psi) + \ldots
\end{equation}
The leading term in the above equation is the equilibrium distribution function, which in this parametrisation reads simply
\be
f_{0}(\upsilon, \psi) = e^{- \upsilon}.
\ee
As it turns out, one can now iteratively solve the RTA Boltzmann equation~(\ref{rtanewvars}) which at each order $n$ gives a linear algebraic relation for $f_{n} (\upsilon, \psi)$ with the result depending on $\pac_{n}$. Imposing now the Landau matching condition, i.e. demanding that a given $f_{n} (\upsilon, \psi)$ does not contribute to the local energy density of plasma for $n > 0$ fixes the corresponding contribution $\pac_{n}$ to the gradient expansion of the pressure anisotropy~$\pa(w)$. In~Ref.~\cite{Heller:2016rtz} an analysis of this type has been performed up to terms having $200$ derivatives revealing vanishing radius of convergence.

Let us now point out a nice analogy between the distribution function in kinetic theory and bulk metric in holography that can be seen here in the context of the boost-invariant flow. In both cases the microscopic dynamics is captured by equations of motion in higher number of dimensions, albeit of a very different mathematical nature. In the context of holography, the additional variable has an interpretation of the energy scale in a hQFT, see \rfs{sect:adscft}. In the context of kinetic theory, there are three additional variables that represent on-shell particle momenta, see \rfs{sect:kt}. Both in the holographic approach considered in the previous section and in the kinetic theory approach  of this section we measure these additional variables in units of the effective temperature, which sets the characteristic near-equilibrium scale in conformal theories. After discarding singular solutions in holography, which in~\rfs{sect:resu:holo} was achieved by using an appropriate numerical representation of bulk metric components, similarly to kinetic theory setup the Landau matching condition fixes the transport coefficients. This analogy, very much inspired by the fluid-gravity duality~\cite{Bhattacharyya:2008jc} discussed in \rfs{sect:matching}, should naturally extend to general hydrodynamic flows.
\begin{figure}
\begin{center}
\includegraphics[height=0.32\textheight]{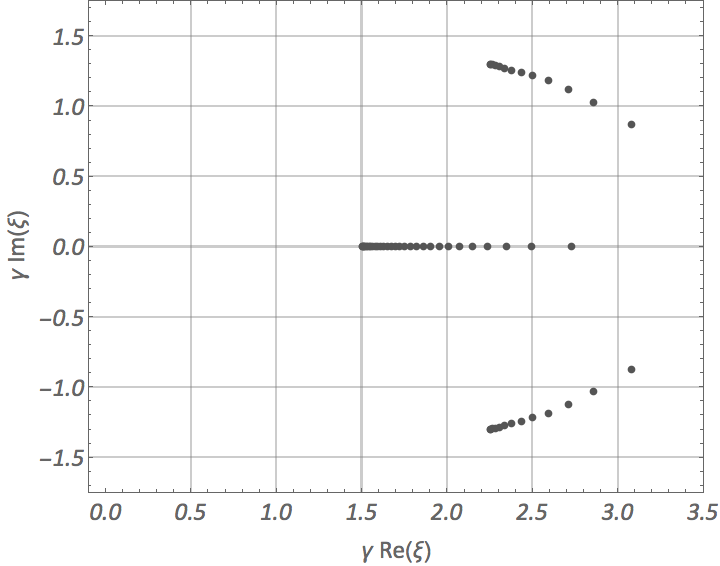}
\caption{The plot shows singularities of the approximate analytic continuation
  of the Borel transform of hydrodynamic gradient expansion in the conformal
  RTA kinetic theory. The beginning of the leading singularity lying on the
  real axis is in an excellent agreement with $3/2 \, i$ times the frequency
  of the transient excitation at vanishing momentum. The major puzzle at the
  moment of writing this review is the presence of other singularities lying
  off real axis and beginning at $\gamma \, A_{\pm} \approx 2.25 \pm 1.3 \,
  i$, since they cannot be matched in a straightforward way with equilibrium
  modes of the model, c.f. \rfs{sect:RTAmodes}.}
\label{fig.borelRTA}
\end{center}
\end{figure}

Applying the Borel transform technique and appropriate analytic continuation to the large-order gradient series in the
conformal RTA theory reveals the structure of singularities displayed in \rff{fig.borelRTA}. One can see there a branch
point singularity located at $3/2\, \gamma^{-1} $, where $\gamma$ is the dimensionless proportionality constant setting
the scaling of the relaxation time $\trelm$ with temperature in conformal RTA, see~\rf{conf-teq}. Following the
intuition gained from the analysis of hydrodynamic theories reviewed earlier in this section, one should seek the origin
of this singularity in terms of the single non-hydrodynamic mode present in the RTA kinetic theory. One subtlety that
needs to be stressed is that in contrast with all the other known models, this mode is a pole only at vanishing $\vk =
0$. Otherwise, it is represented in the retarded two-point function of the energy-momentum tensor as a branch-cut
singularity, see \rfs{sect:RTAmodes}. The consequence of this difference is not understood at the moment of writing the
present review. An even more surprising outcome of the analysis in Ref.~\cite{Heller:2016rtz} is the presence of two
symmetric singularities lying off the real axis on the Borel plane, see Fig.~\ref{fig.borelRTA}. Following the intuition
developed in other studies, one would be tempted to interpret them as a signature of oscillatory modes present in the
expanding plasma of the conformal RTA kinetic theory. The puzzle comes from the fact that no obvious equilibrium mode
can be matched to this singularity and, perhaps, it is an excitation that originates from the underlying expansion of
the system. Somewhat similar in spirit ``emergent'' modes were discussed earlier in Ref.~\cite{Bazow:2015zca} in the
context of kinetic theory with a different, more complicated collisional kernel.

Finally, the last point we wish to bring to the readers' attention is the
question of initial conditions for the kinetic theory and the number of
singularities in the Borel plane. In the hydrodynamic theories reviewed
earlier in this section, a one-to-one correspondence was found between the
number of initial conditions for $\pa(w)$ and the number of singularities in
the Borel plane. The latter was found to match the number of nonhydrodynamic
modes in the system. In the present case, somewhat analogously to holography,
the initial state in kinetic theory is encoded in the initial distribution
function which contains infinitely many parameters. The universal hydrodynamic
gradient expansion does not know about them and one should expect this
information to enter through ``non-perturbative'' transient effects associated
with infinitely many singularities on the Borel plane. This puzzling issue is
currently under investigation~\cite{HellerSvensson}.

\subsubsection{Significance of the asymptotic character of hydrodynamic gradient expansion}
\label{sect:resu:significance}

The results reviewed in this section in the context of one-dimensionally expanding systems in diverse models and
theories point unequivocally to the asymptotic character of hydrodynamic gradient expansion for generic flows as well as
other models (not necessarily conformally-invariant). The phenomenological relevance of this finding stems from the fact
that it makes the phenomenon of hydrodynamization, reviewed in Sec.~\ref{sect:hydrod}, mathematically less
surprising. Since the gradient series is not convergent anyway, smallness of subsequent terms is not a criterion for the
applicability of hydrodynamics, which is rather determined by the separation of the nonhydrodynamic modes whose presence
is -- as we have seen -- intimately connected with the divergence of the gradient series.

%% file: sect-out.tex
\section{Summary and Outlook}
\label{sect:outlook}

\subsection{Key lessons}

The overarching theme of our review is broadly-understood hydrodynamic model
building given insights from {\em ab initio} studies of time-dependent
processes in microscopic frameworks: holography and relativistic kinetic
theory. The first key notion here is the idea of modes of equilibrium plasma
reviewed in Secs.~\ref{sect:micro} and~\ref{sect:qnm}. There are two kinds of
modes: transient ones and the ones which can be made long-lived by lowering
their spatial variations. The latter are called hydrodynamic.

The second fundamental idea for our presentation is the notion of universal dynamics at late times described
in~\rfs{sect:hydrod}. This universality lies in our ability to parametrize the dynamics by a smaller number of functions
than the number of independent components of the expectation value of the energy-momentum tensor. The key to doing this
is the gradient expansion. Bringing in the mode picture, the gradient expansion corresponds to a ``condensation'' of
hydrodynamic modes and this why it is called the hydrodynamic gradient expansion (see \rfs{sect:fund}).

A very surprising feature which becomes apparent when comparing {\em ab initio} solutions for the expectation value of the
energy-momentum tensor with its predicted form within the hydrodynamic gradient expansion is that the latter, when
truncated keeping only the first few orders, can perform very well despite the fact that the leading corrections are
very large. As reviewed in \rfs{sect:hydrod}, this gives rise to the phenomenon of hydrodynamization: the violation of
approximate local thermal equilibrium in the hydrodynamic regime. This means that the solutions of hydrodynamic
equations can often be trusted even if they describe highly nonequilibrium states. The phenomenological relevance of
this finding lies in providing an explanation for the success of hydrodynamic modelling in ultrarelativistic heavy-ion
collisions at RHIC and LHC where spatiotemporal variations and momentum anisotropies are very large
initially~\cite{Strickland:2014pga} and, even more so, in the context of small systems (proton-nuclei and
high-multiplicity proton-proton collisions)~\cite{Bozek:2011if,Habich:2015rtj}.

The fourth key idea (covered in Sections~\ref{sect:effective}
and~\ref{sect:kint}) is to focus on dissipative hydrodynamic theories with a
well-posed initial value problem, which are or can be used in modeling a part
of the evolution of nuclear matter at RHIC and LHC. The feature that we wish 
to emphasize is that each of these models necessarily comes with a set of
transient modes which act as a regulator ensuring
causality~\cite{Spalinski:2016fnj}. Depending on the microscopic dynamics and
phenomenological context one can construct effective theories of hydrodynamics
which try to match the truncated gradient expansion of an underlying theory,
but one may also try to capture some features of the non-hydrodynamic sector.

Finally, in \rfs{sect:resu} we reviewed recent developments demonstrating the asymptotic character of the hydrodynamic
gradient expansion and the subtle interplay between the long-lived and the transient modes. This finding can be seen as
a mathematical reason behind hydrodynamization.

\subsection{Open directions}

The developments presented here are very recent and open many promising directions for future research. Below we mention
four among the ones that we find particularly exciting.

The fundamental notion advocated in this review are the modes of equilibrium systems (see Secs.~\ref{sect:micro}
and~\ref{sect:qnm}). Whereas hydrodynamic modelling in the context of ultrarelativistic heavy-ion collisions provides us
with access to some information about the hydrodynamic sector, one can safely say that nothing is known about the
transient modes. It would be very interesting to investigate them in detail, also in the context of other systems. One
should mention here a recent effort in this direction using data describing damped oscillations of trapped unitary Fermi
gases~\cite{Brewer:2015ipa}.

On a more theoretical side, it would be very illuminating to understand more comprehensively the properties of modes in
relativistic kinetic theory, since the only study available to date and reviewed in \rfs{sect:RTAmodes} concerns perhaps
the simplest available collisional kernel. Related to this are studies of higher curvature corrections in holography and
their influence on the spectrum of quasinormal modes, see~\rfs{sect:qnm}. These developments fall into a broader class
of efforts to bridge strong and weak-coupling results in QFTs, see
e.g.~Refs.~\cite{Keegan:2015avk,Heller:2016rtz,Grozdanov:2016zjj}.

Another promising avenue of research has to do with the breaking of conformal
symmetry. For example, in the context of hydrodynamization, it has been
observed that the bulk viscous term can give a very large contribution to the
energy-momentum
tensor~\cite{Attems:2016tby}\footnote{Ref.~\cite{Attems:2016tby} describes
  this phenomenon as an instance of ``EoSization'', because at sufficiently
  early times the average pressure in the local rest frame is not related to
  the local energy density by the (equilibrium) equation of state. If this can
  be explained at the level of hydrodynamics by the presence of a bulk viscous
  contribution one may view it as another instance of hydrodynamization.}. It
would be interesting to investigate it more comprehensively, also in the
context of second order transport which is very diverse in non-conformal
systems. Furthermore, in our analysis of systems expanding along
one dimension (reviewed in \rfs{sect:hydrod}) a lot of mileage was gained by
considering the pressure anisotropy $\pa$ as a function of the clock variable
$w$. In particular, it is this parametrization that allows one to see the
attractor solution. Trying to understand if a similar construction can be
found when conformal symmetry is broken is certainly a very tempting question
to ask.

In this review we pursued the perspective of attractor solutions for conformal boost-invariant flow seen in various
theories or microscopic models as a notion of hydrodynamics beyond the gradient
expansion~\cite{Heller:2015dha,Romatschke:2017vte,Spalinski:2017mel}.  If this set of ideas is to develop, such
attractor solutions should be found in the absence of conformal symmetry and for less symmetric flows than the Bjorken
expansion -- the slow roll method applied in the context of hydrodynamics in Ref.~\cite{Heller:2015dha} and in kinetic
theory in \rfc{Romatschke:2017vte} may be a good starting point.  It may be that attractor behaviour has already been
observed in holographic studies of planar shockwave collisions~\cite{Chesler:2013lia,Chesler:2015fpa}. For such
non-boost-invariant expanding plasma systems it has been found that at late times the local velocity profile approaches
that of Bjorken flow, and the local energy density profile tends to a universal form. Finally, it is clearly important
to understand the phenomenological utility of attractors far from local equilibrium keeping in mind that they are
different in different theories.

\subsection{Closing words}

The past 15 years, the golden age of relativistic hydrodynamics, brought
numerous insights on how the hydrodynamic regime emerges from microscopic
theories, many of which this review covered in detail. This set of
developments now constitutes a mature discipline, but one cannot escape the
impression that our understanding is very much model-based and will be
superseded by a more comprehensive picture in the future. We hope that our
review will help to inspire fellow researchers to further contribute to
understanding non-equilibrium QFTs in general and QCD in particular.

%% file: sect-app.tex

\begin{appendices}

\label{sect:app}

\section{Acronyms}
\label{app:acro}

\begin{table}[H]
\begin{center}
  \begin{tabular}{ c  c  }
    \hline
    \\
     RHIC & Relativistic Heavy-Ion Collider at Brookhaven National Laboratory \\
     LHC & Large Hadron Collider at CERN \\ \\
     QCD & quantum chromodynamics  \\
     YM  & Yang-Mills \\
     SYM & supersymmetric Yang-Mills  \\
     QGP & quark-gluon plasma \\
     EOS & equation of state \\
     CGC & color glass condensate \\ \\
     (h)QFT & (holographic) quantum field theory \\
     (h)CFT & (holographic) conformal field theory \\
     AdS & anti-de Sitter (spacetime) \\
     AdS/CFT & anti-de Sitter/conformal field theory (correspondence) \\
     QNM & quasinormal mode   \\ \\
     NS & Navier-Stokes (hydrodynamic equations) \\
     MIS &  M{\"u}ller-Israel-Stewart  (hydrodynamics) \\
     BRSSS & Baier-Romatschke-Son-Starinets-Stephanov (hydrodynamics) \\
     DNMR & Denicol-Molnar-Niemi-Rischke (hydrodynamics) \\
     HJSW & Heller-Janik-Spalinski-Witaszczyk (hydrodynamics) \\
     AHYDRO & anisotropic hydrodynamics \\ \\
     KT, EKT & kinetic theory, effective kinetic theory \\
     RTA & relaxation time approximation (for kinetic theory) \\
     RS & Romatschke-Strickland  (ansatz for the distribution function)\\
     RQMD & relativistic quantum molecular dynamics\\
     ODE & ordinary differential equation\\
     PDE & partial differential equation
   \\
    \hline
  \end{tabular}
\end{center}
\caption{List of acronyms}
\label{tab:acronyms}
\end{table}

\section{Notation}
\label{app:nota}

\begin{table}[H]
\begin{center}
  \begin{tabular}{ c  c  }
    \hline
    \\
    EOS & equation of state \\
    EQ & label specifying global thermal equilibrium  \\
    eq  &  label specifying local thermal equilibrium  \\
    $\Tmunu$ & energy-momentum tensor \\
    $T$  & effective temperature \\
    $\umu$ & flow vector defining the Landau hydrodynamic frame \\
    $\Dmunu$ & operator projecting on the space orthogonal to $\umu$ \\
    $\pimunu $ & shear stress tensor \\
    $\sigmamunu $ & shear flow tensor \\
    $\Pi$ & bulk pressure \\
    $\ed$ & energy density \\
    $\peq(\ed)$ & equilibrium pressure corresponding to the energy density $\ed$ \\
                               &  (functional form $\peq(\ed)$ follows form the equation of state) \\
    $\eta$ & shear viscosity \\
    $\zeta$ & bulk viscosity \\
   \\
    \hline
  \end{tabular}
\end{center}
\caption{Symbols denoting physical concepts and variables (part 1)}
\end{table}

\begin{table}[H]
\begin{center}
  \begin{tabular}{ c  c  }
    \hline
    \\
    $\Tmunua$ & leading-order energy-momentum tensor of anisotropic hydrodynamics \\
    $\ximunu$ & anisotropy tensor \\
    $\phi$ & bulk variable \\
    $\pimunut$ & modified shear stress tensor \\
    $\Pit$ & modified bulk pressure \\
   \\
    \hline
  \end{tabular}
\end{center}
\caption{Symbols and concepts used in AHYDRO}
\label{tab:ahydro}
\end{table}

\section{Conventions}
\label{app:conv}

Throughout the paper we use the natural system of units with $c=\hbar=k_B=1$, except for few places where we use explicit notation to demonstrate the dependence of physical quantities on physical constants. Three-vectors are denoted by the bold font, four-vectors are in the standard font, the dot denotes the scalar product of three- or four-vectors. The Minkowski metric is \mbox{$\eta_{\mu\nu}={\rm diag}(-1,+1,+1,+1)$}.

We use the standard parameterizations of the on-mass-shell four-momentum and spacetime coordinates of a particle,
\bel{pandx}
p^\mu &=& \left(E, p^1,p^2, \pl \right) = (E, \vp ) =
\left(m_\perp \cosh y, p^1, p^2, m_\perp \sinh y \right), \nonumber \\
x^\mu &=& \left( x^{0}, x^{1},x^{2}, x^{3} \right) =  \left( t, \, x^{1},x^{2}, z \right) = (t, \vx) = \left(\tau \cosh{Y}, x^{1}, x^{2}, \tau \sinh{Y} \right).
\eel
Here $m_T = \sqrt{m^2 + \pt^2} = \sqrt{m^2 + (p^{1})^2 + (p^{2})^2}$
is the transverse mass,\,\,$\tau=\sqrt{\left( x^{0} \right)^2 - \left( x^{3} \right)^2}$ is the (longitudinal) proper time, \,$y$ is the rapidity
\bel{rap}
y = \frac{1}{2} \ln \frac{E+\pl}{E-\pl},  \label{y}
\eel
and $Y$ is the spacetime rapidity,
%
\bel{srap}
Y = \frac{1}{2} \ln \frac{x^{0}+x^{3}}{x^{0}-x^{3}}.
\eel
The flow of matter is described by the four-vector
\bel{flow}
\umu = \gamma (1, v_1, v_2, v_3)\,, \quad
\gamma=(1-v^2)^{-1}\,, \quad U \cdot U = -1.
\eel
We note that, due to the form of the metric used, the energy of a particle in the frame connected with the fluid element moving with the four-velocity $\umu$ is $-\pdotU$. Most of other symbols and acronyms are listed in Tables \ref{tab:acronyms}--\ref{tab:ahydro}.

\section{Boost invariant hydrodynamics}
\label{app:bif}

For scalar functions of the space-time coordinates (such as the energy density $\ed(x)$, entropy density $\sd(x)$, pressure $\peq(x)$, or temperature $T(x)$) the boost invariance and transverse homogeneity imply that they depend on the longitudinal proper time $\tau = \sqrt{\left( x^{0} \right)^2 - \left( x^{3} \right)^2}$ only. For vector fields (such as the flow four-vector field $\umu(x)$) the situation is a bit
more complicated. Combining the rule for the longitudinal Lorentz
transformation of a four-vector, $U^{\prime \,\mu}(x^\prime) =
L^\mu_{\,\,\,\nu} U^\nu(x)$, with the condition of boost invariance,
$U^{\prime \,\mu}(x^\prime) = U^{\mu}(x^\prime)$, we find that the
boost-invariant flow vector has the form
\begin{equation}
\umu(x) = \left(x^{0}/\tau,0,0,x^{3}/\tau\right).
\label{U}
\end{equation}
The form (\ref{U}) has been specified also by the condition that $\umu$ is timelike and its spatial part vanishes at $x^{3}=0$ (in order to describe the flow in the center-of-mass reference frame). Note also that proper time~$\tau$ - spacetime rapidity $Y$ coordinates are curvilinear and lead to the following form of the Minkowski metric line element:
\be
ds^{2} = -d \tau^{2} + \tau^{2} dY^{2} + \left( dx^{1} \right)^{2} + \left( dx^{2} \right)^{2}.
\ee

For not dissipative systems, the entropy current is conserved. If such systems are boost invariant and transversally homogenous, this property can be expressed by the equation
\begin{equation}
\partial_\mu (\sd \umu) = \frac{d\sd(\tau)}{d\tau} + \frac{\sd(\tau)}{\tau} = 0,
\label{entcon}
\end{equation}
which has a scaling solution
\begin{equation}
\sd(\tau) = \frac{\sd_0 \tau_0}{\tau}.
\label{stau}
\end{equation}
Here $\sd_0$ is the entropy density at the initial proper time $\tau_0$. Equations (\ref{U})--(\ref{stau}) form the foundation of the
renowned Bjorken hydrodynamic model of heavy-ion collisions~\cite{Bjorken:1982qr}. As we have just seen, they can be introduced
as a consequence of boost-invariance and lack of dissipation.

For conformal systems in equilibrium, the entropy density scales with the third power of temperature, $\sd(\tau) \sim T^3(\tau)$, hence, we find
\begin{equation}
T(\tau) = T_0  \left( \frac{\tau_0}{\tau} \right)^{1/3}.
\label{Ttau}
\end{equation}
Similar expressions can be found for the energy density and pressure,
\begin{equation}
\ed(\tau) = \ed_0  \left( \frac{\tau_0}{\tau} \right)^{4/3}, \quad
\peq(\tau) = \peq_0  \left( \frac{\tau_0}{\tau} \right)^{4/3}.
\label{epsPtau}
\end{equation}

\section{Conformal invariance}
\label{app:weyl}

Conformal symmetry a theory is covariance of its equations of motion under Weyl scaling of the metric:
\bel{Weylrescalings}
g_{\mu \nu} \rightarrow e^{- 2\phi} g_{\mu \nu}, \quad
u^{\mu} \rightarrow e^{\phi} u^{\mu}, \quad
T \rightarrow e^{\phi} T ,
\eel
where $\phi$ depends on the coordinates $x^{\mu}$ \cite{Baier:2007ix,Loganayagam:2008is,Bhattacharyya:2008mz}. A quantity which transforms homogeneously with a factor of $e^{s \, \phi}$ is said to transform with Weyl weight $s$.

A beautiful formalism allowing for manifest Weyl covariance in conformal hydrodynamics was introduced by Ref.~\cite{Loganayagam:2008is} and applied to fluid-gravity duality in \rfc{Bhattacharyya:2008mz}. The
basic tool is the Weyl-covariant derivative $\D_{\mu}$, which preserves the Weyl weight of the differentiated tensor. It is constructed using the vector field $\WA_{\nu}$ defined by \cite{Loganayagam:2008is}
\bel{weylconn}
\WA_{\nu} \equiv U^\lambda\p_\lambda U_{\nu}- \f{\p_\lambda  U^\lambda}{3} U_{\nu} \, .
\eel
This quantity is of order one in the gradient expansion and transforms as a
connection under Weyl-transformations
\bel{weylconntrafo}
\WA_{\nu} \rightarrow \WA_{\nu} +\partial_{\nu}\phi \, .
\eel
Due to this property it can be used to compensate for derivatives of
the Weyl factor when differentiating a Weyl-covariant tensor. For instance,
one has
\bel{DT}
\D_{\mu} T &=&  \partial_{\mu} T - \WA_{\mu} T .
\eel
By adding suitable correction terms one can construct a Weyl-covariant derivative of any tensor which transforms homogeneously under Weyl scaling. The case of most interest in the context of this review is the Weyl-covariant derivative of $\pi_{\mu \nu}$, which reads
\be
\D \, \pi_{\mu \nu} = U^{\lambda} (\partial_{\lambda} + 4 \, A_{\lambda}) \, \pi_{\mu \nu} - 2 A^{\lambda} U_{(\mu}\pi_{\nu) \lambda}.
\ee
The same formula applies also to $\frac{1}{T} \D \, \pi_{\mu \nu}$, which provides therefore a recipe for decoding Eqs.~(\ref{dampedosc})~and~(\ref{eqpi2s}). For further details the reader is referred to the original literature cited above.

\end{appendices}